\documentclass[12pt,english,british]{scrartcl}
\usepackage[T1]{fontenc}
\usepackage[latin9]{inputenc}
\usepackage{babel}
\usepackage{graphicx}
\usepackage{subfigure}
\usepackage{hyperref}

\begin{document}
\url{http://www.cambridge.org/aus/catalogue/catalogue.asp?isbn=9780521149310}
\title{Prandtl and the Göttingen School %
\thanks{Draft of our Chapter for "A Voyage Through Turbulence",
edited by, P. A. Davidson, Y Kaneda, H.K. Moffatt \& K.R. Sreenivasan, Cambridge University Press, Oct. 2011, ISBN-13: 9780521149310, \href{http://www.cambridge.org/aus/catalogue/catalogue.asp?isbn=9780521149310}{http://www.cambridge.org/aus/catalogue/catalogue.asp?isbn=9780521149310}
}}

\author{Eberhard Bodenschatz and Michael Eckert}

\maketitle

\begin{figure}[h]
\begin{center}
\includegraphics[scale=0.15]{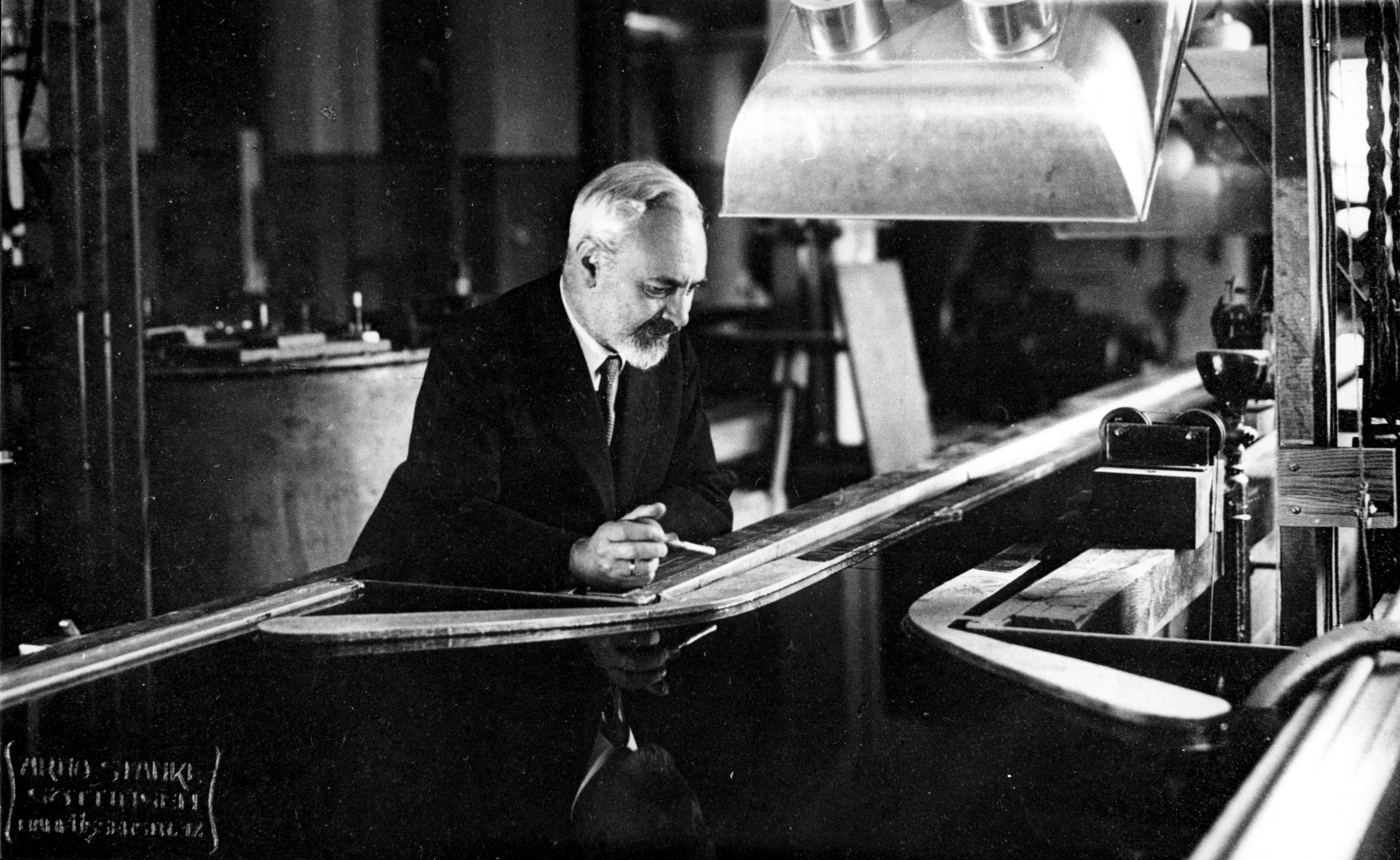}
\caption{Ludwig Prandtl at his water tunnel in the mid to late 1930s (Reproduction from the original photograph DLR: FS-258)}
\label{fig:Prandtl}
\end{center}
\end{figure}

\bigskip
\bigskip
\bigskip
\bigskip
\subsection*{Introduction}
In the early decades of the 20th century Göttingen was the center for
mathematics.   The foundations were laid by Carl Friedrich Gauss  
(1777--1855) who  from 1808 was head of the observatory and professor for
 astronomy at the Georg August University  (founded in 1737). At the turn of the 20th century, 
  the well know mathematician 
 Felix Klein (1849--1925),  who joined the university in 1886, established a research center
 and brought leading scientists to Göttingen.  In 1895 David Hilbert (1862--1943)
  became chair of mathematics and in 1902  Hermann Minkowski (1864--1909)
   joined the mathematics department.   At the time pure and applied mathematics pursued
diverging paths, and mathematicians at Technical Universities were met
with distrust from their engineering colleagues with regard to their
ability to satisfy their practical needs \cite{Hensel:1989}.  Klein was particularly
eager to demonstrate the power of mathematics
in applied fields \cite{Prandtl:1926b,Manegold:1970}. In 1905 he established  an Institute for Applied Mathematics 
and  Mechanics in Göttingen by bringing the 
young Ludwig Prandtl (1875--1953) and the more senior Carl Runge (1856--1927), both from 
the nearby Hanover.  A picture of Prandtl at his water tunnel  around 1935 is shown in Fig.~\ref{fig:Prandtl}.

Prandtl, had studied mechanical engineering at the Technische
Hochschule (TH, Technical University) in Munich in the late 1890s. In his studies he was deeply
influenced by August Föppl (1854-1924), whose
textbooks on technical mechanics became legendary. After finishing his studies as mechanical
 engineer in 1898, Prandtl
became Föppl's assistant and remained closely related to him
throughout his life, intellectually by his devotion to technical mechanics
and privately as Föppl's son-in-law \cite{Vogel-Prandtl:1993}.  
Prandtl wrote his
doctoral dissertation under Föppl's supervision on a problem of technical
mechanics ({}``Kipp-Erscheinungen, ein Fall von instabilem elastischem Gleichgewicht'',  
``On Tilting Phenomena, an Example of Unstable Elastic Equilibrium'').
He could not acquire a doctoral degree from this institution and had to perform the required
academic rituals at the philosophical faculty of the
neighbouring Ludwig Maximilian University of
Munich on January 29, 1900. At the time Technische Hochschulen fought a bitter struggle until they
were granted equal rights with the universities.  The institutional
schism affected in particular the academic disciplines at the interface
of science and engineering, such as applied mathematics and technical
mechanics \cite{Oswatitsch:1987}.     On January 1, 1900,  before receiving his doctorate, Prandtl 
started an engineering position at the Maschinenbaugesellschaft in Nuremberg, which just merged with Maschinenfabriken Augsburg to become MAN  ( Maschinen Fabrik Augsburg N\"urnberg, Machine Works of Augsburg and N\"urnberg).  At MAN he 
was first introduced to problems in fluid dynamics designing a blower. Very shortly thereafter,
he received an offer as the chair of mechanics at Hanover. He left 
Nuremberg on September 30, 1901 to become at age 26 the youngest professor
of Prussia  \cite{Vogel-Prandtl:1993}.  In 1904 Felix Klein was able to convince Prandtl
to take a non-full professor position at Göttingen 
to become head of the department of technical physics at the Institute of Physics with 
 the prospective as a co-director of a new Institute for Applied Mathematics and Mechanics. In the same vein,
Klein had arranged Runge's call to Göttingen. In autumn 1905, Klein's
institutional plans materialized. Göttingen University opened a new
Institute for Applied Mathematics and Mechanics under the common directorship
of Runge and Prandtl. Klein also involved Prandtl as director in the planning
of an extramural aerodynamic research institute, the Motorluftschiffmodell-Versuchsanstalt,
which started its operation with the first Göttingen design windtunnnel in 1907 \cite{Rotta:1990,Oswatitsch:1987}. Klein
regarded Prandtl's {}``strong power of intuition and great originality
of thought with the expertise of the engineer and the mastery of the
mathematical apparatus'' \cite[p. 232]{Manegold:1970} ideal qualities
for what he had planned to establish at Göttingen.

With these institutional measures, the stage was set for Prandtl's
unique career between science and technology--and for the foundation
of an academic school with a strong focus on basic fluid dynamics and 
their applications.
Prandtl directed the Institut für Angewandte Mechanik of Göttingen
University, the Aerodynamische Versuchsanstalt (AVA), as the rapidly
expanding Motorluftschiffmodell-Versuchsanstalt was renamed after
the First World War, and, after 1925, the associated Kaiser-Wilhelm-Institut (KWI)
für Strömungsforschung.  His ambitions and the history leading to the establishment of the KWI 
are well summarized in his opening speech at his institute, which has been translated into English \cite{Prandtl:1925E}. 

During the half century of Prandtl's Göttingen
period, from 1904 until his death, his school extended Göttingen's
fame from mathematics to applied mechanics, a specialty which acquired
in this period the status of a self-contained discipline. Prandtl
had more than eighty doctoral students, among them Heinrich Blasius, 
Theodore von Kármán, Max Munk, Johann Nikuradse, Walter Tollmien,  Hermann Schlichting, Carl Wieghardt, and others
who, like Prandtl, perceived fluid mechanics in general, and turbulence
in particular, as a paramount challenge to bridge the gulf between
theory and practice. Like Prandtl's institutional affiliations, his
approach towards turbulence reflects a broad spectrum of {}``pure''
and {}``applied'' research (if such dichotomies make sense in turbulence
research). We have to consider the circumstances and occasions in
these settings in order to better characterize the approach of the
Göttingen school on turbulence.

\subsection*{The Boundary Layer Concept, 1904--1914}

When Prandtl arrived in Göttingen in autumn 1904, he came with an
asset: the boundary layer concept \cite[chapter 2]{Eckert:2006},
\cite{Meier:2006}. He was led to this concept during his short industrial
occupation when he tried to account for the phenomenon of flow separation
in diverging ducts. Prandtl presented the concept together with photographs
of flow around obstacles in a water trough at the Third International
Congress of Mathematicians in Heidelberg in August 1904 \cite{Prandtl:1905}.
In a summary, prepared at a request from the American Mathematical
Society, he declared that the {}``most important result'' of this
concept was that it offered an {}``explanation for the formation
of discontinuity surfaces (vortex sheets) along continuously curved
boundaries.''%
\footnote{Undated draft in response to a request from 13 August 1904, Blatt
43, Cod. Ms. L. Prandtl 14, Acc. Mss. 1999.2, SUB.%
} In his Heidelberg presentation he expressed the same message in these
words: {}``A fluid layer set in rotational motion by the friction
at the wall moves into the free fluid and, exerting a complete change
of motion, plays there a similar role as Helmholtz' discontinuity
sheets.'' \cite[p. 578]{Prandtl:1905} (On the emergence of Helmholtz's
concept of discontinuity surfaces see \cite[chapter 4.3]{Darrigol:2005}).

According to the recollection of one participant at the Heidelberg congress, Klein recognized the momentousness of Prandtl's method immediately \cite{Sommerfeld:1935}. However, if this recollection from many years later may be trusted, Klein's reaction was exceptional. Prandtl could
not offer more than plausibility arguments. The boundary layer concept
required elaboration before its potential was more widely recognized
\cite{Dryden:1955,Goldstein:1969,Tani:1977,Grossmann:2004}. 
Its modern understanding in terms of singular perturbation theory
\cite{Malley:2010} emerged only decades later. 
The first tangible evidence
that Prandtl's concept provided more than qualitative ideas was offered
by Blasius, who derived in his doctoral dissertation the coefficient
for (laminar) skin friction from the boundary layer equations for
the flow along a flat plate ({}``Blasius flow'', \cite{Blasius:1908},
\cite{Hager:2003}). However, this achievement added little to understand
what Prandtl had considered the {}``most important result'' of his
concept, how vortical motion is created at the boundary, not to speak
about turbulence.

Even before Prandtl arrived in Göttingen, the riddle of turbulence
was a recurrent theme in Klein's lectures and seminars. In a seminar
on hydraulics in the winter semester 1903/04 Klein called it a {}``true
need of our time to bridge the gap between separate developments.''
The notorious gulf between hydraulics and hydrodynamics served to
illustrate this need with many examples. The seminar presentations
were expected to focus on the comparison between theory and experiment
in a number of specific problems with the flow of water, such as the
outflow through an orifice, the flow over a weir, pipe flow, waves,
the water jump ({}``hydraulic jump''), or the natural water flow
in rivers.%
\footnote{Klein, handwritten notes. SUB Cod. Ms. Klein 19 E (Hydraulik, 1903/04),
and the seminar protocol book, no. 20. Göttingen, Lesezimmer des Mathematischen
Instituts. Available online at librarieswithoutwalls.org/klein.html.%
}

In the winter semester 1907/08 Klein dedicated another seminar to
fluid mechanics, this time with the focus on {}``Hydrodynamics, with
particular emphasis of the hydrodynamics of ships.'' With Prandtl
and Runge as co-organizers, the seminar involved again a broad spectrum
of problems from fluid mechanics that Klein and his colleagues regarded
as suitable for mathematical approaches.%
\footnote{Klein's seminar protocol book, no. 27. Göttingen, Lesezimmer des Mathematischen
Instituts. Available online at librarieswithoutwalls.org/klein.html.%
} Theodore von Kármán, who made then his first steps towards an outstanding
career in Prandtl's institute, presented a talk on {}``unsteady potential
motion.'' Blasius, who was accomplishing his dissertation on the
laminar boundary layer in 1907, reviewed in two sessions the contemporary
research on {}``turbulent flows.'' Other students and collaborators
of Prandtl dealt with {}``vortical motion'' (Karl Hiemenz) and {}``boundary
layers and detachment of vortices'' (Georg Fuhrmann). Although little
was published about these themes at the time, Klein's seminar served
as a proving ground for debates on the notorious problems of fluid
mechanics like the creation of vorticity in ideal fluids ({}``Klein's
Kaffeelöffelexperiment'', see \cite{Klein:1910} and \cite[chapter 6]{Saffman:1992}).

With regard to turbulence, the records of Blasius's presentation from
this seminar illustrate what Prandtl and his collaborators must have
regarded as the main problems at that time. After reviewing the empirical
laws like Chezy's law for channel flow and Reynolds's findings about
the transition to turbulence in pipe flow (for these and other pioneering
19th century efforts see \cite[chapter 6]{Darrigol:2005}, Blasius
concluded that the problems {}``addressed to hydrodynamics'' should
be sorted into two categories, {}``I. Explanation of instability,''
and {}``II. Description of turbulent motion'' which had to address
both the dichotomy of {}``hydraulic description'' versus {}``rational
hydrodynamic explanations.'' Concerning the first category, the onset
of turbulence as a result of hydrodynamic instability, Blasius reviewed
Hendrik Antoon Lorentz's recent approach where a criterion for the
instability of laminar flow was derived from a consideration of the
energy added to the flow by a superposed fluctuation \cite{Lorentz:1897,Lorentz:1907}.
With regard to the second category, the description of {}``fully
developed turbulence,'' Blasius referred mainly to Boussinesq's pioneering
work \cite{Boussinesq:1897} where the effect of turbulence was described
as an additional viscous term in the Navier-Stokes equation. In contrast
to the normal viscosity, this additional {}``turbulent'' viscosity
term was due to the exchange of momentum by the eddying motion in
turbulent flow. Boussinesq's concept had already been the subject
of the preceding seminar in 1903/04, where the astronomer Karl Schwarzschild
and the mathematicians Hans Hahn and Gustav Herglotz reviewed the
state of turbulence \cite{HahnHerglotzEtAl:1904}. However, the seminarists's
efforts to determine the (unknown) eddy viscosity of Boussinesq's
approach proved futile. {}``Agreement between this theory and empirical
observations is not achieved,'' Blasius concluded in his presentation.%
\footnote{Klein's seminar protocol book, no. 27, p. 80. Göttingen, Lesezimmer
des Mathematischen Instituts. Available online at librarieswithoutwalls.org/klein.html.%
}

In spite of  the emphasis on the riddles of turbulence in these seminars,
it is  commonly reported that  Prandtl ignored turbulence as a research theme until
many years later. For example, the editors of his Collected Papers dated his first
publication in the category {}``Turbulence and Vortex Formation''
to the year 1921 (see below). The preserved archival sources, however,
belie this impression. Prandtl started to articulate his ideas on
turbulence much earlier. {}``Turbulence I: Vortices within laminar
motion,'' he wrote on an envelope with dozens of loose manuscript
pages. The first of these pages is dated by himself to October 3,
1910, and is headlined with {}``Origin of turbulence.'' Prandtl
considered there {}``a vortex line in the boundary layer close to
a wall'' and argued that such a vortical motion {}``fetches (by
frictional action) something out of the boundary layer which, because
of the initial rotation, becomes rolled up to another vortex which
enhances the initial vortex.'' Thus he imagined how flows become
vortical due to processes that originate in the initially laminar
boundary layer.%
\footnote{Cod. Ms. L. Prandtl 18 ({}``Turbulenz I: Wirbel in Laminarbewegung''),
Acc. Mss. 1999.2, SUB.%
} In the same year he published a paper on {}``A relation between
heat exchange and flow resistance in fluids'' \cite{Prandtl:1910}
which extended the boundary layer concept to heat conduction. Although
it did not explicitly address turbulence--the article is more renowned
because Prandtl introduced here what was later called {}``Prandtl
number''--it reveals Prandtl's awareness for the differences of laminar
and turbulent flow with regard to heat exchange and illustrates from
a different perspective how turbulence entered Prandtl's research
agenda \cite{Rotta:2000}.

\begin{figure}[htp]
\begin{center}
\subfigure[Without a trip wire.]{\includegraphics[scale=0.75]{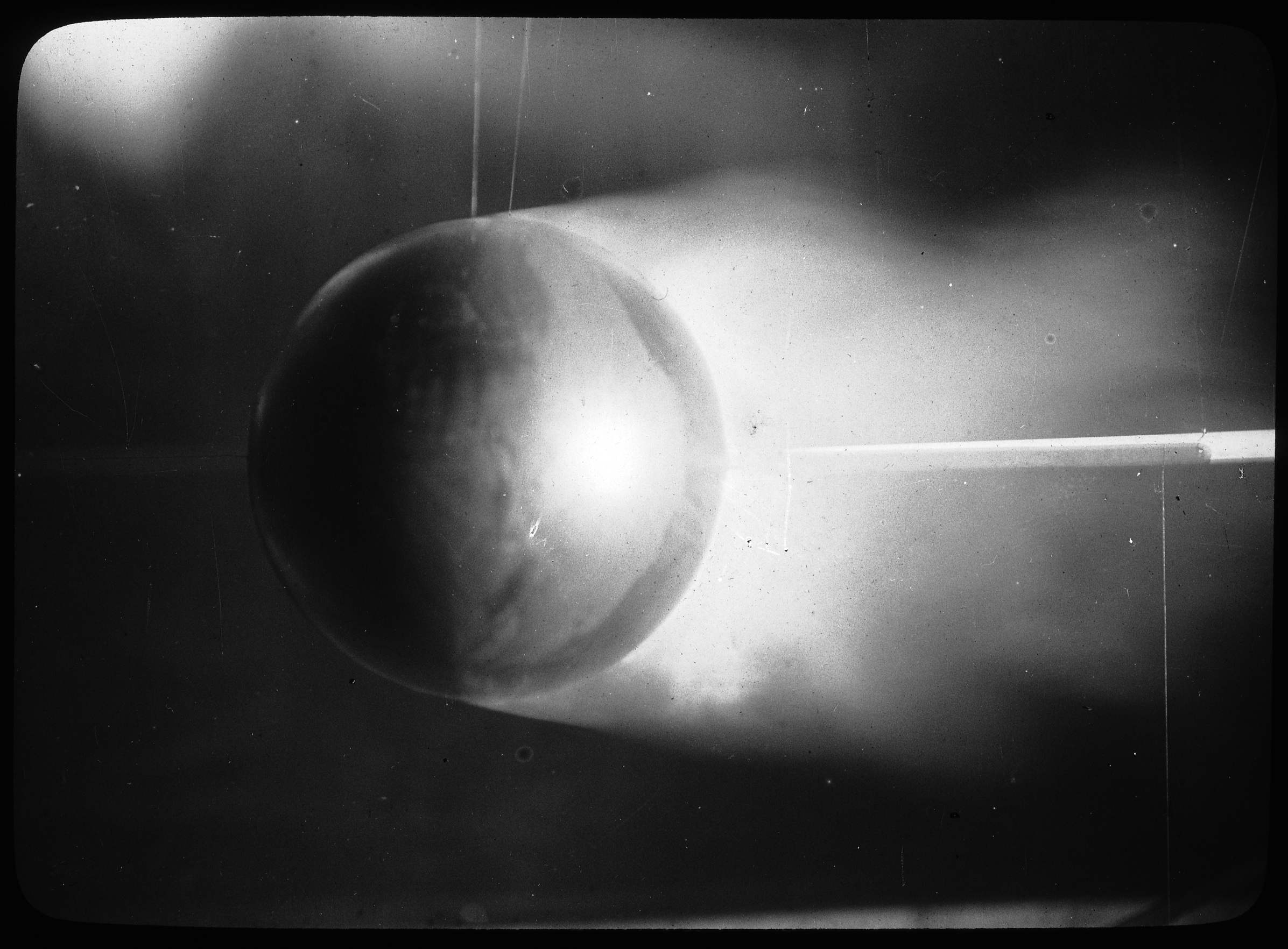}}
\subfigure[With a trip wire.]{\includegraphics[scale=0.75]{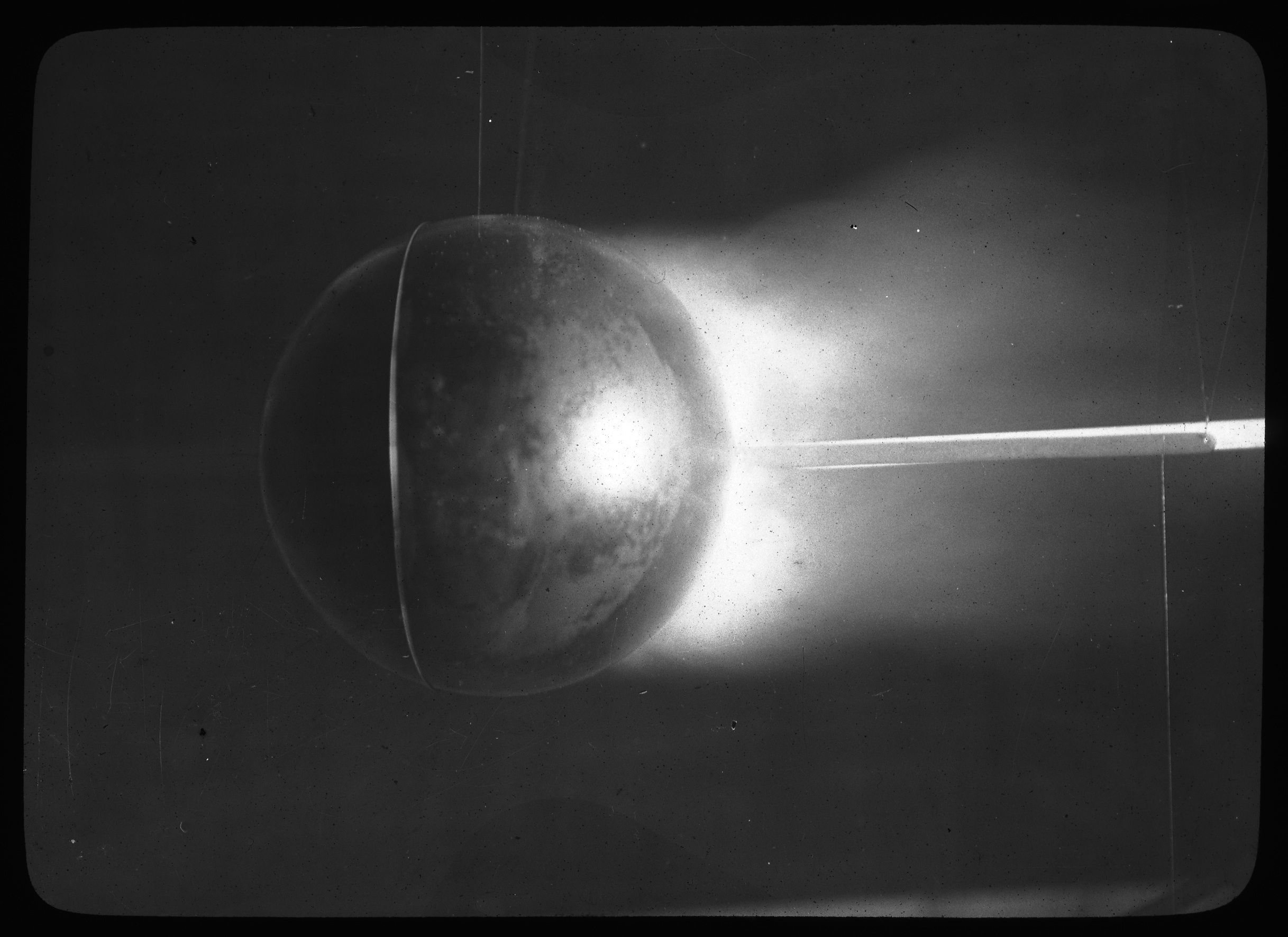}}
\caption{Turbulence behind a sphere made visible with smoke. (Reproduction from the original 1914 photograph  GOAR: GK-0116 and  GK-0118)}
\label{fig:sphere}
\end{center}
\end{figure}

Another opportunity to think about turbulence from the perspective
of the boundary layer concept came in 1912 when wind tunnel measurements
about the drag of spheres displayed discrepant results. When Otto
Föppl (1885-1963), Prandtl's brother-in-law and collaborator at the
airship model test facility, compared the data from his own measurements
in the Göttingen wind tunnel with those from the laboratory of Gustav
Eiffel (1832-1921) in Paris, he found a blatant discrepancy and supposed
that Eiffel or his collaborator had omitted a factor of 2 in the final
evaluation of their data \cite{Foeppl:1912}. Provoked by this claim,
Eiffel performed a new test series and found that the discrepancy
was not the result of an erroneous data evaluation but a new phenomenon
which could be observed only at higher air speeds than those attained
in the Göttingen wind tunnel \cite{Eiffel:1912}. After inserting
a nozzle into their wind tunnel, Prandtl and his collaborators were
able to reproduce Eiffel's discovery: At a critical air speed the
drag coefficient suddenly dropped to a much lower value. Prandtl also
offered an explantion of the new phenomenon. He assumed that the initially
laminar boundary layer around the sphere becomes turbulent beyond
a critical air speed. On the assumption that the transition from laminar
to turbulent flow in the boundary layer is analogous to Reynolds'
case of pipe flow, Prandtl displayed the sphere drag coefficient as
a function of the Reynolds number, $UD/\nu$ (flow velocity $U$,
sphere diameter $D$, kinematic viscosity $\nu$), rather than of
the velocity like Eiffel; thus he demonstrated that the effect occurred
at roughly the same Reynolds number even if the individual quantities
differed widely (the diameters of the spheres ranged from 7 to 28
cm; the speed in the wind tunnel was varied between 5 an 23 m/s).
Prandtl further argued that the turbulent boundary layer flow entrains
fluid from the wake so that the boundary layer stays attached to the
surface of the sphere longer than in the laminar case. In other words,
the onset of turbulence in the boundary layer reduces the wake behind
the sphere and thus also its drag. But the argument that turbulence
decreases the drag seemed so paradoxical that Prandtl conceived an
experimental test: When the transition to turbulence in the boundary
layer was induced otherwise, e. g. with a thin {}``trip wire'' around
the sphere or a rough surface, the same phenomenon occured. When smoke
was added to the air stream, the reduction of drag became visible
by the reduced extension of the wake behind the sphere 
\cite{Wieselsberger:1914,Prandtl:1914} (see Fig.~\ref{fig:sphere}).

\subsection*{A {}``Working Program for a Theory of Turbulence''}

During the First World War turbulence became pertinent in many guises.
Arnold Sommerfeld (1868-1951), theoretical physicist at the University
of Munich, once forwarded to Prandtl a request {}``concerning the
fall of bombs in water and air.'' Sommerfeld was involved at that
time in other war-related research (about wireless telegraphy) and
had heard about this problem from a major whom he had met during a
visit in Berlin. {}``It deals with the drag of a sphere (radius $a$)
moving uniformly through water (density $\rho$) at a velocity $V$.
At $R$ (Reynolds number) $>1000$ the drag is $W=\psi\rho a^{2}V^{2}$.''
By similarity {}``$\psi$ should be universal and also independent
of the fluid,'' Sommerfeld alluded to Prandtl's recent study about
the drag of spheres in air, but according to older measurements of
the friction coefficient $\psi$ for water this was not the case.%
\footnote{Sommerfeld to Prandtl, 9 May 1915. GOAR 2666.%
} Prandtl suspected an error with the assessment of the experimental
measurements. Furthermore, the impact of a falling bomb at the water
surface involved additional effects so that a comparison was difficult.%
\footnote{Prandtl to Sommerfeld, 14 Mai 1915. GOAR 2666.%
} A few months later, the aerodynamics of bomb shapes became officially
part of Prandtl's war work.%
\footnote{He received, for example, contracts from the Bombenabteilung der Prüfanstalt
u. Werft der Fliegertruppen, dated 23 December 1915, concerning {}``Fliegerbombe,
M 237'', and on {}``Carbonit-Bomben, Kugelform,'' dated 1 September
1916. GOAR 2704B.%
}

The turbulence effect as observed with the drag of spheres became
also pertinent for the design of airplanes. The struts and wires which
connected the wings of bi- and triplanes were subject to the same
sudden changes of drag. For this reason, Prandtl's institute was charged
with a systematic wind tunnel investigation of struts and wires. The
goal was to find out how the sudden change of drag could be avoided
by choosing appropriate strut and wire shapes. {}``The critical range
{[}of Reynolds numbers{]} is considered as the interval within which
there are two fundamentally different modes of flow,'' Prandtl's
collaborator, Max Munk, explained in a technical war report on measurements
of the drag of struts. The report also mentioned how this phenomenon
occurred in practice. {}``In particular, a reduction of the speed,
for example, when the plane changes from horizontal flight into a
climb, results in a sudden increase of the drag coefficient, and often
of a considerable increase of the drag itself.'' It was therefore
not sufficient to minimize the drag by streamlining the profile of
a strut, but also to give it a shape that did not experience the sudden
change of drag when the airplane passed through the critical speed
range \cite{Munk:1917}.

In view of such practical relevance, Prandtl sketched in March 1916
a {}``Working program about the theory of turbulence.''%
\footnote{Page 15 (dated 6 March 1916) in Cod. Ms. L. Prandtl, 18, Acc. Mss.
1999.2, SUB.%
} Like Blasius in his presentation in Klein's seminar, Prandtl discriminated
between the {}``onset of turbulence''{}, \emph{i.e.}, the transition from
laminar to turbulent flow, and the {}``accomplished turbulence,''
i. e. fully developed turbulence, as the two pillars of this research
program. The onset of turbulence was generally perceived as the consequence
of a hydrodynamic instability, a problem with a long history of frustrated
efforts \cite{Darrigol:2005}; although it had been revived during
the preceding decade by William McFadden Orr, Sommerfeld, Ludwig Hopf,
Fritz Noether and others, a solution seemed out of sight \cite{Eckert:2010}.
Prandtl sketched plane flows with different piece-wise linear velocity
profiles. The stability of such flow configurations had been extensively
studied since the 1880s by Lord Rayleigh for the inviscid case. Profiles
with an inflection were unstable according to Rayleigh's analysis
\cite{Rayleigh:1887}. Prandtl's strategy seemed clear: He approached
the stability analysis from the limiting case of infinite Reynolds
numbers, i. e. the inviscid case treated by Rayleigh, in order to
derive from this limit approximations for flows of low viscosity.
Like his boundary layer concept this approach would be restricted
to high Reynolds numbers (unlike the Orr-Sommerfeld approach which
applied to the full range of Reynolds numbers). According to his sketches
and somewhat cryptic descriptions, Prandtl attempted to study the
behaviour of {}``a sinusoidal discontinuity (vortex along the boundary
between two stripes'' or {}``a vortex point in an otherwise undisturbed
stripe flow.'' Prandtl's {}``stripes,'' i. e. piece-wise linear
flow profiles, indicate that he aimed at a theory for the onset of
turbulence in the plane flow bounded by two fixed walls and a flow
bounded by a single wall. The latter configuration obviously was perceived
as an approximation of the {}``Blasius flow,'' i. e. the velocity
profile of the laminar boundary layer flow along a flat plate. According
to Rayleigh's inflexion theorem, both flows were stable in the inviscid
case because the curvature of the velocity profile did not change
direction. The focus was on the {}``boundary layer motion with Rayleigh
oscillation,'' as Prandtl concluded this part of his turbulence program.

With regard to the other part of his {}``working program,'' fully
developed turbulence, Prandtl apparently had no particular study in
mind as a starting point. {}``Statistical equilibrium of a set of
vortices in an ideal fluid in the vicinity of a wall,'' he noted
as one topic for future research. For the goal of a {}``complete
approach for very small friction'' he started from the assumption
that vortices from the wall ({}``in the boundary layer'') are swept
into the fluid by {}``disordered motion''. He envisioned a balance
between the vortex creation at the wall and the vortices destroyed
in the fluid as a result of friction. For a closer analysis of the
involved vortex interaction he introduced the {}``rough assumption
that the vortex remains unchanged for a certain time $\sim r^{2}/\nu$
and then suddenly disappears, whereby it communicates its angular
momentum to the mean flow.''%
\footnote{Page 16 in Cod. Ms. L. Prandtl, 18, Acc. Mss. 1999.2, SUB. Apparently
$r$ and $\nu$ are the radius of the vortex and the kinematic viscosity
of the fluid, respectively. Prandtl did not define the quantities
involved here. His remarks are rather sketchy and do not lend themselves
for a precise determination of the beginnings of his future mixing
length approach.%
}

During the war Prandtl had more urgent items on his agenda \cite[pp. 115-198]{Rotta:1990}.
But the riddle of turbulence did not disappear from his mind as a
paramount challenge. Nor from that of his former prodigy student,
Theodore von Kármán, who returned after the war to the Technische
Hochschule Aachen as director of the aerodynamic institute which was
then in its fledgling stages. Both the Aachen and the Göttingen fluid
dynamicists pursued the quest for a theory of turbulence in a fierce
rivalry. {}``The competition was gentlemanly, of course. But it was
first-class rivalry nonetheless,'' Kármán later recalled, {}``a
kind of Olympic Games, between Prandtl and me, and beyond that between
Göttingen and Aachen.'' \cite[p. 135]{Karman:1967} Since they had
nothing published on turbulence, both Prandtl and Kármán pondered
how to ascertain their priority in this quest. In summer 1920, Prandtl
supposed that von Kármán used a forthcoming science meeting in Bad
Nauheim to present a paper on turbulence at this occasion. {}``I
do not yet know whether I can come,'' he wrote to his rival, {}``but
I wish to be oriented about your plans. As the case may be I will
announce something on turbulence (experimental) as well. I have now
visualized turbulence with lycopodium in a 6 cm wide channel.''%
\footnote{Prandtl to Kármán, 11 August 1920. GOAR 1364.%
} The Aachen-Göttingen rivalry did not yet surface publicly at this
occasion. By correspondence, however, it was further developing. The
range of topics encompassed Prandtl's entire {}``working program.''
Early in 1921 Prandtl learned that von Kármán was busy to elaborate
a theory of fully developed turbulence in the boundary layer along
a flat wall--with {}``fabulous agreement with observations.'' Ludwig
Hopf and another collaborator of the Aachen group started by this
time with hot-wire experiments. They attempted to measure in a water
channel {}``the mean square fluctuation and the spectral distribution
of the fluctuations,'' Hopf revealed about the Aachen plans.%
\footnote{Hopf to Prandtl, 3 February 1921. MPGA, Abt. III, Rep. 61, Nr. 704.%
}

Little seems to have resulted from these experiments, neither in Aachen
by means of the hot-wire technique nor in Prandtl's laboratory by
visualizing turbulence with lycopodium. Von Kármán's theoretical effort,
however, appeared promising. {}``Dear Master, colleague, and former
boss,'' Kármán addressed Prandtl in a five-page letter with ideas
for a turbulent boundary layer theory (see below) and about the onset
of turbulence.%
\footnote{Kármán to Prandtl, 12 February 1921. GOAR 3684.%
} The latter was regarded as {}``the'' turbulence problem. The difficulty
to explain the transition from laminar to turbulent flow was rated
as a paramount riddle since the late 19th century. In his dissertation
performed under Sommerfeld in 1909, Ludwig Hopf had titled the introductory
section {}``The turbulence problem,'' because neither the {}``energy
considerations'' by Reynolds and Lorentz nor the stability approaches
by Lord Kelvin and Lord Rayleigh were successful. Hopf was confronted
with the problem in the wake of Sommerfeld's own stability approach
to viscous flows, but {}``the consequent analysis of the problem
according to the method of small oscillations by Sommerfeld is not
yet accomplished.'' \cite[pp. 6-7]{Hopf:1910} In the following decade
the problem was vigorously attacked by this method (later labeled
as the Orr-Sommerfeld method)--with the discrepant result that plane
Couette flow seemed stable for all Reynolds numbers \cite{Eckert:2010}. 

By comparison with these efforts, Prandtl's approach as sketched in
his {}``working program'' appeared like a return to the futile attempts
from the 19th century: {}``At a large Reynolds number the difference
between viscous and inviscid fluids is certainly imperceptible,''
Hopf commented Prandtl's idea to start from the inviscid limit, but
at the same time he regarded it {}``questionable whether one is able
to arrive at a useful approximation that leads down to the critical
number from this end.''%
\footnote{Hopf to Prandtl, 27 October 1919. GOAR 3684.%
} In response to such doubts Prandtl began to execute his {}``working
program'' about the onset of turbulence in plane flows with piecewise
linear flow profiles. {}``Calculation according to Rayleigh's papers
III, p. 17ff,'' he noted on a piece of paper dated {}``5-8. 1. 21,''
followed by several pages of mathematical calculations.%
\footnote{Pages 22-26 in Cod. Ms. L. Prandtl, 18, Acc. Mss. 1999.2, SUB.%
} Despite their initial reservations, the Aachen rivals were excited
about Prandtl's approach. Kármán {}``immediately rushed'' his collaborators
to undertake a stability analysis for certain piecewise linear flow
profiles, Hopf confided to Prandtl.%
\footnote{Hopf to Prandtl, 3 February 1921. MPGA, III, Rep. 61, Nr. 704.%
} Prandtl had by this time already asked a doctoral student to perform
a similar study. {}``Because it deals with a doctoral work, I would
be sorry if the Aachener would publish away part of his dissertation,''
he asked Hopf not to interfere in this effort. Kármán responded that
the Aachen stability study was aiming at {}``quite different goals,''
namely the formation of vortices in the wake of an obstacle (labeled
later as the {}``Kármán vortex street'' after von Kármán's earlier
theory about this phenomenon \cite[chapter 2]{Eckert:2006}). The
new study was motivated by {}``the hope to determine perhaps the
constants that have been left indetermined in my old theory,'' Kármán
calmed Prandtl's worry. Why not arrange a {}``division of labor''
between Göttingen and Aachen, he further suggested, so that his group
deals with these wake phenomena and Prandtl's doctoral student with
boundary layer instability.%
\footnote{Kármán to Prandtl, 12 February 1921. GOAR 3684.%
} Prandtl agreed and asked von Kármán to feel free with his plans.
He explained once more that the focus at Göttingen was to study the
onset of turbulence in the boundary layer. {}``We have now a method
to take into account friction approximately.''%
\footnote{Prandtl to Kármán, 16 February 1921. MPGA, III, Rep. 61, Nr. 792.%
}

A few months later Prandtl reported that the calculations of his doctoral
student were terribly complicated and yielded {}``a peculiar and
unpleasant result.'' If the corresponding flow is unstable according
to Rayleigh's inviscid theory, the instability was not reduced by
taking viscosity into account--as they had expected--but increased.
The calculation was done in first-order approximation, but its extension
to the second-order seemed {}``almost hopeless,'' Prandtl wrote
in frustration, {}``and so, once more, we do not obtain a critical
Reynolds number. There seems to be a very nasty devil in the turbulence
so that all mathematical efforts are doomed to failure.''%
\footnote{Prandtl to Kármán, 14 June 1921. MPGA, III, Rep. 61, Nr. 792.%
}

At this stage Prandtl published his and his doctoral student's, Oskar
Tietjens (1893-1971), effort. In addition to the profiles which were
unstable in the inviscid case, the study was extended to profiles
which were stable in the inviscid case (i. e. profiles without an
inflexion)--with the surprising result that these profiles also became
unstable if viscosity was included. Contrary to the stability deadlock
of the earlier studies concerning plane Couette flow, Prandtl's approach
left the theory in an instability deadlock. {}``We did not want to
believe in this result and have performed the calculation three times
independently in different ways. There was always the same sign which
indicated instability.'' \cite[p. 434]{Prandtl:1921}

Prandtl's paper appeared in a new journal edited by the applied mathematician
Richard von Mises, the \emph{Zeitschrift für Angewandte Mathematik
und Mechanik (ZAMM)} where the turbulence problem was presented as
a major challenge. In his editorial von Mises described the {}``the
present state of the theory'' as completely open. He regarded it
{}``undecided whether the viscous flow approach is able to explain
turbulence at sufficient mathematical depth.'' \cite[p. 12]{Mises:1921}.
Fritz Noether, like Hopf a Sommerfeld disciple who had struggled with
this matter for years, introduced the subject with a review article
titled {}``The Turbulence Problem.'' \cite{Noether:1921} He summarized
the series of futile attempts during the preceding decades and presented
the problem in a generic manner. (Noether presented the {}``stability
equation'' or {}``perturbation differential equation''--to quote
the contemporary designations--in the form in which it became familiar
later as the {}``Orr-Sommerfeld equation''. His paper became the
door-opener for many subsequent studies of the {}``Orr-Sommerfeld
approach''). Noether was also well informed about the Göttingen effort
as is evident from his correspondence with Prandtl. In one of his
letters he expressed some doubts about Prandtl's approach, but he
belittled his dissent and regarded it merely as a {}``difference
of mindset and expression.''%
\footnote{Noether to Prandtl, 29 June 1921. GOAR 3684.%
} Another contributor to the turbulence problem in this first volume
of the \emph{ZAMM} was Ludwig Schiller, a physicist working temporarily
in Prandtl's laboratory; Schiller surveyed the experimental efforts
to measure the onset of turbulence \cite{Schiller:1921}.

The turbulence problem was also discussed extensively in September
1921 at a conference in Jena, where the Deutsche Physikalische Gesellschaft,
the Deutsche Gesellschaft für Technische Physik and the Deutsche Mathematiker-Vereinigung
convened their annual meetings of this year in a common event. At
this occasion, Prandtl's {}``Remarks about the Onset of Turbulence''
caused quite a stir. {}``With regard to the theoretical results which
have always yielded stability it should be noted that these referred
to the so-called Couette case,'' Prandtl explained the difference
between his result and the earlier studies. But Sommerfeld found it
{}``very strange and at first glance unlikely'' that all flows are
unstable except Couette flow. {}``What causes the special position
of Couette flow?'' Kármán pointed to kinks at arbitrary positions
of the piece-wise linear profiles as a source of arbitrariness. Hopf
criticized Prandtl's approximation $R\rightarrow\infty$. \cite[p. 22-24]{Prandtl:1922}.

The Jena conference and the articles on the turbulence problem in
the \emph{ZAMM} from the year 1921 marked the beginning of a new period
of research about the onset of turbulence. Prandtl did not participate
henceforth with own contributions to this research. But he continued
to supervise doctoral dissertations about this part of his {}``working
program.'' Tietjens paved the way \cite{Tietjens:1925} along which
Walter Tollmien (1900-1968), in another Goettingen doctoral dissertation, achieved
the first complete solution of the Orr-Sommerfeld equation for a special
flow \cite{Tollmien:1929}. A few years later, another disciple of Prandtl, Hermann Schlichting
(1907-1982), further extended this theory \cite{Schlichting:1933},
so that the process of instability could be analysed in more detail.
But the {}``Tollmien-Schlichting'' approach remained disputed until
it was experimentally corroborated in World War II \cite{Eckert:2008b}.

\subsection*{Skin friction and turbulence I: the 1/7th law}

Originally, Prandtl's boundary layer concept had focused on laminar
flow. Ten years later, with the interpretation of Eiffel's drag phenomenon
as a turbulence effect, boundary layer flow could also be imagined
as fully turbulent. From a practical perspective, the latter appeared
much more important than the former. Data on fluid resistance in pipes,
as measured for decades in hydraulic laboratories offered plenty of
problems to test theories about turbulent friction. Blasius, who had
moved in 1911 to the Berlin Testing Establishment for Hydraulics and
Ship-Building (Versuchsanstalt für Wasserbau und Schiffbau), published
in 1913 a survey of pipe flow data: When displayed as a function of
the Reynolds number $R$, the coefficient for {}``hydraulic'' (i.
e. turbulent) friction varied proportional to $R^{-1/4}$ (in contrast
to laminar friction at low Reynolds numbers, where it is proportional
to $R^{-1}$) \cite{Blasius:1913}. 

No theory could explain this empirical {}``Blasius law'' for turbulent
pipe flow. But it could be used to derive other semi-empirical laws,
such as the distribution of velocity in the turbulent boundary layer
along a plane smooth wall. When Kármán challenged Prandtl in 1921
with the outline of such a theory, he recalled that Prandtl had told
him earlier how one could extrapolate from pipe flow to the flow along
a plate, and that Prandtl already knew that the velocity distribution
was proportional to $y^{1/7}$, where $y$ was the distance from the
wall. Prandtl responded that he knew this {}``already for a pretty
long time, say since 1913.'' He claimed that he had {}``already
in earlier times attempted to calculate boundary layers in which I
had assumed a viscosity enhanced by turbulence, which I chose for
simplicity as proportional to the distance from the wall and proportional
with the velocity in the free flow.'' But he admitted that Kármán
was further advanced with regard to a full-fledged turbulent boundary
layer theory. \foreignlanguage{english}{\textquotedblleft{}I have
planned something like this only for the future and have not yet begun
with the elaboration.'' Because he was busy with other work he suggested
that Kármán should proceed with the publication of this theory. {}``I
will see afterwards how I can gain recognition with my different derivation,
and I can get over it if the priority of publishing has gone over
to friendly territory.\textquotedblright{}}%
\footnote{Prandtl to Kármán, 16 February 1921. MPGA, Abt. III, Rep. 61, Nr.
792.%
}

\selectlanguage{english}%
Kármán published his derivation without further delay in the first
volume of the \emph{ZAMM} with the acknowledgement that it resulted
from a suggestion \textquotedblleft{}by Mr. Prandtl in an oral communication
in autumn 1920.\textquotedblright{} \cite[p. 76]{Karman:1921} Prandtl\textquoteright{}s
derivation appeared in print only in 1927--with the remark that \textquotedblleft{}the
preceding treatment dates back to autumn 1920.\textquotedblright{}
\cite{Prandtl:1927} Johann Nikuradse (1894-1979), whom Prandtl assigned by that time an
experimental study about the velocity distribution in turbulent flows
as subject of a doctoral work, dated Prandtl's derivation more precisely to \textquotedblleft{}a discourse in Göttingen on November,
5th, during the winter semester of 1920.\textquotedblright{} \cite[p. 15]{Nikuradse:1926}
Indeed, Prandtl outlined this derivation in notices dated (by himself) to
28 November 1920.\footnote{Prandtl, notices, MPGA, Abt. III, Rep. 61, Nr. 2296, page 65. %
} Further evidence is contained in the first volume of the \emph{Ergebnisse der Aerodynamischen Versuchsanstalt zu Göttingen}, accomplished at \textquotedblleft{}Christmas 1920\textquotedblright{} (according to the preface), where Prandtl offered a formula for the friction coefficient proportional to $R^{-1/5}$, with the Reynolds number \emph{R} related to the length of the plate \cite[p. 136]{Prandtl:1921a}. Although Prandtl did not present the derivation, he could not have arrived at this friction coefficient without the 1/7th law for the velocity distribution. (The derivation was based on the assumption that the shear stress
at the wall inside the tube only depends on the flow in the immediate
vicinity; hence it should not depend on the radius of the tube. Under
the additional assumption that the velocity grows according to a power
law with increasing distance from the wall, the derivation was straight
forward.)

Kármán presented his theory on turbulent skin friction again in 1922
at a conference in Innsbruck \cite{Karman:1924}. He perceived it
only as a first step on the way towards a more fundamental understanding
of turbulent friction. The solution, he speculated at the end of his
Innsbruck talk, would probably come from \textquotedblleft{}a statistical
consideration.\textquotedblright{} But in order to pursue such an
investigation \textquotedblleft{}a fortunate idea\textquotedblright{}
was necessary, \textquotedblleft{}which so far has not yet been found.\textquotedblright{}
\cite[p. 167]{Karman:1924} (On the quest for a statistical theory
in the 1920s see \cite{Battimelli:1984}). Prandtl, too, raised little
hope for a more fundamental theory of turbulence from which empirical
laws, such as that of Blasius, could be derived from first principles.
\textquotedblleft{}You ask for the theoretical derivation of Blasius\textquoteright{}s
law for pipe friction,\textquotedblright{} Prandtl responded to the
question of a colleague in 1923. \textquotedblleft{}The one who will
find it will thereby become a famous man!\textquotedblright{}%
\footnote{Prandtl to Birnbaum, 7. Juni 1923, MPGA, Abt. III, Rep. 61, Nr. 137.%
} 

\selectlanguage{british}%

\subsection*{The mixing length approach}

Prandtl's ideas concerning fully developed turbulence remained the
subject of informal conversations and private correspondence for several
more years after 1921. {}``I myself have nothing brought to paper
concerning the $1/7$-law,'' Prandtl wrote to Kármán in continuation
of their correspondence about the turbulent boundary layer theory
in summer 1921.%
\footnote{Prandtl to Kármán, 14 June 1921. MPGA, Abt. III, Rep. 61, Nr. 792.%
} A few years later, t\foreignlanguage{english}{he velocity distribution
in the turbulent boundary layer of a smooth plate in a wind tunnel
was measured directly in Johannes M. Burgers\textquoteright{}s (1895-1981)
laboratory in Delft by the new method of hot wire anemometry \cite{Burgers:1925}.
Prandtl had hesitated in 1921 to publish his derivation of the 1/7th
law because, as he revealed in another letter to his rival at Aachen, he aimed at a theory \textquotedblleft{}in which the experimental
evidence will play a crucial role\textquotedblright{}.}%
\footnote{\selectlanguage{english}%
Prandtl to Kármán, 16 February 1921. MPGA, Abt. III, Rep. 61, Nr. 792.\selectlanguage{british}
}\foreignlanguage{english}{ Four years later, with the data from Burgers's
laboratory, Nikuradse\textquoteright{}s dissertation \cite{Nikuradse:1926}
and other investigations about the resistance of water flow in smooth
pipes \cite{JakobErk:1924} this evidence was available. The experiments
confirmed the 1/7th law within the range of Reynolds numbers for which
the Blasius law had been ascertained. But they raised doubts whether
it was valid for higher Reynolds numbers. Prandtl, therefore, attempted
to generalize his theoretical approach so that he could derive from
any empirical resistance law a formula for the velocity distribution.
\textquotedblleft{}I myself have occupied myself recently with the
task to set up a differential equation for the mean motion in turbulent
flow,\textquotedblright{} he wrote to Kármán in October 1924, \textquotedblleft{}which
is derived from rather simple assumptions and seems appropriate for
very different cases. ... The empirical is condensed in a length which
is entirely adjusted to the boundary conditions and which plays the
role of a free path length.\textquotedblright{}}%
\footnote{\selectlanguage{english}%
Prandtl to Kármán, 10 October 1924. MPGA, Abt. III, Rep. 61, Nr. 792.\selectlanguage{british}
}\foreignlanguage{english}{ }

\selectlanguage{english}%
Thus he referred to the {}``mixing length'' approach, as it would
be labeled later. He published this approach together with the derivation
of the 1/7th law. Prandtl's basic idea was to replace the unknown
eddy viscosity $\epsilon$ in Boussinesq\textquoteright{}s formula
for the turbulent shear stress, $\tau=\rho\epsilon\frac{dU}{dy}$,
by an expression which was amenable to plausible assumptions and could
be tested by experiments. The dimension of $\epsilon$ is $\frac{m^{2}}{s}$,
i. e. the product of a length and a velocity. Prandtl made the Ansatz
$\epsilon=l\cdot l\left|\frac{dU}{dy}\right|$, with $l\left|\frac{dU}{dy}\right|$as
the mean fluctuating velocity with which a \textquotedblleft{}fluid
eddy\textquotedblright{} (\textquotedblleft{}Flüssigkeitsballen\textquotedblright{})
caused a lateral exchange of momentum. Formally, the approach was
analogous to the kinetic gas theory, where a particle could travel
a mean free path length before it exchanged momentum with other particles.
In turbulent flow, however, the exchange process was less obvious.
Prandtl visualized \emph{l} first as a \textquotedblleft{}braking
distance\textquotedblright{} \cite[p. 716]{Prandtl:1925,Prandtl:1949E} then as
a \textquotedblleft{}mixing length.\textquotedblright{} \cite[p. 726]{Prandtl:1926}
He made this approach also the subject of his presentation at the
Second International Congress of Applied Mechanics, held in Z\"urich from 12-17 September 1926 \cite{Prandtl:1927a}.

\selectlanguage{british}%
The historic papers on turbulent stress and eddy viscosity by Reynolds
and Boussinesq were of course familiar to Prandtl since Klein's seminars
in 1904 and 1907, but without further assumptions these approaches
could not be turned into practical theories. \foreignlanguage{english}{At
a first look, Prandtl had just exchanged one unknown quantity ($\epsilon$)
with another ($l$). However, in contrast to the eddy viscosity $\epsilon$,
the mixing length \emph{l} was a quantity which, as Prandtl had written
to Kármán, {}``is entirely adjusted to the boundary conditions''
of the problem under consideration.}%
\footnote{\selectlanguage{english}%
Prandtl to Kármán, 10 October 1924. MPGA, Abt. III, Rep. 61, Nr. 792.\selectlanguage{british}
}\foreignlanguage{english}{ The problem of turbulent wall friction,
however, required rather sophisticated assumptions about the mixing
length. {}``More simple appear turbulent motions in which no wall
interactions have to be taken into account,'' Prandtl resorted to
other phenomena for illustrating the mixing length approach, such
as the mixing of a turbulent jet ejected from a nozzle into an ambient
fluid at rest. In this case the assumption that the mixing length
is proportional to the width of the jet in each cross section gave
rise to a differential equation from which the broadening of the jet
behind the nozzle could be calculated. The theoretical distribution
of mean flow velocities obtained by this approach was in excellent
agreement with experimental measurements \cite{Prandtl:1927a}, \cite{Tollmien:1926}.}

\begin{figure}
\begin{center}
\includegraphics[scale=.3]{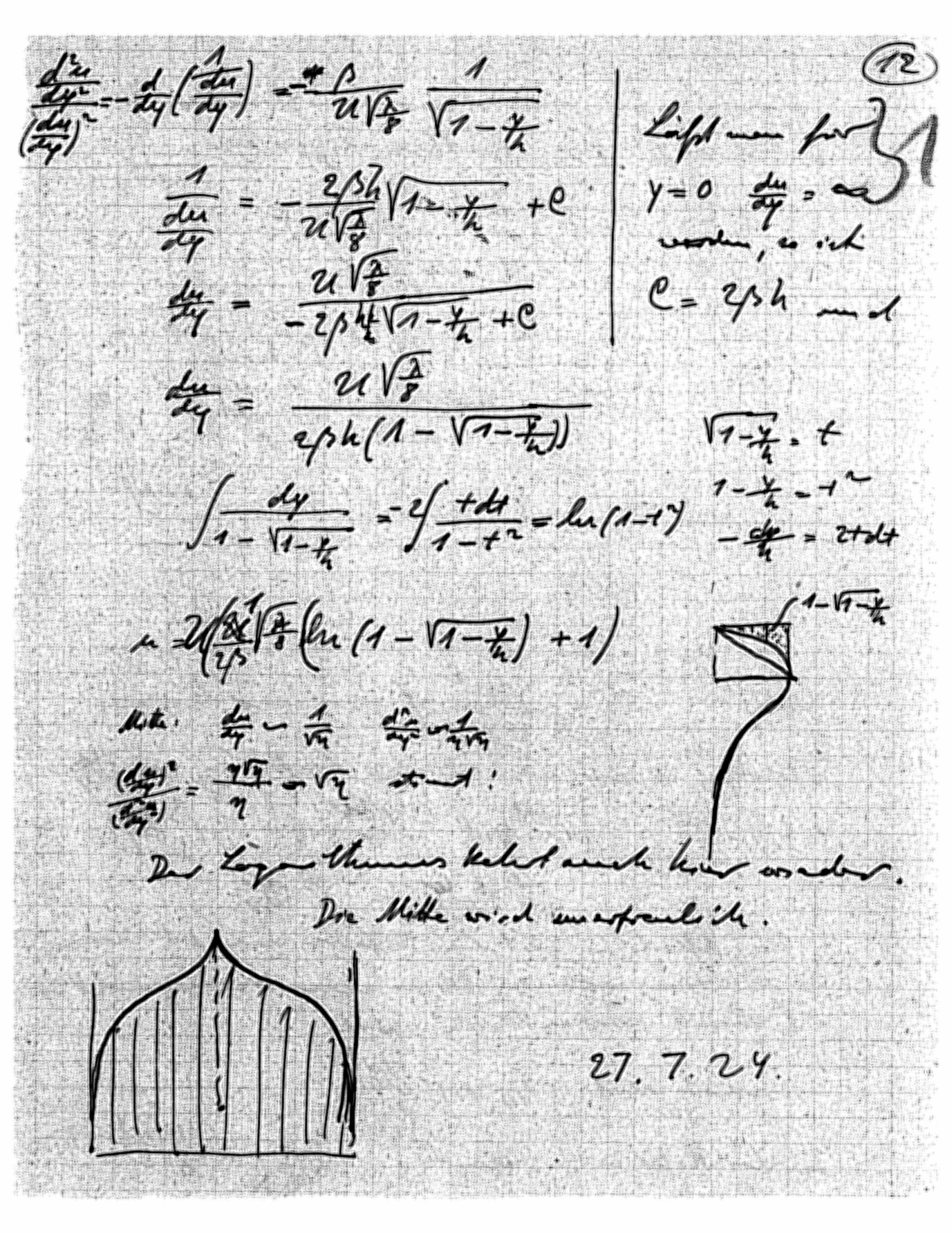}
\caption{Excerpt of Prandtl's back of the envelope calculations from 1924.}
\label{fig:loglaw}
\end{center}
\end{figure}

\selectlanguage{english}%
For the turbulent shear flow along a wall, however, the assumption of
proportionality between the mixing length \emph{l} and the distance
\emph{y} from the wall did not yield the 1/7th law as Prandtl had hoped. 
Instead, when he attempted to derive the distribution of velocity for plane channel flow, he arrived at a logarithmic law--which he dismissed because of {}``unpleasant'' behaviour at the centerline of the channel (see Fig.~\ref{fig:loglaw}).\footnote{Prandtl, notices, MPGA, Abt. III, Rep. 61, Nr. 2276, page 12.} From his notices in summer 1924 it is obvious that he struggled hard to derive an appropriate distribution of velocity from one or another plausible assumption for the mixing length--and appropriate meant to him  that the mean flow $U(y) \propto y^{1/7}$, not some logarithmic law.

Three years later, in a lecture
in Tokyo in 1929, he dismissed the logarithmic velocity distribution again. He argued that {}``\emph{l} proportional \emph{y}
does not lead to the desired result because it leads to $U$
prop. log $y$, which would yield $U=-\infty$ for $y=0$.''
\cite[p. 794]{Prandtl:1930} 
This provided an opportunity for Kármán to win the next round in their
{}``gentlemanly'' competition.

\selectlanguage{british}%

\subsection*{Skin friction and turbulence II: the logarithmic law and beyond}

\selectlanguage{english}%
In June 1928, Walter Fritsch, a student of Kármán, published the results of an experimental study of turbulent channel flow with different wall surfaces \cite{Fritsch:1928}. He found that the velocity profiles line up with each other in the middle parts if they are shifted parallel. This suggested that the velocity distribution in the fluid depends only on the shear stress transferred to the wall and not on the particular wall surface structure. Kármán derived from this empirical observation a similarity approach. In a letter to Burgers he praised this approach for its simplicity: \textquotedblleft{}The
only important constant thereby is the proportionality factor in the
vicinity of the wall.\textquotedblright{} 
As a result, he was led to logarithmic laws both for the velocity
distribution in the turbulent boundary layer and for the turbulent
skin friction coefficient. \textquotedblleft{}The resistance law fits very well
with measurements in all known regions,\textquotedblright{} he concluded
with a hint to recent measurements.%
\footnote{Kármán to Burgers, 12 December 1929. TKC 4.22.%
} 

The recent measurements to which Kármán alluded where those of Fritsch in Aachen and Nikuradse in Göttingen. The latter, in particular, showed a marked deviation
from Blasius's law, and hence from the 1/7th law for the distribution
of velocity, at higher Reynolds numbers. Nikuradse had presented some of his results in June 1929 at a conference in Aachen \cite{Nikuradse:1930}; the comprehensive study appeared only in 1932 \cite{Nikuradse:1932}. By introducing a dimensionless wall distance $\eta = v_{*}y/\nu $ and velocity $\varphi = u/v_{*}$, where $v_{*}=\sqrt{\tau_0/\rho }$ is the friction velocity, $\tau_{0}$ the shear stress at the wall and $\rho$ the density, Nikuradse's data suggested a logarithmic velocity distribution of the form $\varphi = a + b$ log $\eta $. 

Backed by these results from Prandtl's laboratory, Kármán submitted a paper titled
\textquotedblleft{}Mechanical Similarity and Turbulence\textquotedblright{}
to the Göttingen Academy of Science. Unlike Prandtl, he introduced
the mixing length as a characteristic scale of the fluctuating velocities
determined by $l=k U'/U''$, where \emph{k} is a dimensionless
constant (later called \textquotedblleft{}Kármán constant\textquotedblright{})
and $U'$, $U''$ are the first and second derivatives of
the mean velocity of a plane parallel flow in \emph{x}-direction with
respect to the perpendicular coordinate \emph{y.} He derived this
formula from the hypothesis that the velocity fluctuations are similar
anywhere and anytime in fully developed turbulent flow at some distance
from a wall. He had plane channel flow in mind, because he chose his
coordinate system so that the x-axis coincided with the centerline
between the walls at $y=\pm h$. The approach would fail both at the
center line and at the walls, but was supposed to yield reasonable
results in between. (For more detail on Kármán's approach, see the chapter by Leonard and Peters on Kármán). Where Prandtl\textquoteright{}s approach required a further
assumption about the mixing length, Kármán\textquoteright{}s \emph{l}
was an explicit function of \emph{y} at any point in the cross section
of the flow. Kármán obtained a logarithmic velocity distribution and a logarithmic formula for the turbulent friction
coefficient \cite{Karman:1930}.

A few months later, Kármán presented his theory at the Third International
Congress of Applied Mechanics, held in Stockholm from 24-29 August 1930. For this occasion he also
derived the resistance formula for the turbulent skin friction of
a smooth plate. \textquotedblleft{}The resistance law is no power
law,\textquotedblright{} he hinted at the earlier efforts of Prandtl
and himself. \textquotedblleft{}I am convinced that the form of the
resistance law as derived here is irrevocable.\textquotedblright{}
He presented a diagram about the plate skin friction where he compared
the \textquotedblleft{}Prandtl v. Kármán 1921\textquotedblright{}
theory with the \textquotedblleft{}new theory\textquotedblright{}
and with recent measurements from the Hamburgische Schiffbau-Versuchsanstalt.
{}``It appears to me that for smooth plates the least mismatch between
theory and experiment has disappeared,'' Kármán concluded his Stockholm
presentation \cite{Karman:1930a}. 

Prandtl was by this time preparing a new edition of the \emph{Ergebnisse
der Aerodynamischen Versuchsanstalt zu Göttingen} and eager to include
the most recent results.%
\footnote{Prandtl to Kármán\textquoteright{}s colleagues at Aachen, 30 October
1930. TKC 23.43; Prandtl to Kármán, 29 November 1930; Kármán to Prandtl,
16 December 1930. MPGA, Abt. III, Rep. 61, Nr. 792.%
} The practical relevance of Kármán's theory was obvious. In May 1932,
the Hamburgische Schiffbau-Versuchsanstalt convened a conference where
the recent theories and experiments about turbulent friction were
reviewed. Kármán was invited for a talk on the theory of the fluid
resistance, but he could not attend so that he contributed only in
the form of a paper which was read by another attendee \cite{Karman:1932}.
Franz Eisner, a scientist from the Preussische Versuchsanstalt für
Wasserbau und Schiffbau in Berlin, addressed the same theme from a
broader perspective, and Günther Kempf from the Hamburg Schiffbau-Versuchsanstalt
presented recent results about friction on smooth and rough plates
\cite{Eisner:1932}, \cite{Kempf:1932}. Prandtl and others were invited
to present commentaries and additions \cite{PrandtlETAL:1932}. By
and large, this conference served to acquaint practitioners, particularly
engineers in shipbuilding, with the recent advances achieved in the
research laboratories in Göttingen, Aachen and elsewhere.

Two months after this conference, the Schiffbautechnische Gesellschaft
published short versions of these presentations in its journal \emph{Werft,
Reederei, Hafen}. From Eisner\textquoteright{}s presentation a diagram
about plate resistance was shown which characterized the logarithmic
law \textquotedblleft{}after Prandtl (Ergebnisse AVA Göttingen, IV.
Lieferung 1932\textquotedblright{} as the best fit of the experimental
values. According to this presentation, the \textquotedblleft{}interregnum
of power laws\textquotedblright{} had lasted until 1931, when Prandtl
formulated the correct logarithmic law \cite{Eisner:1932a}. When
Kármán saw this article he was upset. He felt that his breakthrough
for the correct plate formula in 1930 as he had presented it in Stockholm
was ignored. From the article about the Hamburg conference \textquotedblleft{}it
looks as if I had given up to work on this problem after 1921,\textquotedblright{}
he complained in a letter to Prandtl, \textquotedblleft{}and that
everything has been done in 1931/32 in Göttingen.\textquotedblright{}
He asked Prandtl to correct this erroneous view in the Göttingen \emph{Ergebnisse},
which he regarded as the standard reference work for all future reviews.
\textquotedblleft{}I write so frankly how I think in this matter because
I know you as the role model of a just man,\textquotedblright{} he
appealed to Prandtl\textquoteright{}s fairness. But he had little
sympathy for \textquotedblleft{}your lieutenants who understandably
do not know other gods beside you. They wish to claim everything for
Göttingen.\textquotedblright{}%
\footnote{Kármán to Prandtl, 26 September 1932. MPGA, Abt. III, Rep. 61, Nr.  793.%
} He was so worried that he also sent Prandtl a telegram with the essence
of his complaint.%
\footnote{Kármán to Prandtl, 28 September 1932. MPGA, Abt. III, Rep. 61, Nr.  793.%
} 

Prandtl responded immediately. He claimed that he had no
influence on the publications in \emph{Werft,
Reederei, Hafen}. With regard to the \emph{Ergebnisse der Aerodynamischen
Versuchsanstalt zu Göttingen} he calmed Kármán\textquoteright{}s worries.
In this publication \textquotedblleft{}of course we refer to your
papers.\textquotedblright{}%
\footnote{Prandtl to Kármán, 29 September 1932. MPGA, Abt. III, Rep. 61, Nr. 793.%
} Like in the preceeding volumes of the \emph{Ergebnisse}, the emphasis
was on experimental results. The news about the logarithmic laws were
presented in a rather short theoretical part (12 out of 148 pages)
titled \textquotedblleft{}On turbulent flow in pipes and along plates.\textquotedblright{}
By and large, Prandtl arrived at the same results as Kármán. He duly
acknowledged Kármán\textquoteright{}s publications from the year 1930,
but he claimed that he had arrived at the same results \textquotedblleft{}at
a time when Kármán\textquoteright{}s papers had not yet been known,
so that once more, like ten years ago with the same problem, the thoughts
in Aachen and Göttingen followed parallel paths.\textquotedblright{}
\cite[p. 637]{Prandtl:1932} For the Hamburg conference proceedings,
Prandtl and Eisner formulated a short appendix where they declared
\textquotedblleft{}that the priority for the formal (\textquotedblleft{}formelmässige\textquotedblright{})
solution for the resistance of the smooth plate undoubtedly is due
to Mr. v. Kármán who has talked about it in August 1930 at the Stockholm
Mechanics Congress.\textquotedblright{} \cite{PrandtlEisner:1932}

When Kármán was finally aware of these publications, he felt embarrassed:
\textquotedblleft{}I hope that there will not remain an aftertaste
from this debate,\textquotedblright{} he wrote to Prandtl.%
\footnote{Kármán to Prandtl, 9 December 1932. MPGA, Abt. III, Rep. 61, Nr. 793.%
} Prandtl admitted that he had \textquotedblleft{}perhaps not without
guilt\textquotedblright{} contributed to Kármán\textquoteright{}s
misgiving. But he insisted that his own version of the theory of plate
resistance was better suited \textquotedblleft{}for the practical
use\textquotedblright{}.%
\footnote{Prandtl to Karman, 19 December 1932. MPGA, Abt. III, Rep. 61, Nr. 793.%
} 
Although the final results of Prandtl's and Kármán's approaches agreed with each other, there
were differences with regard to the underlying assumptions
and the ensuing derivations. Prandtl did not start from a similarity
hypothesis like Kármán. There was no \textquotedblleft{}Kármán\textquoteright{}s
constant\textquotedblright{} in Prandtl's version. Instead, when Prandtl
accepted the logarithmic law as empirically given, he used the same dimensional considerations from which he had derived the 1/7th law from Blasius\textquoteright{}s
empirical law. In retrospect, with the hindsight of Prandtl's notices\footnote{Prandtl, notes, MPGA, Abt. III, Rep. 61, Nr. 2276, 2278.}, it is obvious that he came close to Kármán's reasoning--but the problem how to account for the viscous range close to the wall (which Kármán bypassed by using the centerline of the channel as his vantage point) prevented a solution. In his textbook presentations \cite{Prandtl:1931}, \cite{Prandtl:1942a}, Prandtl avoided the impression of a rivalry about the ``universal wall law'' and duly acknowledged Kármán's priority. 

\begin{figure}[h]
\begin{center}
\includegraphics[width=5in]{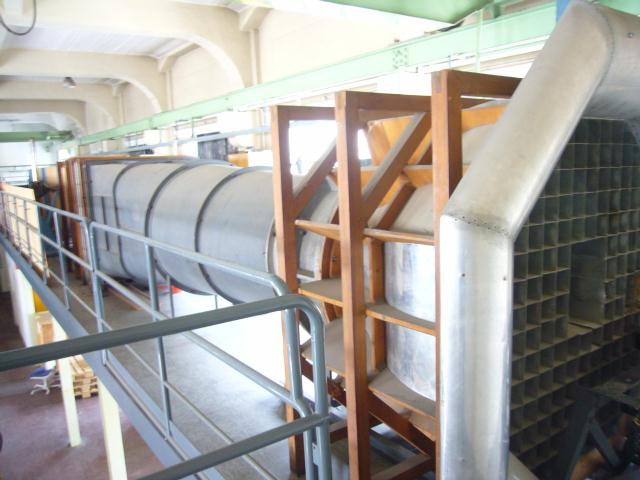}
\caption{Picture of the Rauhigkeitskanal at the Max Planck Institute for Dynamcis and Self-Organization. It was built in 1935 and reconstituted by  Helmut Eckelmann and  James Wallace in the 1970s.}
\label{fig:tunnel}
\end{center}
\end{figure}

But the rivalry between Prandtl and Kármán did not end with the conciliatory
exchange of letters in December 1932. Kármán, who had moved in 1933 permanently to
the USA, presented his own version of
\textquotedblleft{}Turbulence and Skin Friction\textquotedblright{}
in the first issue of the new \emph{Journal of the Aeronautical Sciences}
\cite{Karman:1934}. The Göttingen viewpoint was presented in textbooks
such as Schlichting's \emph{Boundary Layer Theory}, which emerged
from wartime lectures that were translated after the war and first
published as Technical Memoranda of the NACA \cite{Schlichting:1949}.)
The Göttingen school was also most active in elaborating the theory
for practical applications which involved the consideration of pressure
gradients \cite{Gruschwitz:1931} and roughness \cite{Nikuradse:1933},
\cite{Prandtl:1933,Prandtl:1933E1}, \cite{PrandtlSchlichting:1934}, \cite{Prandtl:1934}. 
After these basic studies, the turbulent boundary continued to be a major concern
at Göttingen.  By 1937-1938  the engineer Fritz Schultz-Grunow had joined the institute  and built a special  \textquotedblleft{}Rauhigkeitskanal (see Fig.~\ref{fig:tunnel}) -- roughness channel\textquotedblright{}  for the study of airplane surfaces \cite{Schultz-Grunow:1940},
which was used subsequently for a variety of war-related turbulence
research \cite{Wieghardt:1947,Prandtl:1948a}. This windtunnel, which became the workhorse at the KWI for measurements from 1939-45 \cite{Wieghardt:1942,Wieghardt:1943,Wieghardt:1944,WieghardtTillmann:1944,WieghardtTillmann:1951E} is the only tunnel that survived  the dismantling at the end of the war as it was part of the KWI and not the AVA.\footnote{private communication  Helmut Eckelmann,  and as commented in the British Intelligence Objectives Sub-Committee report 760  that summarizes a visit at the KWI 26--30 April 1946. \textquotedblleft{}Much of the equipment of the A.V.A. has been or is in process of being shipped to the U.K. under M.A.P. direction, but the present proposals for the future of the K.W.I G\"ottingen, appear to be that it shall be reconstituted as an institute for fundamental research in Germany under allied control, in all branches of physics, not solely in fluid motion as hitherto. Scientific celebrities now at the K.W.I. include Profs. Planck, Heisenberg, Hahn and Prandtl among others. In the view of this policy, it is only with difficulty that equipment can be removed from the K.W.I. The K.W.I records and library have already been reconstituted \textquotedblright{}}

\bigskip

\subsection*{Fully Developed Turbulence I - 1932 to 1937}

 Prandtl's interest in fully developed turbulence \footnote{It is important to note that all the research on turbulence in G\"ottingen was conducted at Prandtl's Kaiser Willhem Institute and not at the more technically oriented AVA.} --beyond the quest for a ``universal wall law''--started in the  early 1930s and is best captured following the regular correspondence he had with G.I. Taylor. Starting in 1923, Prandtl was regularly communicating with Taylor on topics in turbulence and instabilities.  In 1923, after reading Taylor's seminal  paper on Taylor-Couette flow \cite{Taylor:1923}, Prandtl in his reply sent him a package of iron-glance powder (hematite)  for flow visualization. It was the same material Prandtl had used for his visualization studies  that led him to his 1904 discovery.  He proposed to Taylor to use it in his experiments,\footnote{Prandtl to Taylor, 25 April 1923. MPGA, Abt. III, Rep. 61, Nr. 1653.} which Taylor immediately and successfully did.  This initial contact led later to a very close relationship between the two giants of fluid mechanics. Although their relationship broke 15 years later on a disagreement about the politics of the Third Reich, their close relationship and the openness with which they communicated is impressive.  Prandtl and Taylor were sometimes exchanging letters weekly. Prandtl visited Cambridge three times. The first time in 1927 was on the invitation by Taylor, the second time in 1934 for the Fourth International Congress of Applied Mechanics and the last time in 1936 to receive a honorary doctorate from Cambridge University.  Prandtl would usually write in typewritten German (Taylor's wife was fluent in German),\footnote{Prandtl to Taylor,  5 June 1934. MPGA, Abt. III, Rep. 61, Nr. 1653.} while Taylor would reply in handwriting in English.  

A letter from 1932 from Prandtl to Taylor marks a new stage with regard to turbulence: \textquotedblleft{}Your new theory of the wake behind a body\textquotedblright{}, Prandtl referred to Taylor's recent work  following the measurements of Fage and Falkner (see the chapter on Taylor by Sreenivasan),  \textquotedblleft{}and the experimental statement by Mr. Fage and Mr. Falkner  on this matter reveals a very important new fact concerning turbulence. It demonstrates, that there are two different forms of turbulence, one belonging to the fluid motions along walls and the other belonging to mixture of free jets. In the first the principal axis of vorticity is parallel to the direction of the main flow, in the other this direction is perpendicular to the flow.\textquotedblright{} He continued that they found agreement with Taylor'a theory at Göttingen by measuring the flow of a cold jet of air through a warm room. He closed the letter with the following footnote: \textquotedblleft{}In the last weeks I studied your old papers of 1915 and 1922 with greatest interest. I think, that if I had known these papers, I would have found the way to turbulence earlier.\textquotedblright{}\footnote{Prandtl to Taylor, 25 July 1932. MPGA, Abt. III, Rep. 61, Nr. 1653.}  

Thus Prandtl concluded that there are two kinds of turbulence, one being wall-turbulence and the other  jet-turbulence \cite{Prandtl:1933}. A third kind of fully developed turbulence, the turbulence in a wind tunnel, had appeared on Prandtl's agenda as early as in 1921, when Hugh Dryden from the National Bureau of Standards in Washington, D.C., had asked him about ``a proper method of defining numerically the turbulence of tunnels and your idea as to the physical conception of the turbulence.''\footnote{Dryden to Prandtl, 6 March 1921. MPGA, Abt. III, Rep. 61, Nr. 362.}  Already then Prandtl was considered the pioneer in wind tunnel design as is reflected in his instructions that he wrote in 1932 in the Handbook of Experimental Physics and that were translated shortly thereafter into Englisch \cite{Prandtl:1933E2}. 

In his response Prandtl had pointed to the vortices in the turbulent air stream that ``are carried with the flow and are in time consumed by the viscosity of the air. In a turbulent flow the velocity of the flow is changing in space and time. Characteristic quantities are the average angular velocity of the  vortex and the diameter, whereby one has to think of a statistical distribution, in which vortices of different sizes and intensity coexist next to each other.''\footnote{Prandtl to Dryden, 20 April 1921. MPGA, Abt. III, Rep. 61, Nr. 362.} However, without appropriate means to measure these quantities, the problem disappeared again from his agenda--until the 1930s, when the isotropic turbulence behind a grid in a wind tunnel was measured with sophisticated new techniques.  In 1932 Fage and Townend \cite{Fage-Townend:1932,Collar:1978} had investigated the full three dimensional mean flow and the associated average three dimensional average velocity fluctuations in turbulent channel and pipe flow using particle tracking streak images of micron size tracers with a microscope. In addition, Dryden and Kuethe in 1929 \cite{Dryden-Kuethe:1929,Kuethe:1988} invented the compensated hotwire measurement, which were to revolutionize the field of turbulence measurements.  This set the stage for spectral measurements of turbulent velocity fluctuations.

From two letters between Taylor  and Prandtl in August and December 1932, following the correspondence discussed above\footnote{Prandtl to Taylor, 25 July 1932. MPGA, Abt. III, Rep. 61, Nr. 1653.}  it is apparent that both had started to conduct hot-wire experiments to investigate the turbulent velocity fluctuations. Taylor in collaboration with  researchers at the National Physical Laboratory (NPL)--most likely Simmons and Salter \cite{Simmons-Salter-Taylor:1938})--and Prandtl with Reichardt \cite{Reichardt:1933,Prandtl-Reichardt:1934}.  Taylor responded with suggestions for pressure correlation measurements and argued: \textquotedblleft{}The same kind of analysis can be applied to hotwire measurements and I am hoping to begin some work on those lines. In particular the "spectrum of turbulence" has not received much attention.\textquotedblright{}\footnote{Taylor to Prandtl, 18 August 1932. MPGA, Abt. III, Rep. 61, Nr. 1653}  Prandtl replied\footnote{Prandtl to Taylor,  23 December 1932. MPGA, Abt. III, Rep. 61, Nr. 1653}: \textquotedblleft{}I do not believe that one can achieve a clear result with pressure measurements, as there is no instrument that can measure these small pressure fluctuations with sufficient speed.  Instead hotwire measurements should lead to good results. We ourselves have conducted an experiment, in which two hotwires are placed at larger or smaller distances from each other and are with an amplifier connected to a cathode ray tube that the fluctuations of the one hotwire appear as horizontal paths and the one of the other as perpendicular paths on the fluorescent screen\footnote{This way of showing correlations was used at the KWI since 1930\cite{Reichardt:1938b}}... To measure also the magnitude of the  correlation my collaborator Dr. Reichardt built an electrodynamometer  with which he can observe the mean of $u_1', u_2'$ and $u_1'u_2'$. In any case, I am as convinced as you that from the study of those correlations as well as between the direction and magnitude fluctuations, for which we have prepared a hotwire setup, very important insights into turbulent flows can be gained.\textquotedblright{}  In the same letter  Prandtl sketched  three pictures of the deflections of the oscilloscope  that are also published in  the 1934 article \cite{Prandtl-Reichardt:1934}. In this article Prandtl and Reichardt reported that the hotwire measurements leading to the figures had been finished in August 1932 (date of the letter of Taylor to Prandtl), and that in October 1933  a micro-pressure manometer had been developed to measure the very weak turbulent fluctuations.\footnote{see the paper \cite{Reichardt:1934} which was at that time in preparation.} It is very remarkable that it took less than a year for Prandtl and Reichardt to pick up the pressure measurement proposal by Taylor.  It also shows the technical ingenuity and the excellent mechanics workshop at the Göttingen KWI.  The micro-pressure  gauge first described in \cite{Reichardt:1935, Reichardt:1948E} is still a very useful design. 

This exchange of letters marks the beginning of the  correlation and spectral analysis of turbulent fluctuations that are at the foundation of turbulence research even today.   Only three weeks later, Taylor replied from a skiing vacation in Switzerland.\footnote{Taylor to Prandtl,  14 January 1933. MPGA, Abt. III, Rep. 61, Nr. 1653} He relied on the NPL with regard to windtunnel measurements. They measured the spectrum of turbulence behind a screen of equally spaced rods  and found it to settle down to a time error function for which he had no theoretical explanation. Prandtl suggested in his reply that the frequency spectrum behind a grid made of rods may be attributable to von Kármán vortices. He added that \textquotedblleft{}apart from this one needs to wait for the publication.\textquotedblright{}. Finally he asked whether Taylor could have his letters rewritten by someone else in more legible writing as he had problems to decipher Taylor's handwriting.\footnote{Prandtl to Taylor,  25 January 1933. MPGA, Abt. III, Rep. 61, Nr. 1653} This seems to have caused an interruption of their communication on turbulence for a while. 

The next letter in the MPG-Archive is from June 1934.\footnote{Taylor to Prandtl,  1 June 1934. MPGA, Abt. III, Rep. 61, Nr. 1653} Taylor invited Prandtl to stay in his house during the upcoming Fourth International Congress for Applied Mechanics, held at Cambridge from 3-9 July, 1934. Prandtl answered in a quite formal and apologetic manner: \textquotedblleft{}In reply to your exceedingly friendly lines from 1.6.34 I may reply to you in German, as I know that your wife understands German without difficulty.\textquotedblright{}\footnote{Prandtl to Taylor,  5 June 1934. MPGA, Abt. III, Rep. 61, Nr. 1653} 

The discussion on turbulence came back to full swing after Prandtl's 60th birthday on 4 February 1935, with almost weekly correspondence.  1935 was the year in which Taylor published what one may regard as his most important papers in turbulence \cite{Taylor:1935a,Taylor:1935b,Taylor:1935c,Taylor:1935d}. The correspondence between the two in 1935 seems to have greatly influenced those papers. Taylor contributed as the only non German scientist to the Festschrift published in ZAMM and handed to Prandtl at the occasion of his birthday. In his article Taylor compared his calculation of the development of turbulence in a contraction with independent measurements by Salter using a hot wire, as well as photographs by Townend of spots of air heated by a spark and by Fage using his ultramicroscope \cite{Taylor:1935}. In other words, the best English fluid-dynamicists contributed to this Festschrift. 

Prandtl thanked Taylor immediately asking him about details of the paper.\footnote{Prandtl to Taylor,  28 February 1935. MPGA, Abt. III, Rep. 61, Nr. 1654} The reply from Taylor convinced Prandtl of the correctness of Taylor's work. As a sideline, Taylor also mentioned that turbulence after a constriction \textquotedblleft{}readjusts itself into a condition where the turbulent velocities are much more nearly equally distributed in space\textquotedblright{}.  (This was later investigated in detail by Corrsin\cite{Comte-Bellot-Corrsin:1966}.) In the same letter Taylor informed Prandtl that he \textquotedblleft{}lately has been doing a great deal of work on turbulence ... In the course of my work I have brought out two formulae which seem to have practical interest. The first concerns the rate of decay of energy in a windstream ... and I have compared them with some of Drydens's measurements behind a honeycomb -- it seems to fit. It also fits Simmons' measurements with turbulence made on a very different scale...\textquotedblright{}.\footnote{Taylor to Prandtl,  2 March  1935. MPGA, Abt. III, Rep. 61, Nr. 1654} The second formula was concerned with the \textquotedblleft{}theory of the critical Reynolds number of a sphere behind a turbulence-grid\textquotedblright{}, as Prandtl replied in his letter pointing him to his own experimental work from 1914.\footnote{Prandtl to Taylor,  12 March  1935. MPGA, Abt. III, Rep. 61, Nr. 1654}  Prandtl  also mentioned that measurement  from G\"ottingen see a signature of the grid. Taylor interpreted this as the \textquotedblleft{}shadow of a screen\textquotedblright{}, which according to Dryden's experiments dies away after a point, where the turbulence is still fully developed.\footnote{Taylor to Prandtl, 14 March  1935. MPGA, Abt. III, Rep. 61, Nr. 1654}  It was this region where Taylor expected his theory to apply.  Taylor submitted his results in four consecutive papers "On the Statistics of Turbulence" on 4 July 1935 \cite{Taylor:1935a,Taylor:1935b,Taylor:1935c,Taylor:1935d} Later Prandtl rederived Taylor's  decay law of turbulence \cite{Wieghardt:1941,Wieghardt:1942E,Prandtl-Wieghardt:1945}.  A detailed discussion of the physics of the decay law of  grid generated turbulence can be found in the  the chapter on Taylor by Sreenivasan. 

A month later Taylor\footnote{Taylor to Prandtl, 21 April 1935. MPGA, Abt. III, Rep. 61, Nr. 1654} thanked Prandtl for sending him the 1934 paper with Reichardt \cite{Prandtl-Reichardt:1934} on measurements of the  correlations of turbulent velocity 
fluctuations that Prandtl had already referred to in his letter in 1932\footnote{Prandtl to Taylor,  23 December 1932. MPGA, Abt. III, Rep. 61, Nr. 1653}. Taylor needed these data for \textquotedblleft{}comparison with my theory of energy dissipation\textquotedblright{}. 

Again the correspondence with Prandtl significantly enhanced Taylor's understanding and finally led to the third paper in the 1935 sequence \cite{Taylor:1935c}.  After returning from the 5th Volta Congress in Rome (on high speeds in aviation), which both attended,  Prandtl mentioned to Taylor that  \textquotedblleft{}Mr. Reichardt conducts new correlation measurements this time correlations between $u{^,}$ and $v{^,}$. The results we will send in the future.\textquotedblright{}\footnote{Prandtl to Taylor, 12 November 1935. MPGA, Abt. III, Rep. 61, Nr. 1654} Again, just as in 1933\footnote{Prandtl to Taylor,  25 January 1933. MPGA, Abt. III, Rep. 61, Nr. 1653}, where Prandtl used a similar formulation, the  correspondence does not return to the matter of turbulence until more than a year later.  

They resumed the discussion again when Taylor sent Prandtl a copy of his 1937 paper on \textquotedblleft{}Correlation Measurements in a Turbulent Flow through a Pipe\textquotedblright{}\cite{Taylor:1936}.  Prandtl responded  that \textquotedblleft{}currently we are most interested in the measurement of correlations between locations in the pipe\textquotedblright{}; he suggested that Taylor may consider measurements away from the center of the pipe and mentioned  that \textquotedblleft{}the measurements in Fig. 4 agree qualitatively well with our $u^,$ measurements. A better agreement is not to be expected as we measured in a rectangular channel and you in a round pipe\textquotedblright{}.\footnote{Prandtl to Taylor,  9 January 1937. MPGA, Abt. III, Rep. 61, Nr. 1654}  Taylor replied on 11 January 1937 and also 23 January 1937 when he sent Prandtl  \textquotedblleft{}our best measurements so that you may compare with your measurements in a flat pipe\textquotedblright{}.\footnote{Taylor to Prandtl,  23 January 1937. MPGA, Abt. III, Rep. 61, Nr. 1654}

Thus by 1937 the stage was set at G\"ottingen and Cambridge for the most important measurements about the statistics of turbulent fluctuations.  At the same time, Dryden and his coworkers at the National Bureau of Standards in Washington measured the decay of the longitudinal correlations behind grids of different mesh sizes $M$  \cite{Dryden:1937} and calculated from it by integration of the correlation function the integral scale of the flow, what Taylor called the \textquotedblleft{}the scale of turbulence\textquotedblright{} \cite[p. 296]{Taylor:1938b}. By using different grids they were able to show that the grid mesh size $M$ determined the large scale $L$ of the flow, just as Taylor had assumed in 1935.  They also found that the relative integral scale $L/M$ increased with the relative distance $x/M$  from the grid independently of $M$. This was later analyzed in more detail by Taylor with data from the National Physical Laboratory in Teddington \cite{Taylor:1938b}.  The 1937 paper by Dryden and collaborators \cite{Dryden:1937} was very important for the further development of turbulence research, as it was data from this paper that Kolmogorov used in 1941 to compare his theory with (see the chapter on on the Russian School by Falkovich).

\subsection*{Fully Developed Turbulence II -- 1938}

After Taylor's paper on the \textquotedblleft{}Spectrum of Turbulence\textquotedblright{} appeared \cite{Taylor:1938a}, Taylor answered  to a  previous  letter by Prandtl.\footnote{Taylor to Prandtl, 18 March 1938. MPGA, Abt. III, Rep. 61, Nr. 1654. Prandtl's letter which prompted this response has not yet been found.} He first thanked Prandtl for sending him Reichardt's $u^{,}v^{,}$ correlation data in a channel flow \cite{Reichardt:1938a,Reichardt:1938b,Reichardt:1951E}, which he regarded as  \textquotedblleft{}certainly of the same type\textquotedblright{} as those of Simmons for the round pipe. Then he answered to a question of Prandtl about the recent paper on the spectrum of velocity fluctuations \cite{Taylor:1938a} and explained to him, what we now know as \emph{Taylor's frozen flow hypothesis}, \emph{i.e.}, \textquotedblleft{}that the formula depends only on the assumption that $u$ is small compared to $U$ so that the succession of events at a point fixed in the turbulent stream is \underline{assumed}  to be related directly to the Fourier analysis of the $(u,x)$ curve obtained from simultaneous measurements of  $u$ and $x$ along a line parallel to the direction of $U$.\textquotedblright{} It is interesting to note that Taylor had a clear concept of the self-similarity of grid generated turbulence:  \textquotedblleft{}The fact that increasing the speed of  turbulent motion leaves the curve $\{U~F(n), n/U\}$ unchanged except at the highest levels of $n$ means that an increase in the \textquotedblleft{}Reynolds number of turbulence\textquotedblright{} leaves the turbulence pattern unchanged in all its features except in the components of the highest frequency\textquotedblright{}.  Six months after this exchange, Prandtl and Taylor met in Cambridge, Mass., for the Fifth International Congress for Applied Mechanics, held at Harvard University and the Massachusetts Institute of Technology from 12--16 September 1938. By now the foundations  for most important discoveries in turbulence research were set.  For the next 60 years, experimental turbulence research was dominated by the Eulerian approach,  \emph{i.e.}, spatial and equal time measurements of turbulent fluctuations as introduced in the period from 1932 -- 1938.  
 
At the Fifth International Congress for Applied Mechanics turbulence was the most important topic. \textquotedblleft{}In the view of the great interest in the problem of turbulence at the Fourth Congress and of the important changes in accepted views since 1934 it was decided to hold a Turbulence Symposium at the Fifth Congress. Professor Prandtl kindly consented to organize this Symposium and ... The Organizing committee is grateful to Professor Prandtl and considers his Turbulence Symposium not only the principal feature of this Congress, but perhaps the Congress activity that will materially affect  the orientation of future research\textquotedblright{} wrote  Hunsaker and von Kármán  in the \textquotedblleft{}Report of the Secretaries\textquotedblright{} in the conference proceedings. Prandtl had gathered the leading turbulence researchers of his time to this event. The most important talks, other than the one by Prandtl, were the overview lecture by  Taylor on \textquotedblleft{}Some Recent Developments in the Study of Turbulence\textquotedblright{} \cite{Taylor:1938b} and Dryden's presentation \cite{Dryden:1938} with his measurements of the energy spectrum. It is interesting to note, that in concluding his paper, Dryden presented a single hotwire technique that he intended to use to measure the turbulent shearing stress $u^{'}v^{'}$.  Prandtl's laboratory under Reichardt's leadership already had found a solution earlier\footnote{Taylor to Prandtl, 18 March 1938. MPGA, Abt. III, Rep. 61, Nr. 1654.}  and Prandtl  presented this data in his talk.  Reichardt used hot wire anemometry with a probe consisting of three parallel wires, where the center wire was  mounted a few millimeters downstream and used as a temperature probe\cite{Reichardt:1938a,Reichardt:1938b,Reichardt:1951E}.  The transverse component of velocity was sensed by the wake of one of the front wires.  This probe was calibrated  in oscillation laminar flow.  At the discussion of  Dryden's paper Prandtl made the following important remark concerning the turbulent boundary layer: \textquotedblleft{}One can assume that the boundary zone represents the true `eddy factory' and the spread towards the middle would be more passive.\textquotedblright{} In his comment, he also showed a copy of the fluctuation measurements conducted by Reichardt and Motzfeld \cite[here Fig. 3]{Reichardt:1938b} in a windtunnel of 1m width and 24cm height. 

Prandtl's paper \cite{Prandtl:1938} at the turbulence symposium deserves a closer review because, on the one hand,  it became the foundation for his further work on turbulence and, on the other hand, as an illustration of Prandtl's style. It reflects beautifully and exemplarily, what von Kármán, for example, admired as Prandtl's \textquotedblleft{}ability to establish systems of simplified equations which expressed the essential physical relations and dropped the nonessentials\textquotedblright{}; von Kármán regarded this ability \textquotedblleft{}unique\textquotedblright{} and even compared Prandtl in this regard with \textquotedblleft{}his great predecessors in the field of mechanics -- men like Leonhard Euler (1707--1783)  and d'Alembert (1717--1783).\textquotedblright{}\footnote{\cite{Karman:1957,Anderson:2005}; see also Prandtl's own enlightening contribution to this topic \cite{Prandtl:1948b}}
   
In this paper, Prandtl distinguished four types of turbulence: wall turbulence, free turbulence, turbulence in stratified flows (see also \cite{Prandtl-Reichardt:1934}), and the decaying isotropic turbulence. He first considered the decay law of turbulence behind a grid using his mixing length approach by assuming that the fluctuating velocity is generated at a time $t^,$ and then decays: 
$u^{,2}= \int_{-\infty}^t \! \frac{\mathrm{d}{t^,}}{T}~[ (l_1  \frac{dU}{dy})_{t^,} \times f(\frac{t-t^,}{T})]^2$ with $f(\frac{t-t^,}{T})\approx~\frac{T}{T+t-t^,}$ justified by Dryden's measurements. With the shear stress $\tau = \rho u{^,}l_2 \frac{dU}{dy}$ and $l_2 = k m$, where $m$ is the grid spacing and $k \approx 0.103$, he  derived  $u{^,}= \frac{const}{t+T} = \frac{c U_m}{x}$ where  $x$ is the downstream distance from the grid, $c$ is related to the thickness of the rods, and $U_m$ is the mean flow. From the equation of motion to lowest order, $U_m \frac{\partial U}{\partial x} = \frac{1}{\rho}\frac{\partial \tau}{\partial y}$, and the Ansatz  $U = U_m + A x{^{-n}} cos(\frac{2 \pi y}{m})$, he obtained $n = 4 \pi{^2} k \frac{c}{m}$.  By analyzing the largest frequency component of  $U-U_m$ (from data provided by Dryden) he determined $n \approx 4.5$.  This result led him to assume a transition from anisotropic flow near the grid to isotropic turbulence further downstream. How this transition occurs was left open. 

As a next item, Prandtl  considered the  change of a wall-bounded, turbulent flow at the transition from a smooth to a rough wall and vice versa. He derived model equations and found reasonable agreement with measurements of his student Willi Jacobs.

\begin{figure}
\begin{center}
\includegraphics[scale=0.35]{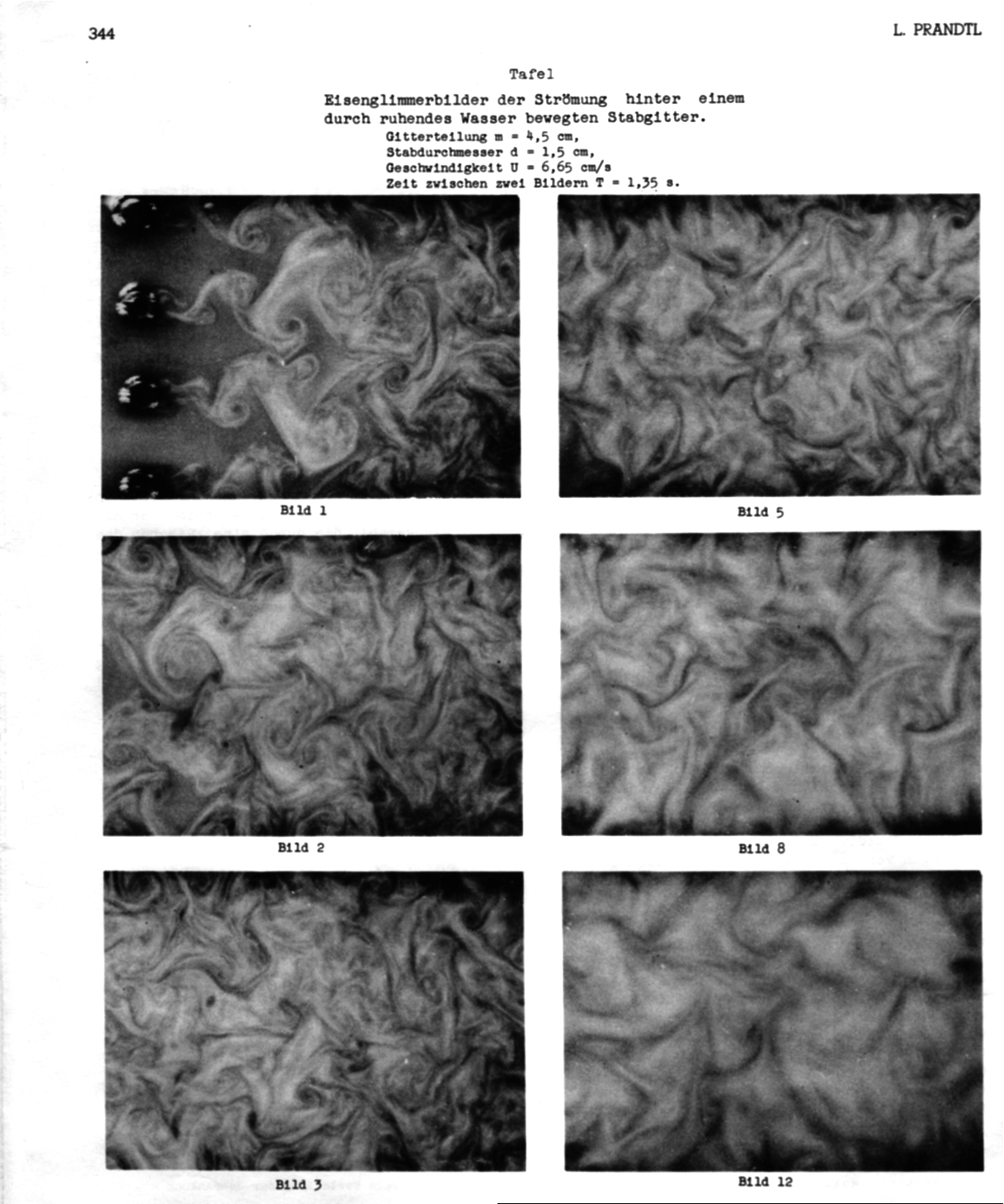}
\bigskip
\caption{Prandtl's visualization of the development of grid generated isotropic turbulence. Pictures were taken at relative grid spacings of  $2,4,6,10,16$ and $24$.}
\label{fig:pict}
\end{center}
\end{figure}

The third item of Prandtl's conference paper concerned an ingeniously simple experiment. By visualizing the flow with iron-glance flakes (the same one he had used in his 1904 experiments) he measured what he preceived as the {}``Taylor-scale''{} of turbulence in a grid generated turbulent water flow (see Fig.~\ref{fig:pict}; in the region of large shear the flakes align and make visible the eddies in the turbulent flow). From the surface area per eddy as a function of mesh distances behind the grid (see Fig. 1 in \cite{Prandtl:1938}) he found that these areas grew linearly starting from about 16 mesh distances  downstream from the grid. From this observation Prandtl concluded that the Taylor scale $\lambda$ \cite{Taylor:1935b} increases like $(x-x_0)^{0.5}$, where $x_0 \approx 10$ is  the \textquotedblleft{} starting length\textquotedblright{}. This was in contradiction with Taylor's own result, but agreed with the prediction by von Kármán and Howard \cite{KarmanHoward:1938} from eight months earlier. (It is not clear whether Prandtl knew about their work -- it is not credited in his paper. Later the  von Kármán and Howard prediction was quantitatively measured with hot-wires  by Comte-Bellot and Corrsin \cite{Comte-Bellot-Corrsin:1966}).

\begin{figure}
\begin{center}
\subfigure[Fluctuating quantities measured in a flow between two parallel plates for $Re_{max} = 17500$.]{\includegraphics[scale=0.31]{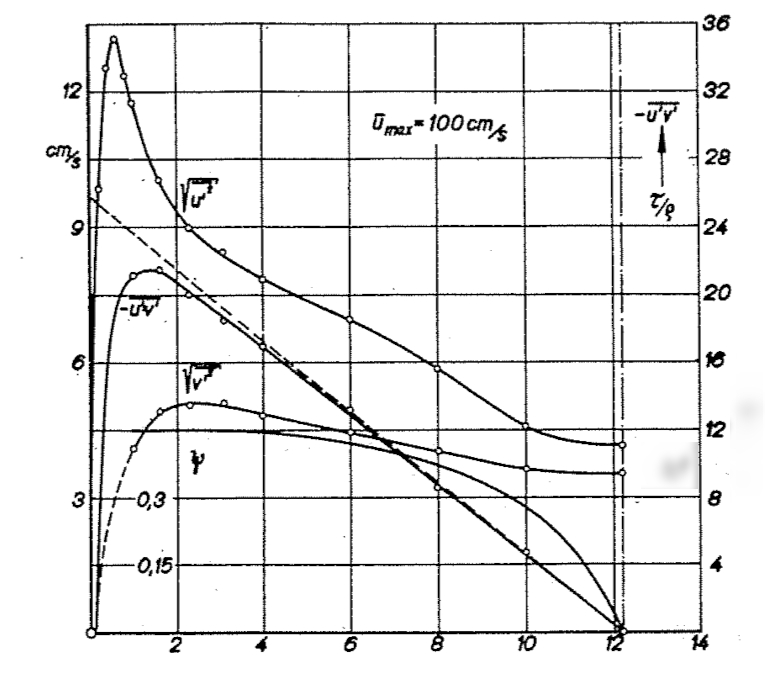}}
\subfigure[Velocity spectra for different distances from the wall in turbulent channel flow (Wandturbulenz).]{\includegraphics[scale=0.32]{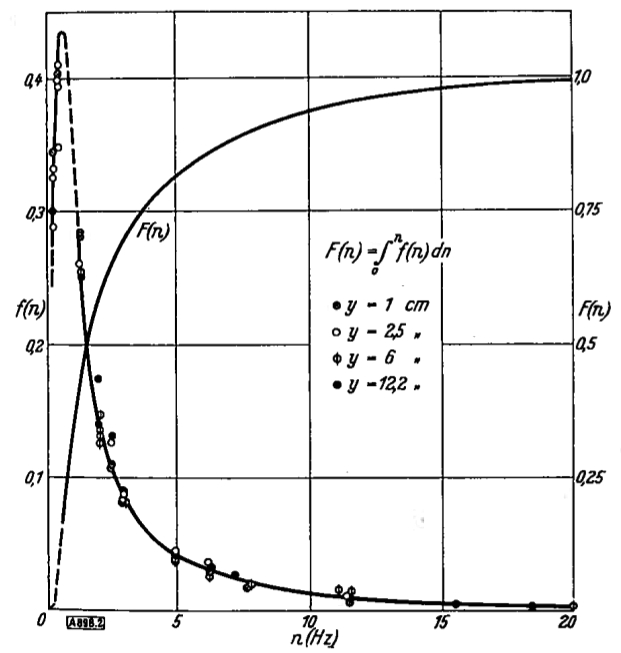}}
\bigskip
\bigskip
\subfigure[Log/lin plot of the  spectrum in (a)]{\includegraphics[scale=0.21]{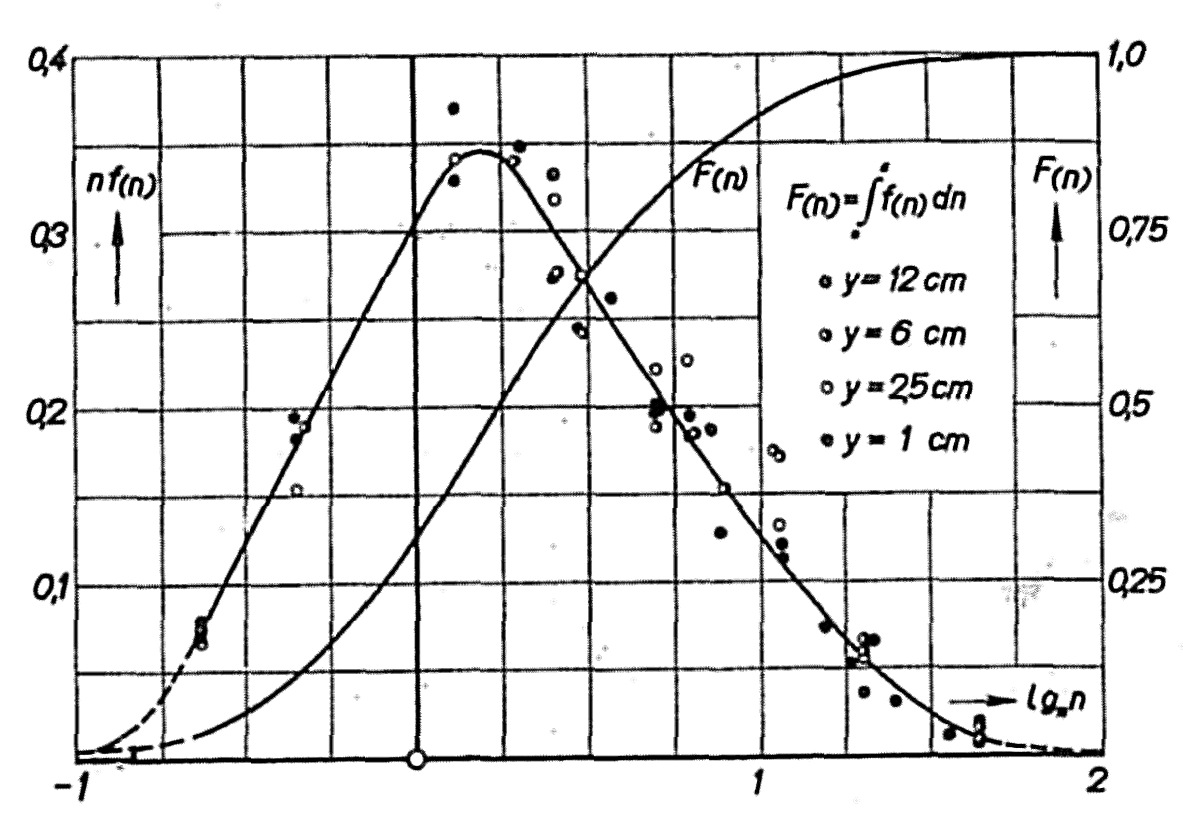}}
\subfigure[Velocity spectrum compared between shear--generated (Wandturbulenz -- wall turbulence) and grid--generated decaying turbulence (Freie Turbulenz -- free turbulence )]{\includegraphics[scale=0.35]{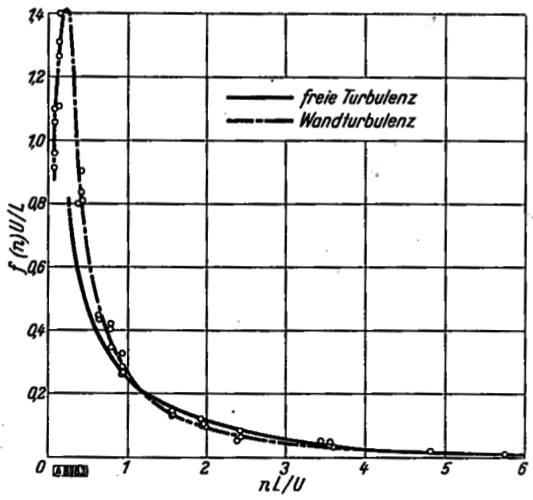}}
\bigskip
\caption{Turbulence spectra as measured by Motzfeld and Reichardt in 1938.}
\label{fig:camdata}
\end{center}
\end{figure}

Finally, and in retrospect most importantly, Prandtl discussed Reichardt's and Motzfeld's measurements of wall generated turbulence in channel-flow \cite{Reichardt:1938a,Reichardt:1938b,Reichardt:1951E,Motzfeld:1938}.   In Fig.~\ref{fig:camdata}(a)  we reproduce his Fig 2. It  displays the  mean fluctuating quantities $\sqrt{\overline{u{^{,2}}}}$,  $\sqrt{\overline{v{^{,2}}}}$, $\overline{v{^,}u{^,}}$, and $ \Psi =\overline{v{^,}u{^,}}/(\sqrt{\overline{u{^{,2}}}}\, \sqrt{\overline{v{^{,2}}}})$ as a function of distance from the wall to the middle of the tunnel at 12 cm.  Here  $u$ is the streamwise, $v$ the wall normal, and $w$ wall parallel velocity.

\bigskip

\begin{figure}
\begin{center}
\includegraphics[scale=0.18]{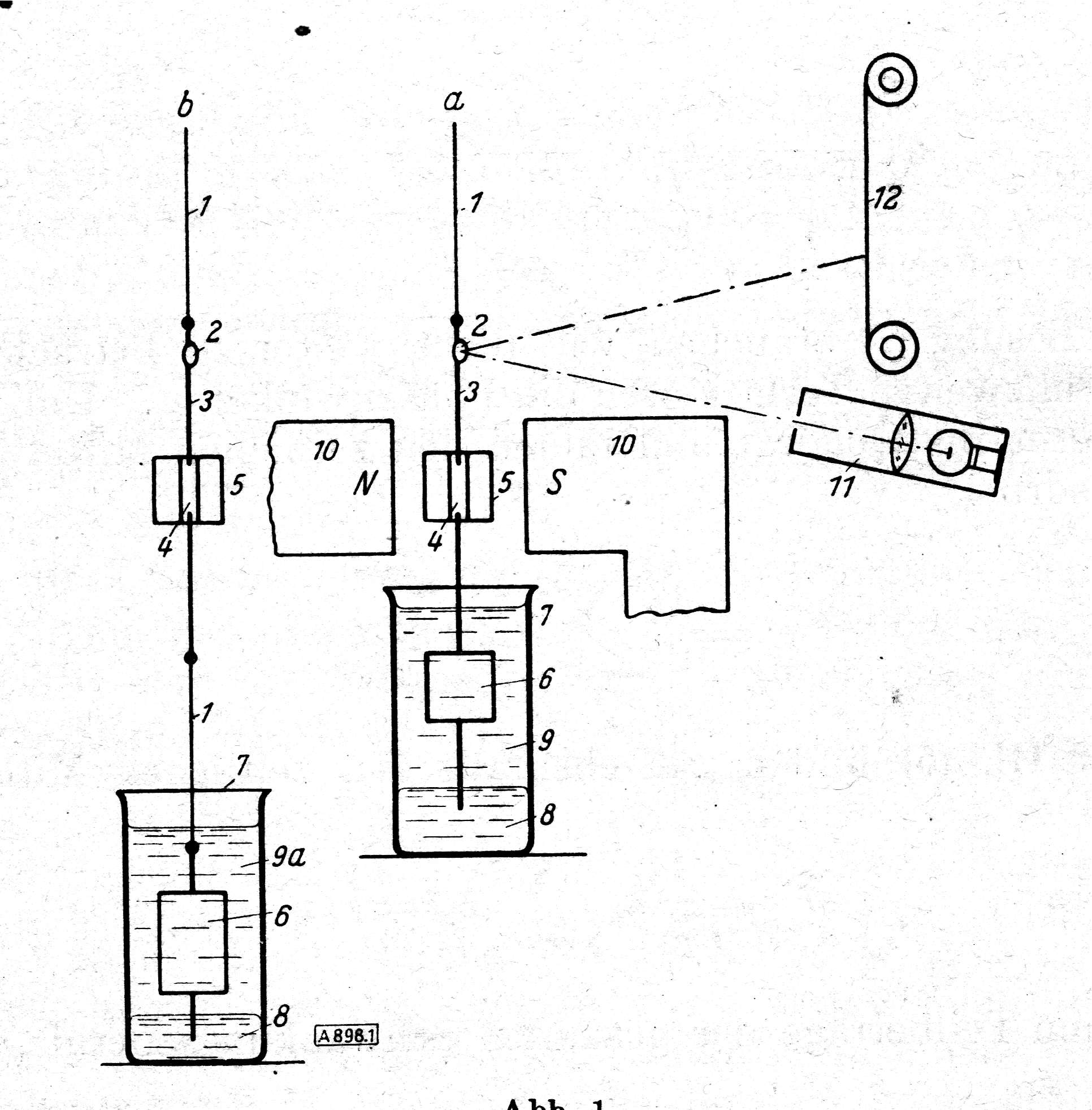}
\caption{Schematic of the electromechanical spectral analysis system used by Motzfeld in 1937/38. Two different designs of a damped torsion pendulum were used: (a) below 20Hz and (b) above 20Hz.  The design (a) consisted of a  torsion wire (1), a thin metal rod (3) with mirror (2), an insulating glass rod (4) around which a coil was wound (5), a swinger consisting of a thin metal rod with a cylindrical body to add inertia (6). The swinger was placed in a beaker (7) that was filled on  the top with a damping fluid (9) and on the bottom with mercury (8).  Electric currents could flow from (1) into the coil and from there to (8). The electromagnet was placed into  a permanent magnetic field.  The deflections of the wire were recorded on  photographic film that was transported with a motor. In the alternative design (b)  for more than 20 Hz the swinger was replaced with a torsion wire (1) and a weight (6). The weight was placed into very viscous oils so that it did not move (9a). Otherwise the design was the same. Altogether 10 swingers were used with 6 of kind (a) and 4 of kind (b). The resonance frequencies were between 0.2Hz and 43Hz.}
\label{fig:camdata2}
\end{center}
\end{figure}

The spectral analysis of the streamwise velocity fluctuation $u$ as a function of frequency revealed that the frequency powerspectra were indistinguishable for distances of 1 cm to 12 cm   from the wall , although, $\overline{u^,}$  decreased by more than a factor of two over the same range (see Fig.~\ref{fig:camdata}(a), reproduced from Motzfeld \cite{Motzfeld:1938}). Prandtl showed in his Fig. 4 (here Fig.~\ref{fig:camdata}(b) the semi log plot of $n f(n)$, which displays  a maximum at the largest scales of the flow.  Later the location of this maximum was proposed as a surrogate for the integral scale \cite{LumleyPanofsky:1964}.  As shown in Fig.~\ref{fig:camdata}(c), Motzfeld also compared his data with the 1938  windtunnel data by  Simmons and Salter \cite{Simmons-Salter-Taylor:1938} by rescaling both datasets with the mean velocity $U$ and the channel height $L$ or, for the windtunnel, with the grid spacing $L$.  As we can see both datasets agree reasonable well.  Prandtl remarked about these surprising collapse of the data in Fig.~\ref{fig:camdata}(a) (Fig. 3 in \cite{Prandtl:1938}): \textquotedblleft{}The most remarkable about these measurements is that de facto the same frequency distribution was found.\textquotedblright{} From the perspective of experiments, the electromechanical measurement technique employed by Motzfeld and Reichardt is also remarkable. As shown in Fig.~\ref{fig:camdata2}, they used  the amplitude of an electromechanically driven and viscously damped torsion wire resonant oscillator to measure by tuning resonance frequencies and damping the frequency components of the hot-wire signal.   

As we will see, these results would lead Prandtl in 1945 \cite{Prandtl-Wieghardt:1945} not only to derive what is now known as the \textquotedblleft{}One Equation Model\textquotedblright{} \cite{Spalding:1991}, but also to assume an universal energy cascade of turbulence cut  off at the dissipation scale \cite{Prandtl:1945,Prandtl:1948a}. Thus, at age 70, Prandtl had finally found what he was looking for all his life albeit at the worst time--when the Second World War ended and  he was not allowed to conduct scientific research.\footnote{Prandtl to Taylor, 18 July 1945 and 11 Oktober 1945. MPGA, Abt. III, Rep. 61, Nr. 1654} 

Prandtl's sojourn in the U.S.A. in September 1938 was also remarkable in another regard -- because it marked the beginning of his, and for that matter Germany's, alienation from the international community. When he tried to convince the conference  committee to have him organize the next congress in Germany, he encountered strong opposition based on political and humanitarian reasons. Against many of his foreign colleagues, Prandtl defended Hitler's politics and actions. Taylor attempted to cure Prandtl from his political views.\footnote{Taylor to Prandtl, 27  September 1938. MPGA, Abt. III, Rep. 61, Nr. 1654.}  As discussed also in the  chapter on Taylor by Sreenivasan, Taylor was concerned with the humanitarian situation of the Jewish population and the political situation in general. However, Taylor's candor (he called Hitler \textquotedblleft{}a criminal lunatic\textquotedblright{}) did not bode well with Prandtl, who responded again with defending German politics.\footnote{Prandtl to Taylor, 29 October 1938. MPGA, Abt. III, Rep. 61, Nr. 1654.} Only a few days before (on 18 October 1938)  12,000 Polish-born Jews were expelled from Germany. On 11 November 1938,  the atrocities of the \textquotedblleft{}Kristallnacht (the Night of Broken Glass)\textquotedblright{}  started the genocide and holocaust \cite{Gilbert:2006}.  Taylor replied on 16 November 1938 with a report about the very bad experiences which his own family members had in Germany.\footnote{Taylor to Prandtl, 16 November 1938. GOAR 3670-1} Nevertheless he ended his letter still quite friendly: \textquotedblleft{}You will see that we are not likely to agree on political matters so it would be best to say no more  about them.  Fortunately there is no reason, why people who do not agree politically should not be best friends.\textquotedblright{} Then he continued to make a remark that he does not understand why Prandtl plotted $n f(n)$ instead of $f(n)$ (shown in Fig. 5b) (Fig.4 in \cite{Prandtl:1938}).  As far as we know Prandtl never replied. After this correspondence Prandtl  wrote one more letter to Mrs. Taylor.\footnote{Prandtl to Mrs. Taylor, 5 August 1939. MPGA, Abt. III, Rep. 61, Nr. 1654.} Only a month later WWII started and cut off their communication. Prandtl tried to resume the contact with Taylor after the war,\footnote{Prandtl to Taylor, 18 July 1945 and 11 Oktober 1945. MPGA, Abt. III, Rep. 61, Nr. 1654} but there is no evidence that Taylor ever responded to this effort. 

\subsection*{Fully Developed Turbulence III - 1939 to 1945}

With the beginning of WWII  on 1 September 1939, German research in fluid dynamics became isolated from the rest of the world.  This may explain why the very important discovery by  Motzfeld and Reichardt was not recognized abroad. We have found no reference to Motzfeld's 1938 publication other than in the unpublished 1945 paper by Prandtl.  As described above their discovery showed that the spectrum of the streamwise velocity fluctuation in a channel flow did not depend on the location of the measurements in  the channel  and did agree qualitatively with those by  Simmons and Salter for decaying isotropic turbulence.   The  1938 G\"ottingen results  show beautifully the universal behavior that Kolmogorov postulated  in his revolutionizing 1941 work (see chapter on the Russian School by Falkovich).

Prandtl and his coworkers were not aware of the developments  in Russia  and continued their program in turbulence at a slower pace. According to a British Intelligence report after the war, based on an interrogation of Prandtl, \textquotedblleft{}due to more urgent practical problems little fundamental work, either experimental or theoretical,  had been conducted during the war. No work had been done in Germany similar to that of G.I. Taylor or Kármán and Howarth on the statistical theory of turbulence. Experiments had been planned on the decay of turbulence behind grids in a  wind tunnel analogous to those undertaken by Simmons of the National Physical Laboratories, but these were shelved at the outbreak of the war.\textquotedblright{}\footnote{British Intelligence Objectives Sub-Committee report 760  that summarizes a visit at the KWI 26--30 April 1946.}  

Indeed as far as the fully developed turbulence was concerned the progress was mostly theoretical and mostly relying on measurements before the war.  In his response to the military interrogators Prandtl  was very modest. From late autumn of 1944 till the middle of 1945 he worked on the theory of fully developed turbulence almost daily (see Fig.~\ref{fig:calender}). This was his most active period in which he pulled together the threads outlined earlier.

We will now review briefly the development from 1939 to 1944 that led to this stage. The status of the knowledge of turbulence in 1941 is well summarized in \cite{Wieghardt:1941,Wieghardt:1942E}, that between 1941 and 1944 in Prandtl's FIAT article entitled \emph{Turbulenz} \cite{Prandtl:1948a}.  Prandtl reviewed on twenty three tightly written pages the work at the KWI in chronological order and by these topics:

\begin{enumerate}
  \item Turbulence in the presence of walls
  \begin{enumerate}
    \item  Pipeflow
    \item  Flat plates
    \item  Flow along wall with pressure increase and decrease
  \end{enumerate}
   \item Free Turbulence
  \begin{enumerate}
    \item  General laws
    \item  Special tasks
    \item  Properties of  jets in jet engines
  \end{enumerate}
 \item Various investigations
  \begin{enumerate}
    \item  Turbulence measurement technologies
    \item  Heat exchange
    \item  Geophysical applications
    \item  Fundamental questions
  \end{enumerate}
\end{enumerate}

Prandtl identified as fundamental and important in particular the work by  Schultz-Grunow \cite{Schultz-Grunow:1940,Schultz-Grunow:1941E} and Wieghardt \cite{Wieghardt:1944} on the measurements of the turbulent boundary layer. Even today, these very careful and now classical experiments provide the data for quantitative comparisons \cite{Nagib:2007}. 

 Furthermore, Prandtl singled out the investigations on heat transfer in turbulent boundary layers by Reichardt \cite{Reichardt:1944}, who applied ideas from earlier papers on turbulent transport of momentum in a free jet \cite{Reichardt:1941,Reichardt:1942}. Reichardt had found experimentally for a planar jet that the PDF of transverse variations of the streamwise velocity profile in the middle of a  jet was Gaussian. In the middle of such flows $\frac{\partial \overline{u}}{\partial y} = 0$, where the mixing length approach failed by design,  as Prandtl  had noted already in 1925, where he suggested another way around this problem \cite{Prandtl:1925}). Based on the observation of the Gaussian distribution he conjectured inductively that the transfer of momentum was similar to that of heat.  By neglecting viscosity he wrote the 2d planar momentum equation  $\frac{\partial}{\partial x} (p/\rho + \overline{u^2})  + \frac{\partial \overline{(uv)}}{\partial y} = 0$ and $\overline{uv} = - \lambda \frac{\partial \overline{u^2}}{\partial y}$, with $\lambda$ as \emph{\"Ubertragunsgr\"osse} (transfer quantity).  Reichardt calculated some examples and showed that  the new theory worked reasonably well. Prandtl had published about it already in 1942 in ZAMM \cite{Prandtl:1942c}. He showed that if the pressure term in lowest order is zero the two equations by Reichardt lead to $\frac{\partial}{\partial x}\overline{u^2} = - \lambda \frac{\partial^2\overline{u^2}}{\partial y^2}$. In a subsequent paper Henry G\"ortler applied the theory to four cases:  the plane mixing layer, the plane jet, the plane wake and the plane grid \cite{Goertler:1942}. He compared the first two cases with the measurements by Reichardt and found good agreement.  

 As another important result Prandtl highlighted improvements of the hot-wire measurement system by H. Schuh who found a method to circumvent the otherwise very tedious calibration of each new hotwire probe in a calibration tunnel \cite{Schuh:1945,Schuh:1946}. With regard to theoretical achievements, Prandtl reported about the work of the mathematician Georg Hamel  who had proven von Kármán's 1930 similarity hypothesis for the two-dimensional flow in a channel as well as Prandtl's log law \cite{Hamel:1943,Prandtl:1925}.

At the end of the FIAT paper, Prandtl mentioned rather briefly with what he was so deeply engaged from the autumn of 1944 to summer of 1945. In only a little more than a page he summarized his energy model of turbulence (the {}``One Equation Model''{} \cite{Spalding:1991}) and his own derivation of the Kolmogorov length scales, for which he used a cascade model of energy transfer to the smallest scales. The latter he attributed to his unpublished manuscript from 1945 (see the discussion below). Then he reviewed the  work by Weizs\"acker \cite{Weizs\"acker:1948} and Heisenberg \cite{Heisenberg:1948,Heisenberg:1958E} that both conducted while detained in England from July 1945 to January 1946.\footnote{Their work has also been reviewed by Batchelor in December 1946 together with the work by Kolmogorov and Onsager \cite{Batchelor:1946}. Of course Batchelor had no knowledge of the fact that Prandtl had derived the same results based on similar reasoning already in January 1945.}.  Weizs\"acker's work was similar, but Prandtl considered it to be mathematically more rigorous than his phenomenologically driven approach.  In addition to the results that Prandtl had obtained,   Weizs\"acker calculated  from the energy transport  also the $k^{-5/3}$ scaling of the energy spectrum. He then reviewed the  Fouriermode analysis by Heisenberg and commented on the good agreement with experiments. He closed with a hint at a paper in preparation by Weizs\"acker concerning the influence of turbulence on cosmogony.

\subsection*{Prandtl's Two Manuscripts on Turbulence, 1944--1945}

When the American troops occupied  G\"ottingen  on 8 April 1945,  Prandtl had already published the {}``One Equation Model''{}  \cite{Prandtl-Wieghardt:1945}  and drafted a first typewritten manuscript of a paper entitled {}``The Role of Viscosity in the Mechanism of Developed  Turbulence''{} that was last dated by him 4 July 1945 \cite{Prandtl:1945} (see Fig.~\ref{fig:calender}). In this paper he derived from a cascade model the dissipation length scale, \emph{i.e.}, the {}``Kolmogorov length''{}.  Before we describe in more detail his discoveries, it is important to ask  why he did not publish this work. Clearly this was an important discovery and  would have retrospectively placed him next to Kolmogorov in the {}``remarkable series of coincidences''{} \cite[p. 883]{Batchelor:1946} now  known as the K41 theory.

His drafting of the paper fell right into the end of  WWII. By July1945 the institute was under British administration and had \textquotedblleft{}many British and American visitors.\textquotedblright{}\footnote{Prandtl to Taylor, 18 July 1945. MPGA, Abt. III, Rep. 61, Nr. 1654.} Prandtl was still allowed \textquotedblleft{}to work on some problems that were not finished during the war and from which also reports were expected. Starting any new work was forbidden.\textquotedblright{}\footnote{Prandtl to Taylor, 18 July 1945. MPGA, Abt. III, Rep. 61, Nr. 1654. See also Prandtl to the President of the Royal Society, London, 11 October 1945. MPGA, Abt. III, Rep. 61, Nr. 1402.} By 11 October 1945 the chances for publication were even worse because \textquotedblleft{}all research was shelved completely\textquotedblright{} and \textquotedblleft{}any continuation of research was forbidden by the Director of Scientific Research in London.\textquotedblright{}\footnote{Prandtl to Taylor, 18 October 1945. MPGA, Abt. III, Rep. 61, Nr. 1654.} So it seems that  as of August 1945 Prandtl followed orders and stopped writing the paper and stopped working on turbulence (see Fig.~\ref{fig:calender}). In addition, in that period the AVA was being disassembled and the parts were sent to England. Then in January 1946,  Heisenberg and Weizs\"acker returned to G\"ottingen from being interned in England  and brought along  their calculations that superseded Prandtl's work. So by January 1946 the window of opportunity for publication had passed. In addition, he was busily writing the FIAT report on turbulence--and that is where he at least mentioned his work.

Let us now discuss briefly  Prandtl's last known work on turbulence. His very carefully written notes\footnote{GOAR 3727} cover the period from 14 October 1944 until 12 August 1945; they allow us to better understand his achievement.  These notes comprise 65 numbered pages, 5 pages on his talk in a Theory Colloquium on 4 January 1945 where he presented his energy equation of turbulence, 7 pages on a trial to understand the distribution function of velocity from probability arguments and 19 pages of sketches and calculations.  Figure~\ref{fig:calender} summarizes the days he entered careful handwritten notes in his workbook. From these entries we see that he devoted a large part of his time to the study of turbulence. It seems remarkable how much time he was able dedicate to this topic, considering that he also directed the research at his institute and that he was engaged as an advisor for the Air Ministry with the direction of aeronautical war research. How much effort he dedicated to the latter activity is open to further historical inquiry.  

In order to discern the subsequent stages of Prandtl's approach we proceed chronologically:

\begin{figure}
\begin{center}
\includegraphics[scale=0.8]{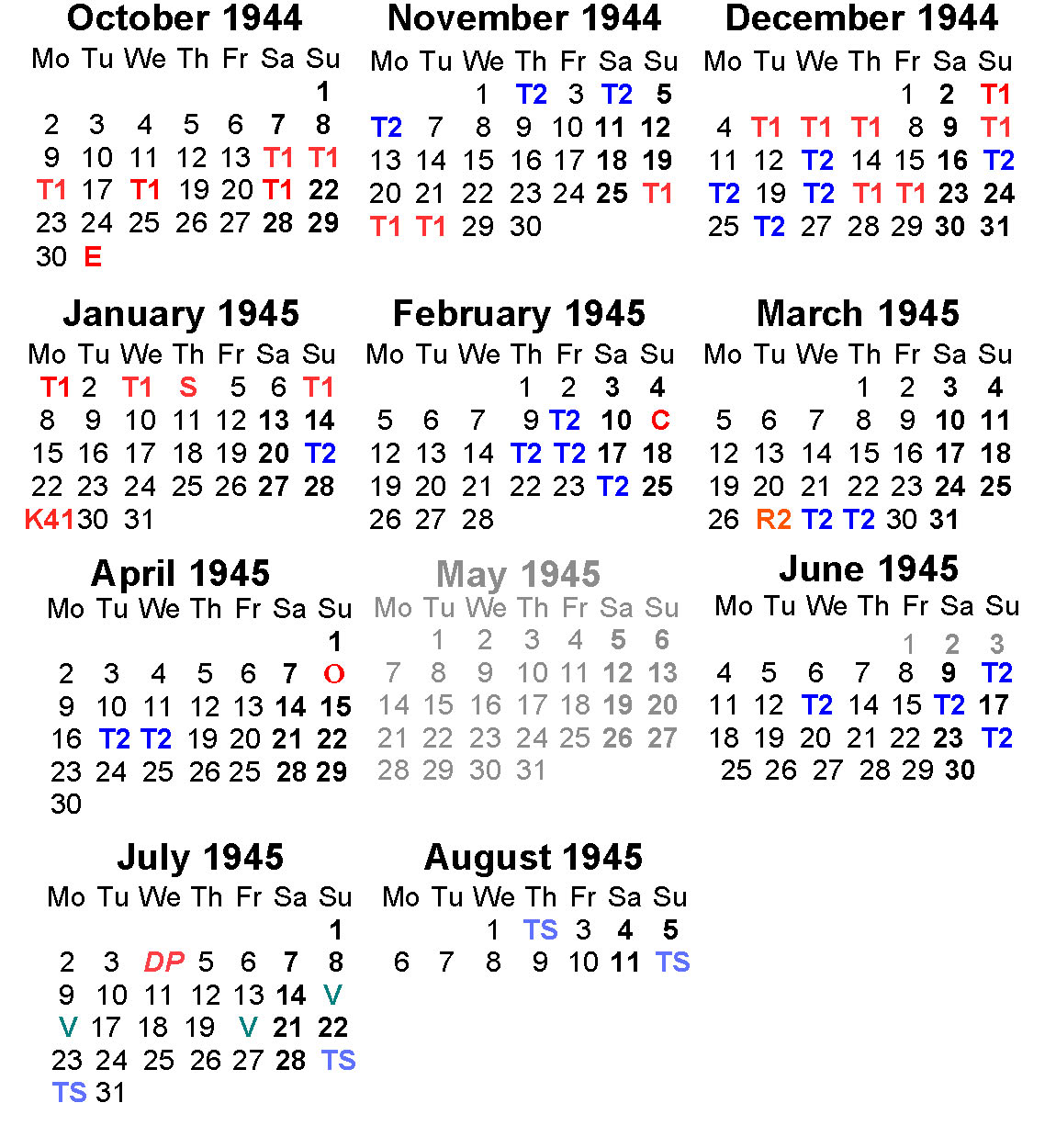}
\caption{Prandtl worked continuously on the topic of fully developed turbulence. T1 marks Prandtl's work on the energy equation of turbulence, T2 his investigations on the effect of dissipation, V a derivation of the vorticity equation in a plane shear flow, and TS  his attempts to develop a statistical theory of velocity fluctuations. The other letters mark important dates: on 31 October 1944 he formulated for the first time the {}``One Equation Model''{} (E); on 4 January 1945 he presented it at a theory seminar (S) and  on 26 January 1945 at a meeting of the Göttingen Academy of Science (A); 29 January 1945 marks his discovery of what is known as the  Kolmogorov length scale (K41); on 4 February 1945 he had his 70s birthday (B); on 11 February 1945 he formulated for the first time his cascade model (C); on 27 March 1945 he is reworking the draft for the paper on dissipation (R2); on 8 April 1945 G\"ottingen was occupied by American forces; on 4 July 1945 Prandtl entered remarks on the already type written draft revisions of the dissipation paper. The period in May, where he had no access to the institute as it was used by  American forces is  light gray -- the institute reopened on 4 June 1945 to close again briefly thereafter.}
\label{fig:calender}
\end{center}
\end{figure}

\begin{figure}
\begin{center}
\includegraphics[scale=0.9]{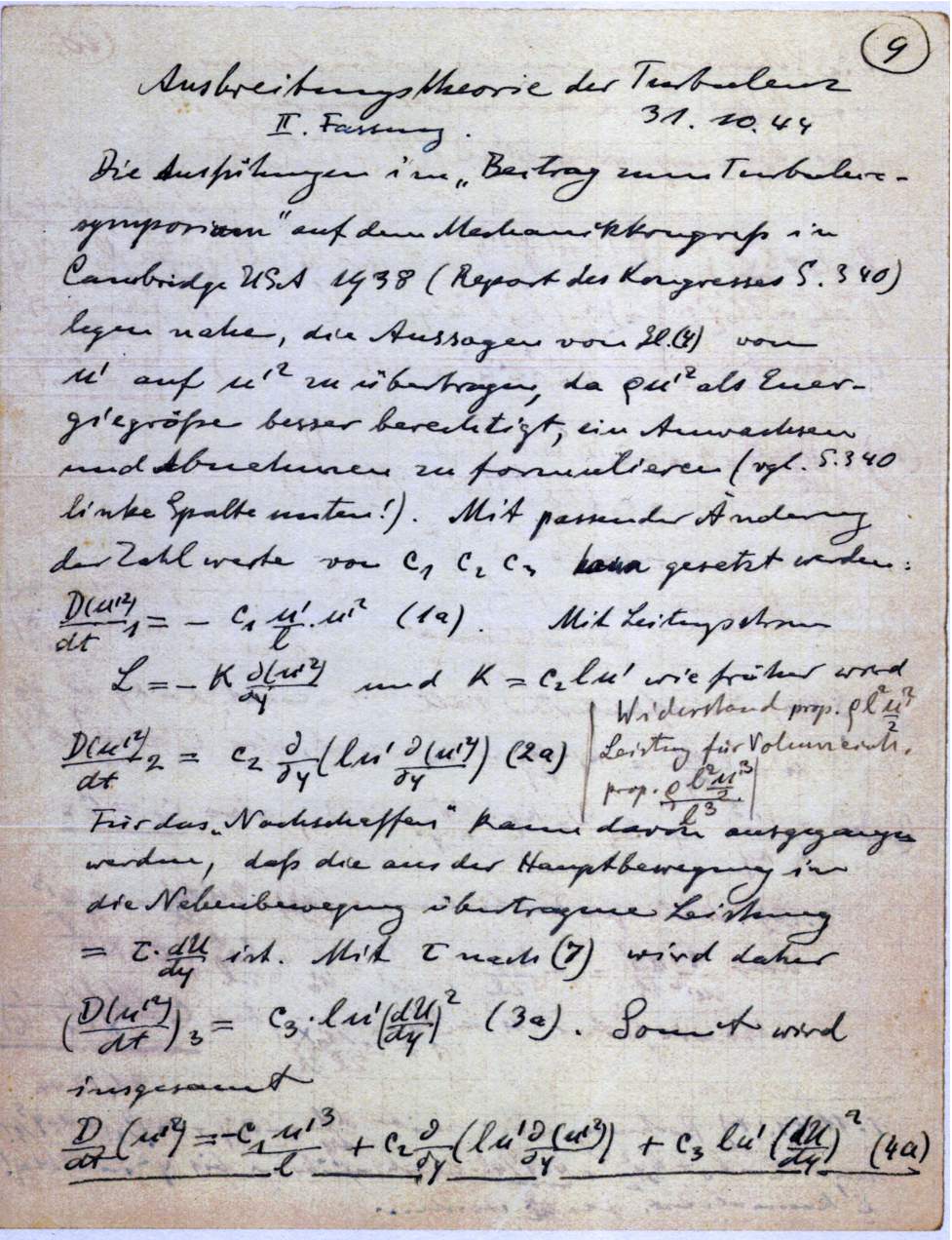}
\caption{ Derivation of energy-balance equation without dissipation }
\label{fig:energy}
\end{center}
\end{figure}

\begin{figure}
\begin{center}
\includegraphics[scale=0.9]{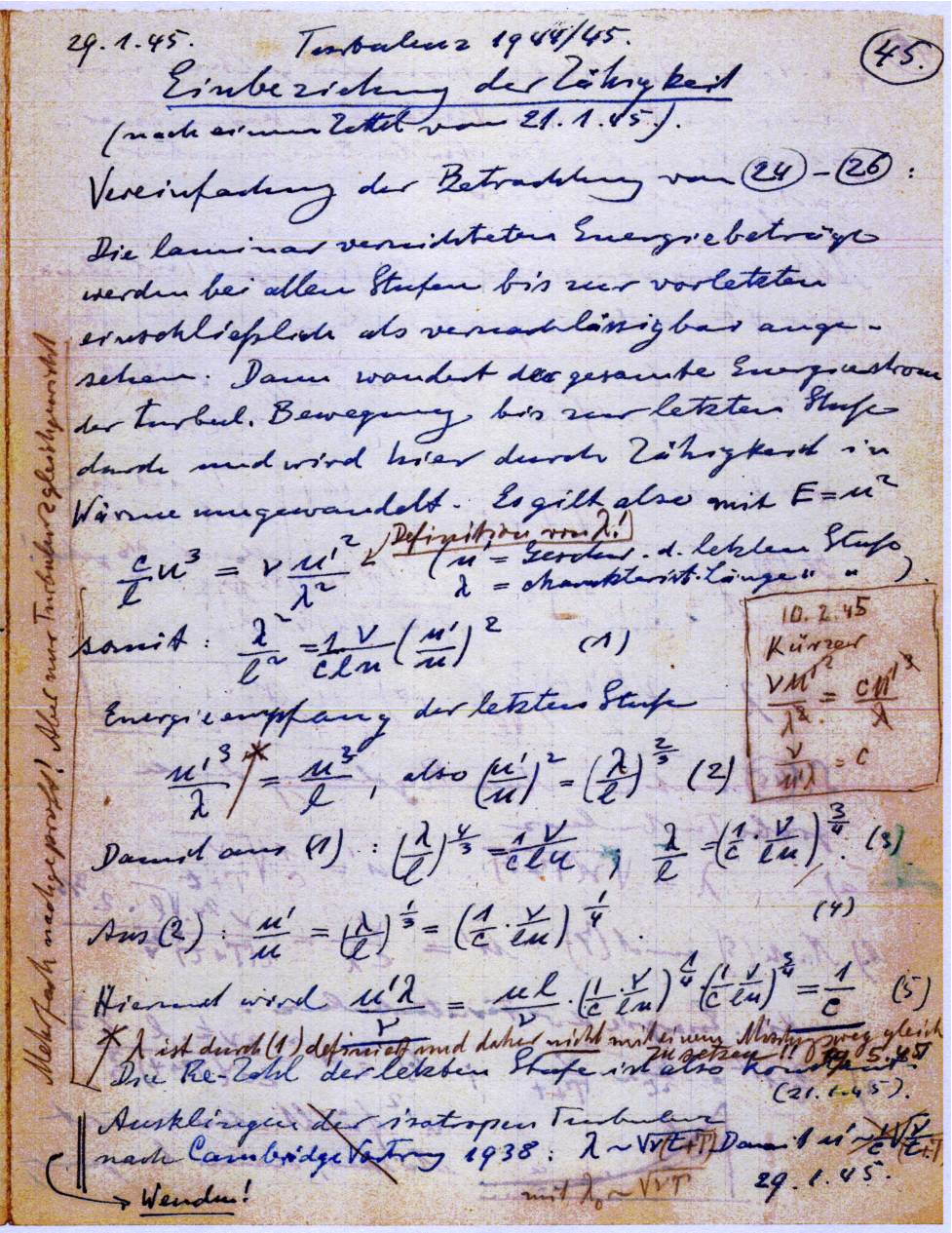}
\caption{First known derivation of the Kolmogorov length scale (here called $\lambda$)}
\label{fig:Kolmlength}
\end{center}
\end{figure}

\begin{figure}
\begin{center}
\includegraphics[scale=0.9]{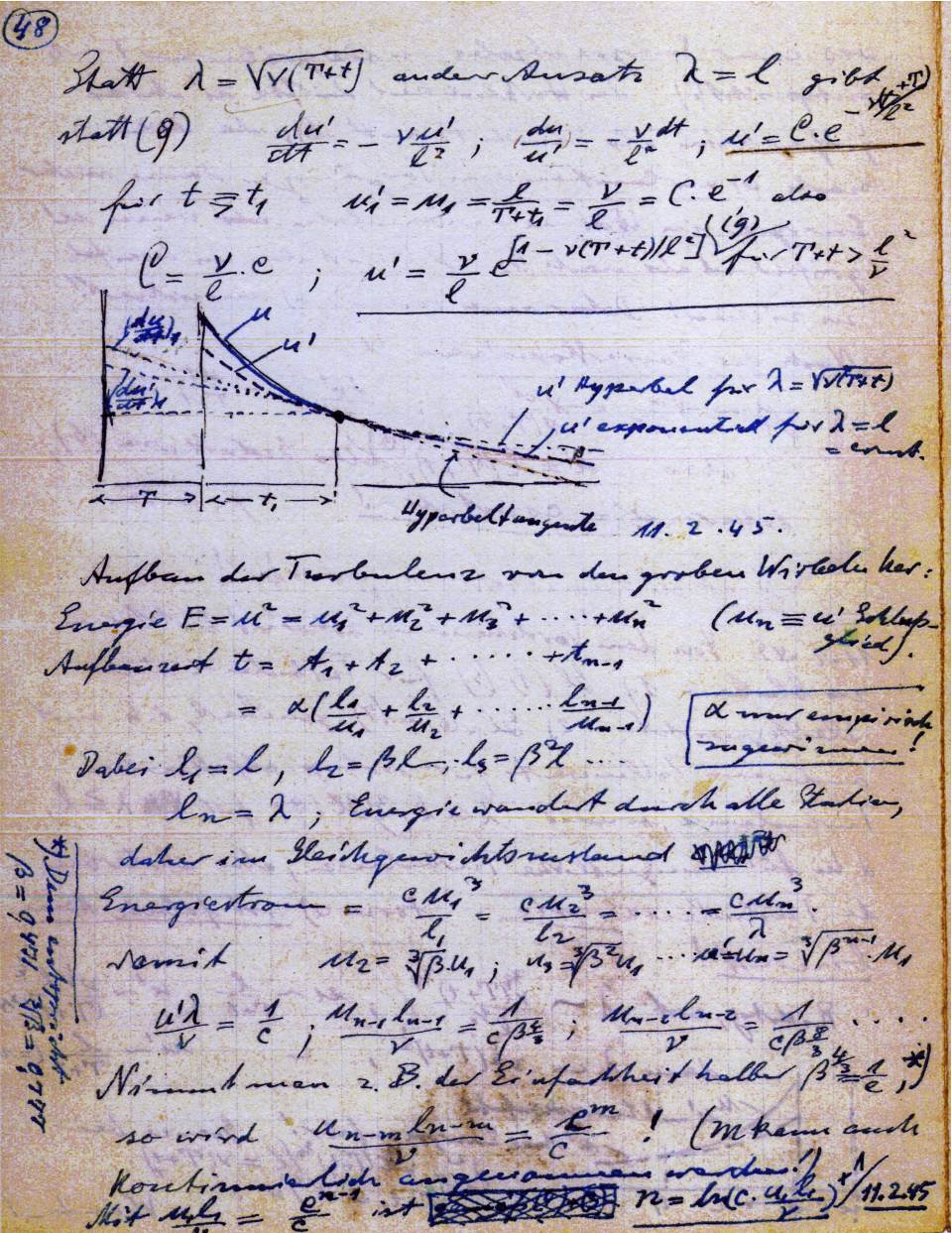}
\caption{Prandtl's cascade model for the fluctuating velocities at different steps in the cascade}
\label{fig:Kolmlength1}
\end{center}
\end{figure}

On 31 October 1944 we find the first complete formulation of the {}``One Equation Model''{} for the evolution of turbulent kinetic energy per unit volume in terms of the square of the fluctuating velocity (see Fig.~\ref{fig:energy}). The same formula now written in terms of kinetic energy per unit volume  was  presented by him in his paper at the  meeting of the G\"ottingen Academy of Science on 26 January 1945 \cite{Prandtl-Wieghardt:1945}. There Wieghardt  also presented his determination of  the parameters from measurements in  grid turbulence and channel flow and found good agreement with the  theory.  His differential equation marked as Eq. 4a of  Fig.~\ref{fig:energy} determined the change of energy per unit volume from three terms: the first term on the right hand side gives the turbulent energy flux for a {}``bale of turbulence''{} (in German: Turbulenzballen) of size $l$ (which he equated with a mixing length), the second term represents the  diffusion of turbulence in the direction of the  gradient of turbulent energy and the third term represents the source of turbulent energy from the mean shear.  $cu{^,} l$ is the eddy viscosity. Please note that this equation is what is now called a k-model.  This equation was independently derived later by Howard Emmons in 1954 and by Peter Bradshaw in 1967 \cite{Spalding:1993}. Prandtl and Wieghardt also pointed out the deficiencies of the model, namely the role of viscosity at the wall and for the inner structure of a bale of turbulence. For the latter, Prandtl argued that, as long as the Reynolds number of each bale of turbulence is large, a  three dimensional version of his equation should also be applicable to the inner dynamics of the bale of turbulence. He then introduced the idea of bales of turbulence  within bales of turbulence, which we now know as the turbulent cascade. He called them steps (in German: Stufen) in the sequence that goes from large to small. He pointed out that by going from step to step the Reynolds number will decrease with increasing number of the step (decreasing size size of the turbulent eddy) until viscosity is dominant and all energy is transformed to heat. Finally he conjectures that a general understanding of the process can be obtained.

And indeed he discovered it within a short time.  On  29 January 1945, only three days after the presentation at the Academy, he entered in his notes the derivation of the Kolmogorov length scale that at this time he called in analogy to Taylor's smallest length scale $\lambda$ (Fig.~\ref{fig:Kolmlength}). In Eq. 1 the decay reate of the kinetic energy per unit mass is equated with the dissipation at the smallest scales. Eq. 2 connects the final step of the cascade process with the Kolmogorov velocity. By putting (1) and (2) together Prandtl arrived at the Kolmogorov length scale given by eq. (3). This seemed to him so remarkable that he commented on the side of the page {}``Checked it multiple times! But only equilibrium of turbulence.''{}

At this stage he was almost done, but as we can tell from his typewritten manuscript \cite{Prandtl:1945} and from his notes he was not satisfied. He had to put this on more formal grounds. So only  two weeks later, as shown in Fig.~\ref{fig:Kolmlength1} he used a cascade model for each step of turbulence, with $\beta$ as the ratio in length scale from step to step. This allowed him to derive the Kolmogorov length more rigorously from a geometrical series.

This became the content of his draft paper from 1945 that we shall discuss in detail in a separate publication. It is clear that Motzfeld's spectral data \cite{Motzfeld:1938} were instrumental for his progress (those which Prandtl had presented at the Cambridge congress in 1938 and which are reproduced above in Fig.~\ref{fig:camdata}). Here we close our review with a quote from the introduction to his unpublished paper on {}``The Role of Viscosity in the Mechanism of Developed  Turbulence''{} \cite{Prandtl:1945} which  beautifully reflects his thinking and needs no further analysis.  This is only a short excerpt from the introduction of the paper. A full translation of the paper is in preparation and will be published elsewhere. Also, our translation is very close to the original German text and therefore some sentences are a bit long. 

\begin{quote}

The following analyses, which consider in detail the inner processes of a turbulent flow, will prove that the solution for $\lambda$ by Taylor that he obtained from energy considerations is not yet giving the smallest element of turbulence. The mechanism of turbulence generation is not resolved in all details. So much is however known (here Prandtl referred in a footnote to work by Tollmien published in G\"ottinger Nachr. Heft 1 (1935) p. 79) that flows with an inflection point in the velocity profile may become unstable at sufficiently large Reynolds numbers.  Therefore one has to expect that at sufficiently high Reynolds number $\frac{u{^,} l}{\nu}$ the motion of an individual bale of turbulence is by itself turbulent, and that for this secondary turbulence the same is true, and so on. Indeed one observes already at very modest Reynolds numbers a frequency spectrum that extends over many decades. That mostly the smallest eddies are responsible for the conversion of the energy of main motion into heat can easily be understood, as for them the deformation velocities  $\frac{\partial u}{\partial y}$, etc. are the largest.

The earlier discussion is the simplest explanation for the fact that in the turbulent motion always the smaller eddies are present next to the larger ones. G. I. Taylor, 1935, used a different explanation. He pointed out that according to general statistical relations the probability of two particles separating in time is larger than for them to come closer, and he applied this relationship to two particles on a vortex line. From the well-known Helmholtz's theorem it would follow that -- as long as the viscosity does not act in the opposing sense -- the increase of the angular velocity of the vortex line is more probable than its decay. He shows this tendency with an example, whose series expansion clearly shows the evolution towards smaller eddies. However, it could not be continued, so the processes could only be followed for short time intervals. One can counter Taylor's deductions insofar that through the increase of the angular velocities pressure fields develop, which oppose a further increase of the vortex lines. It can thus not be expected that the extension would reach the expected strength. It seems, however, that the action in the sense of Taylor is surely present, if though,  with weaker magnitude than expected from a purely kinematic study. 

For the development of smaller eddy diameters in the turbulence, one can also note that the wall turbulence starts with thin boundary layers and that the free turbulence has  equally thin separating sheets. Therefore, in the beginning only smallest vortices are present and the larger ones appear one after another. Opposing this, however, is the result that in the fully developed channel flow the frequency spectrum de facto does not depend on the distance from the wall \cite{Motzfeld:1938}. This one would not expect, if all of the fine turbulence would originate at the wall. This strongly supports the validity of the conjecture for the stationary turbulence presented here. A further support is given by investigations conducted later, which concerned isotropic, temporally decaying turbulence and which have been quite satisfactorily justified by experiments. The two descriptions of the recreation of the smaller eddies by the turbulence of second and higher order and the one that relates to the Helmholtz theorem, are, by the way, intricately related, they are both, so to say,  descriptions that elucidate one and the same process only from different perspectives.

In the following, initially, a temporally stationary turbulence may be assumed, as it is found, for example, in a stationary channel or pipe flow. Of the dissipated power $D$ in a unit volume per unit time, a very small fraction $\mu (\frac{\partial U}{\partial y})^2$ will be dissipated immediately into heat ($U$ velocity of the mean flow); the rest, which one may call $D_1$, increases the kinetic energy of the turbulent submotion (Nebenbewegung) and generates accord to Taylor the secondary turbulence ...  

(Here we leave out some equations.)

We now  establish corresponding equations for turbulence of second step (third step etc.). Instead of the velocity $U$ here a suitably smoothed velocity of the turbulence of first step, second step etc. must be used. The instantaneous values of the velocity $u$, which is used as a representative for the triple (u, v, w) for simplicity, will thus be separated into a sum of partial velocities of which $u_1$ is the smoothest main part of u and represents the {``first step''},  correspondingly the smaller, but finer structured part $u_2$ the second step etc.; the n--th order shall be the last one in the series that will no longer become turbulent (here Prandtl added in a footnote: The separation into steps thereby creates difficulties, that the elements of the first step do not all have the same size and that in the following the differences may increase even further. As the purpose of the analysis is only a rough estimate one may conjecture that the elements in each step have the same well-defined size. A more detailed analysis by considering the statistical ensemble of turbulence elements is an aim for the future.)

Motivated by the way the $u_i$  are introduced, it seems natural to assume that their effective values $u_i^2$ build a geometric series, at least with the exclusion of the last members of the series, for which viscosity is already noticeable. For a first approximation one may assume, that also the last members of the series, other than the very last one, are members the a geometric series. 
\end{quote} 

By this reasoning Prandtl  ended the geometric series by closing it with the  last step at  which all energy dissipation occurs. His final derivation of the Kolmogorov length scale is then quite similar  to what he calculated on 29 January 1945 (see Fig.~\ref{fig:Kolmlength}).
\begin{figure}
\begin{center}
\includegraphics[scale=0.9]{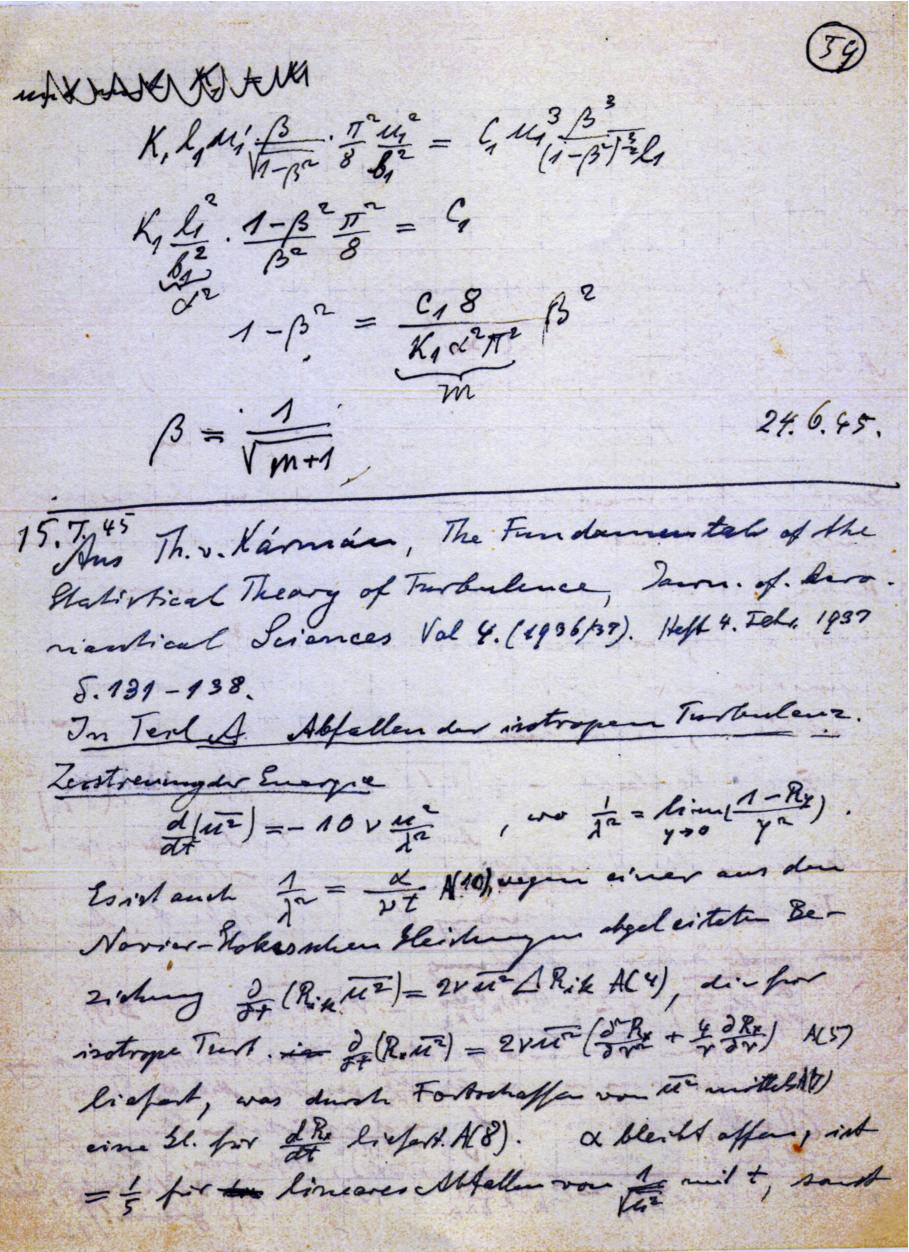}
\caption{Calculation of the mixing length from the vorticity equation 1 of 4 }
\label{fig:vort1}
\end{center}
\end{figure}

\begin{figure}
\begin{center}
\includegraphics[scale=0.9]{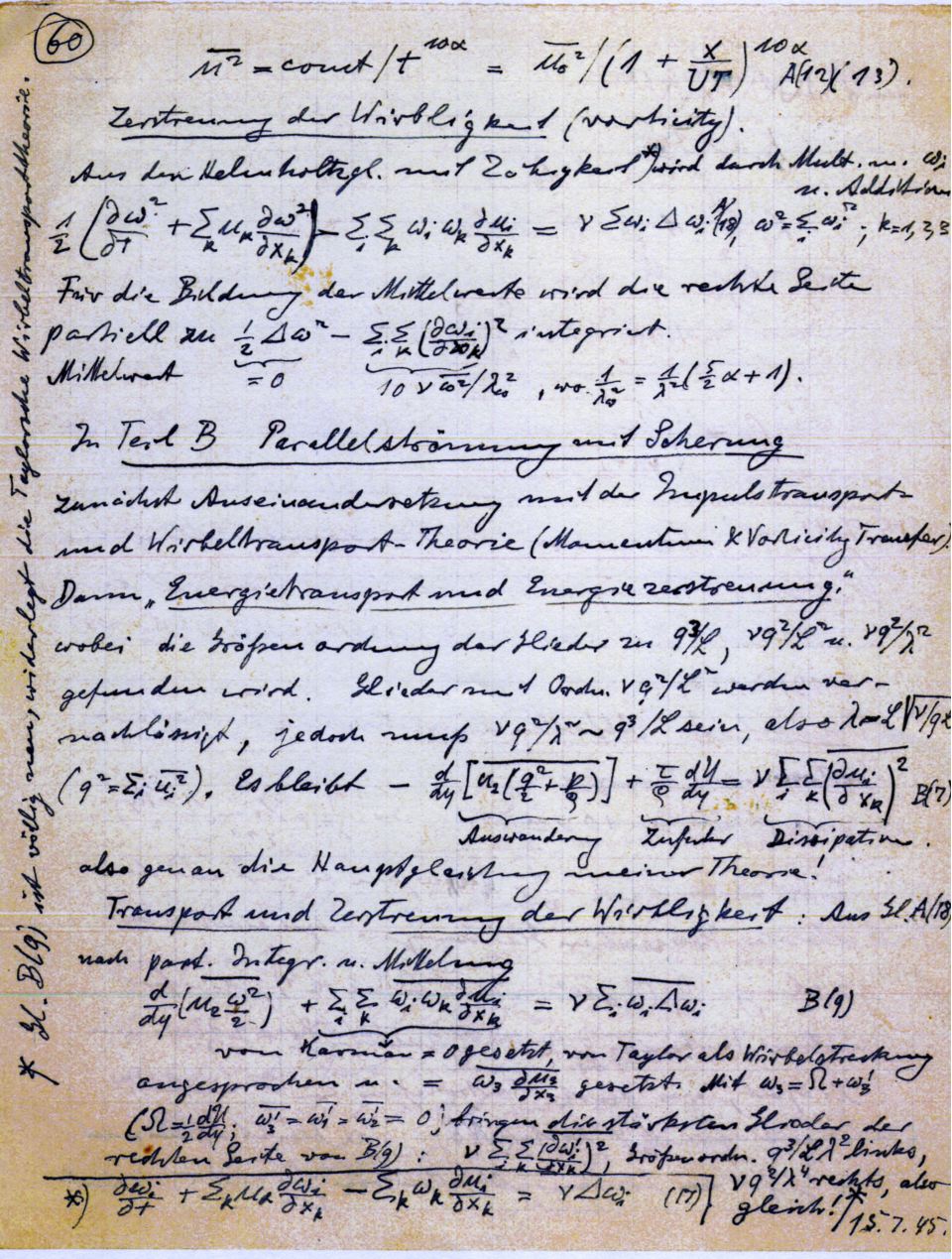}
\caption{Calculation of the mixing length  from the vorticity equation 2 of 4 }
\label{fig:vort2}

\end{center}
\end{figure}
\begin{figure}
\begin{center}
\includegraphics[scale=0.9]{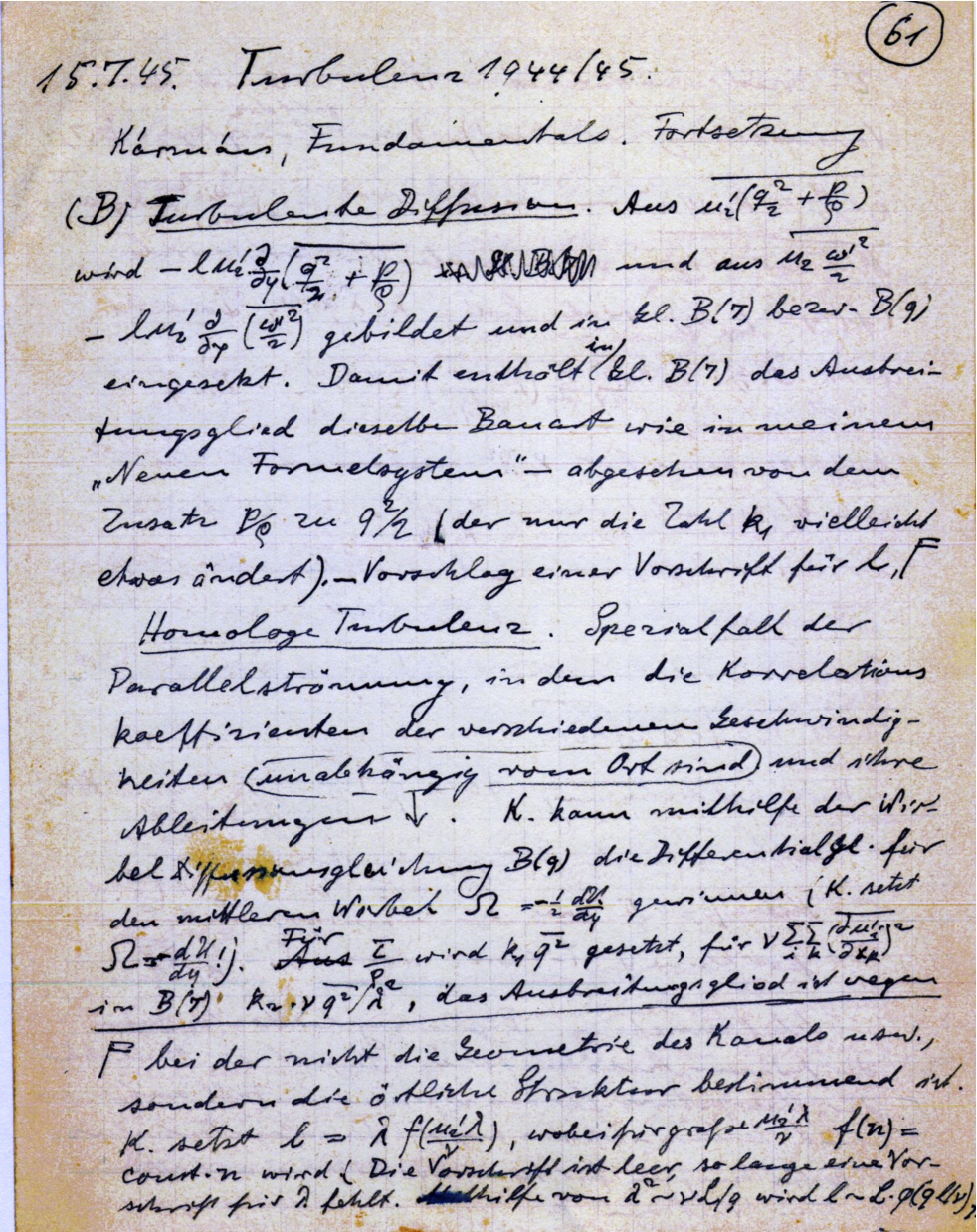}
\caption{Calculation of the mixing length from the vorticity equation 3 of 4 }
\label{fig:vort3}
\end{center}

\end{figure}
\begin{figure}
\begin{center}
\includegraphics[scale=0.9]{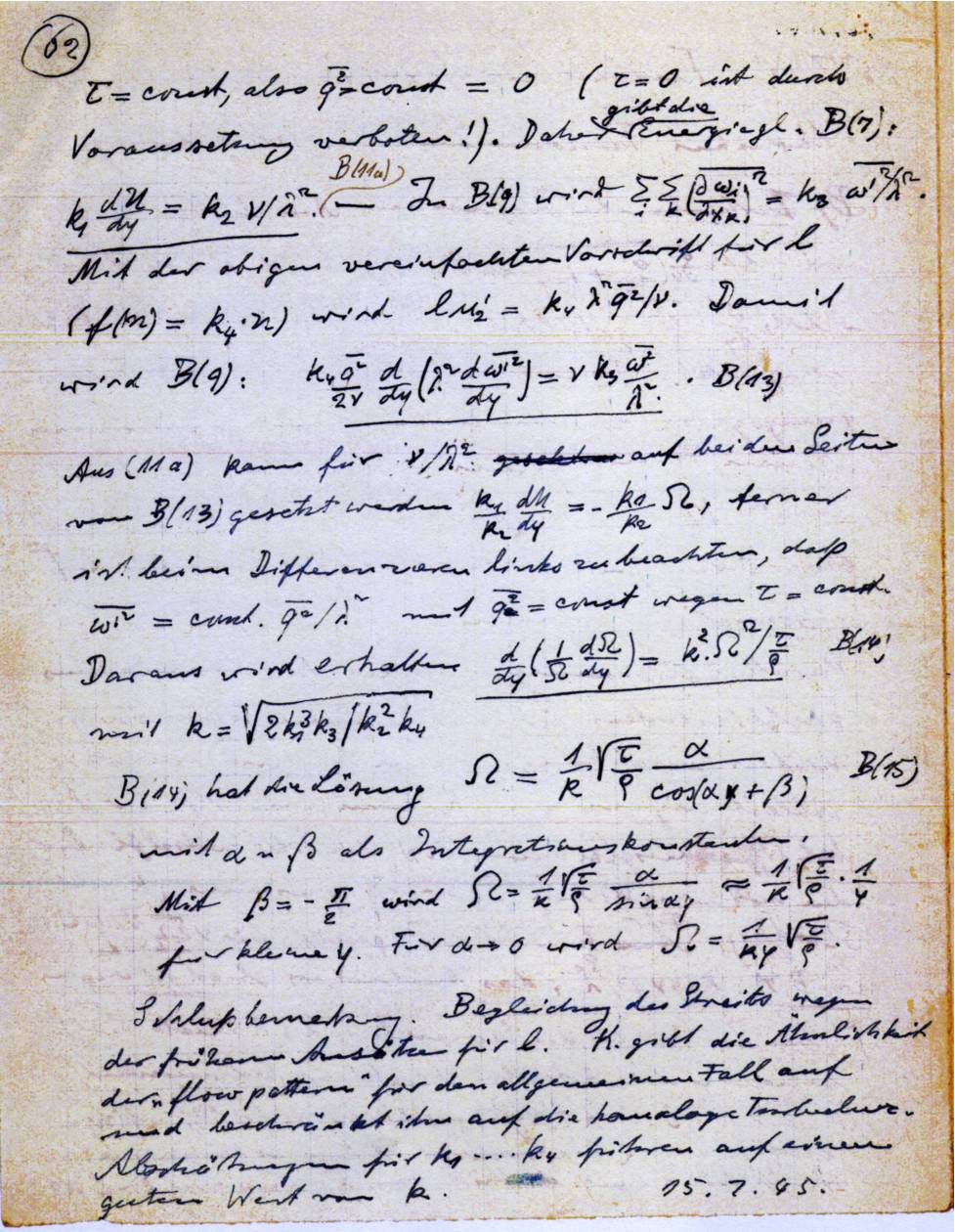}
\caption{Calculation of the mixing length from the vorticity equation 4 of 4 }
\label{fig:vort4}
\end{center}
\end{figure}

\bigskip

This did not conclude Ludwig Prandtl's quest for the understanding of turbulence.  In mid July 1945  he had realized that his one equation model was missing a second equation that allowed him to determine the mixing length. Therefore he resorted to the vorticity equation that  Karman had investigated.  As shown in Figs.~\ref{fig:vort1}--~\ref{fig:vort4} he calculated with help of his vorticity equation B9 (see Fig.~\ref{fig:vort2}) for the case of plane shear flow under the assumption of {}``homologue''{} turbulence (for which the correlation coefficients of the velocity components are independent of space) the mixing length and $\frac{dU}{dy}$. 
\bigskip

So from October 1944 to August 1945, Prandtl had returned to his life long quest for the understanding of turbulence. On 17 September 1945 the Georg August University was reopened as the first German University and Prandtl taught again. By January 1946,  Otto Hahn,  Werner Heisenberg, Max von Laue and Carl Friedrich von Weizs\"acker returned from England to Prandtl's institute that was reopened on 1 August 1946.  Max Planck became interim president of the Kaiser-Wilhelm-Society, which now had its head quarters in the buildings of Prandtl's institute.  On 11 September 1946 the Max Planck Society was founded as the successor of the Kaiser-Wilhelm-Society in Bad Driburg. On 26 February 1948 the Max Planck Society convened its constitutional meeting in the cafeteria of Prandtl's institute.   

Prandtl himself retired from the university and institute's directorship in the fall of 1946 and continued working on problems in meteorology until his death on 15 August 1953.  

\subsection*{Conclusion}

Prandtl's achievements in fluid mechanics generally, and in turbulence in particular, are often characterized by the label `theory'. However, it is important to note that he did not perceive himself as a theoretician. When the German Physical Society of the British Zone awarded him two years after the war the honorary membership, he used this occasion to clarify his research style in a lecture entitled ``My approach towards hydrodynamical theories.'' With regard to the boundary layer theory, for example, he argued that he was guided by a ``heuristic principle'' of this kind: ``If the whole problem appears mathematically hopeless, see what happens if an essential parameter of the problem approaches zero.'' \cite[p. 1606]{Prandtl:1948b} His notes amply illustrate how he used one or another assumption, often combined with clever dimensional arguments, in order to single out those features of a problem which he regarded crucial. He always attempted to gain ``a thorough visual impression'' about the problems with which he was concerned. ``The equations come later when I think that I have grasped the matter.'' \cite[p. 1604]{Prandtl:1948b}

By the same token, Prandtl's approach to theory relied heavily on practice. For that matter, practice could be an observation of flow phenomena in a water channel, an ``experimentum crucis'' like the trip-wire test, or a challenge posed by practical applications such as skin friction. Prandtl's FIAT review on turbulence, in particular, illustrates how his theoretical research was motivated and guided by practice. As we have seen, Prandtl named explicitly, among others, \cite{Schultz-Grunow:1940}, \cite{Wieghardt:1944}, \cite{Reichardt:1944} and \cite{Schuh:1945} as important roots for the theoretical insight expressed in  \cite{Prandtl-Wieghardt:1945}. Prandtl's closest collaborator for the fundamental studies on fully developed turbulence, Wieghardt, was by the same time elaborating a technical expertise on the skin friction of rubber with regard to a possible use for the hull of submarines \cite[p. 58]{Prandtl:1948a}. These and other war-related studies were based on experimental turbulence measurements in the same ``roughness tunnel'' that provided the data for the more fundamental inquiries.

Prandtl's style as well as the closeness of theory and practice is also reflected in the third edition of Prandtl's famous ``Essentials of Fluid Mechanics'' \cite{Prandtl:1948c}. In a paragraph about the onset of turbulence, for example, Prandtl reported about the recent confirmation of the Tollmien-Schlichting theory by the experiments in Dryden's laboratory at the National Bureau of Standards in Washington. Turbulent jets and turbulent shear flow along walls were discussed in terms of the mixing length approach \cite[pp. 115-123]{Prandtl:1948c}. Isotropic turbulence was summarized rather cursorily, with a reference to his FIAT review and the recent work by Weizsäcker and Heisenberg \cite[p. 127]{Prandtl:1948c}. In general, Prandtl  preferred textual and pictorial presentations supported by experiments over sophisticated mathematical derivations.    

For a deeper understanding of Prandtl's and his Göttingen school's contributions to turbulence it would be necessary to account for the broader research conducted at the KWI and the AVA, which covered a host of fundamental and applied topics, from solid elasticity to gas dynamics and meteorology. Research on turbulence was never pursued as an isolated topic. But in view of its ultimate importance for engineering, turbulence always remained an important and challenging problem.  Among the variety of research problems dealt with at Göttingen in the era of Prandtl, turbulence may be regarded as the one with the longest tradition -- from Klein's seminar in 1907 to the climax of Prandtl's unpublished manuscripts in 1945. 

A number of questions have been left unanswered. The timing of Prandtl's breakthrough during the last months of the Second World War, in particular, suggests further inquiries: To what extent was fundamental research on turbulence interrupted during the war by Prandtl's involvement as an scientific advisor for the ministry of aviation (Reichsluftfahrtsministerium)with regards to aeronautical war research?\footnote{from 6 July 1942 Prandtl became the chair of the scientific  research council of the ministry of aviation led by G\"oring and was pushing for fundamental  research in war related matters\cite{Maier:2007}}  Or, was the renewed interest in the basic riddles of turbulence sparked by the wartime applications? Or, to the contrary,  did Prandtl at the end of the war find the time to work on what he was really interested in? 

Both Prandtl's advisory role as well as his local responsibilities for fluid dynamics research at G\"ottingen came to a sudden stop when the American and British troops occupied his institute and prevented further research -- a prohibition which Prandtl perceived as unwarranted. Not only did he write to Taylor for help\footnote{Prandtl to Taylor, 18 October 1945. MPGA, Abt. III, Rep. 61, Nr. 1654.}, but also did he request help from the President of the Royal Society, where he was a foreign member since 1928.\footnote{Prandtl to Royal Society, 18 October 1945. MPGA Abt.III, Rep. 61, Nr. 1402.} ``The continuation of the research activity that had to be shelved during the war should not be hindered any more!''{}, Prandtl demanded in this letter. His request remained unanswered.  

This correspondence provokes further questions regarding Prandtl's political attitude. Biographical knowledge on Prandtl has been provided by his family \cite{Vogel-Prandtl:1993},  by admiring disciples \cite{Flugge:1973,Oswatitsch:1987}, and by reviews on German wartime aeronautical research \cite{Trischler:1994}; a more complete view based on the rich sources preserved in the archives in Göttingen, Berlin and elsewhere seems expedient.\footnote{Almost a complete set of his correspondence is preserved.} Recent historical studies on the war research at various Kaiser-Wilhelm-Institutes (see, for example, \cite{Maier:2002,Schmaltz:2005,SachseWalker:2005,Maier:2007,Heim:2009,Gruss:2011}) call for further inquiries into Prandtl's motivations for turbulence research. An important question of course is, what we can learn form the position of great men like Prandtl and others in the political web of Nazi Germany? What consequences do arise for the responsibilities of scientist or engineers?  Another lacuna which needs to be addressed  in greater detail concerns the relationship of Prandtl with his colleagues abroad and in Germany, in particular with von Kármán,  Taylor,  Sommerfeld and Heisenberg.  Last, but not least, one may ask for the fate of turbulence research at Göttingen under Prandtl's successors after the war. We leave these and many other questions for future studies.

\selectlanguage{british}%

\bigskip{}

\subsection*{Acknowledgement}
It is our pleasure to express our gratitude to a large number of people without whom this work would not have been possible. We are very thankful to the science historian Florian Schmaltz for sharing his insights and for giving us copies of documents from his  collections. We express our gratitude to  the  Max Planck Society Archives, especially Lorenz Beck, Bernd Hofmann, Susanne \"Ubele und Simone Pelzer;  and to the Archives of the German Aerospace Center (DLR) , especially Jessika Wichner and Andrea Missling. We are also very grateful to K. Sreenivasan, N. Peters and G. Falkovich for sharing their insights on this topic and to the editors of this book for motivating us to write this review and helpful suggestions for improvements of the text.  EB also thanks Haitao Xu and Zellman Warhaft for valuable comments, and Helmut Eckelmann for sharing his memories about the institute. We are most grateful for the understanding and support from our families that were missing their husband and father for long evenings and weekends. This work was generously supported by the Max Planck Society and the Research Institute of the German Museum in Munich. Part of this work was written at  the Kavli Institute for Theoretical Physics and was supported in part by the National Science Foundation under Grant No. NSF PHY05-51164.

\subsection*{Abbreviations}

\noindent DMA: Deutsches Museum, Archiv, München.

\noindent GOAR: Göttinger Archiv des DLR, Göttingen.

\noindent LPGA: Ludwig Prandtls Gesammelte Abhandlungen, herausgegeben
von Walter Tollmien, Hermann Schlichting und Henry Görtler. 3 Bände,
Berlin u. a. 1961.

\noindent MPGA: Max-Planck-Gesellschaft, Archiv, Berlin.

\noindent RANH: Rijksarchief in Noord-Holland, Haarlem.

\noindent SUB: Staats- und Universitätsbibliothek, Göttingen.

\noindent TKC: Theodore von Kármán Collection, Pasadena.


\begin{thebibliography}{75}
\providecommand{\natexlab}[1]{#1}
\providecommand{\url}[1]{\texttt{#1}}
\expandafter\ifx\csname urlstyle\endcsname\relax
  \providecommand{\doi}[1]{doi: #1}\else
  \providecommand{\doi}{doi: \begingroup \urlstyle{rm}\Url}\fi


\bibitem[Anderson(2005)]{Anderson:2005}
John D. Anderson.
\newblock {Ludwig Prandtl's Boundary Layer}.
\newblock \emph{Physics Today}, 58:\penalty0 42--48, 2005.

\bibitem[Batchelor(1946)]{Batchelor:1946}
G. K. Batchelor.
\newblock {Double Velocity Correlation Function in Turbulent Motion}.
\newblock \emph{Nature}, 158, 56:\penalty0 883 -- 884, 1946.



\bibitem[Battimelli(1984)]{Battimelli:1984}
Giovanni Battimelli.
\newblock {The mathematician and the engineer: Statistical theories of
  turbulence in the 20's}.
\newblock \emph{Rivista di storia della scienza}, 1:\penalty0 73--94, 1984.

\bibitem[Blasius(1908)]{Blasius:1908}
Heinrich Blasius.
\newblock {Grenzschichten in Flüssigkeiten bei kleiner Reibung}.
\newblock \emph{Zeitschrift für Mathematik und Physik}, 56:\penalty0 1--37,
  1908.

\bibitem[Blasius(1913)]{Blasius:1913}
Heinrich Blasius.
\newblock {Das Ähnlichkeitsgesetz bei Reibungsvorgängen in Flüssigkeiten}.
\newblock \emph{Forschungsarbeiten auf dem Gebiete des Ingenieurwesens}, 131,
  1913.



\bibitem[Boussinesq(1897)]{Boussinesq:1897}
M.~J. Boussinesq.
\newblock \emph{Theorie de l' ecoulement tourbillonnant et tumultueux des
  liquides dans les lits rectilignes a grandes sections}.
\newblock Gauthiers-Villars et fils, Paris, 1897.


\bibitem[Burgers(1925)]{Burgers:1925}
J.~M. Burgers.
\newblock {The Motion of a Fluid in the Boundary Layer along a Plane Smooth
  Surface}.
\newblock \emph{Proceedings of the First International Congress for Applied
  Mechanics Delft 1924, edited by C. B. Biezeno and J. M. Burgers}, pages
  113--128, 1925.
  
\bibitem[Collar(1978)]{Collar:1978}
A.~R. Collar.
\newblock {Arthur Fage. 4 March 1890-7 November 1977}.
\newblock \emph{Biographical Memoirs of Fellows of the Royal Society}, pages
33--53, 1978.


\bibitem[Comte-Bellot and Corrsin(1966)]{Comte-Bellot-Corrsin:1966}
G.~Comte--Bellot and S.~Corrsin.
\newblock \emph{The use of a contraction to improve the isotropy of grid-generated turbulence}.
\newblock \emph{J. Fluid Mech.}, 25:\penalty0 667--682, 1966. 
 

\bibitem[Darrigol(2005)]{Darrigol:2005}
Olivier Darrigol.
\newblock \emph{Worlds of flow}.
\newblock Oxford University Press, Oxford, 2005.

\bibitem[Dryden and Kuethe (1929)]{Dryden-Kuethe:1929}
Hugh~L.~Dryden, A.~M.~Kuethe 
\newblock {The measurement of fluctuations of air speed by the hot-wire anemometer}.
\newblock \emph{NACA Report}, 320:\penalty0 357--382, 1929.


\bibitem[Dryden et al. (1937)]{Dryden:1937}
Hugh~L.~Dryden, G.~B.~Schubauer, W.~C.~Mock, and H.~K.~Skramstadt
\newblock {Measurements of intensity and scale of wind-tunnel turbulence and their relation to the critical Reynolds number of spheres}.
\newblock \emph{NACA Report}, 581:\penalty0 109--140, 1937.

\bibitem[Dryden(1938)]{Dryden:1938}
Hugh~L.~Dryden.
\newblock {Turbulence Investigations at the National Bureau of Standards}.
\newblock \emph{Proceedings of the Fifth International Congress on Applied
  Mechanics, Cambridge Mass.}, edited by J.P. Den Hartog and H. Peters, John Wiley, New York :\penalty0 362--368, 1938.


\bibitem[Dryden(1955)]{Dryden:1955}
Hugh~L.~Dryden.
\newblock {Fifty Years of Boundary-Layer Theory and Experiment}.
\newblock \emph{Science}, 121:\penalty0 375--380., 1955.

\bibitem[Eckert(2006)]{Eckert:2006}
Michael Eckert.
\newblock \emph{The dawn of fluid dynamics}.
\newblock Wiley-VCH, Weinheim, 2006.

\bibitem[Eckert(2008)]{Eckert:2008b}
Michael Eckert.
\newblock {Theory from Wind Tunnels: Empirical Roots of Twentieth Century Fluid
  Dynamics}.
\newblock \emph{Centaurus}, 50:\penalty0 233--253, 2008.

\bibitem[Eckert(2010)]{Eckert:2010}
Michael Eckert.
\newblock {The troublesome birth of hydrodynamic stability theory: Sommerfeld
  and the turbulence problem}.
\newblock \emph{European Physical Journal, History}, 35:1:\penalty0 29--51,
  2010.

\bibitem[Eiffel(1912)]{Eiffel:1912}
Gustave Eiffel.
\newblock {Sur la r\'{e}sistance des sph\`{e}res dans l'air en mouvement}.
\newblock \emph{Comptes Rendues}, 155:\penalty0 1597--1599, 1912.

\bibitem[Eisner(1932{\natexlab{a}})]{Eisner:1932a}
F.~Eisner.
\newblock {Reibungswiderstand}.
\newblock \emph{Werft, Reederei, Hafen}, 13:\penalty0 207--209,
  1932{\natexlab{a}}.

\bibitem[Eisner(1932{\natexlab{b}})]{Eisner:1932}
Franz Eisner.
\newblock {Reibungswiderstand}.
\newblock \emph{G. Kempf, E. Foerster (Hrsg.): Hydromechanische Probleme des
  Schiffsantriebs. Hamburg}, pages 1--49, 1932{\natexlab{b}}.


  
\bibitem[Fage-Townend(1932)]{Fage-Townend:1932}
A.~Fage and H.~C.~H.~Townend.
\newblock {An Examination of Turbulent Flow with an Ultramicroscope}.
\newblock \emph{Proc. R. Soc. Lond. A},135 :\penalty0 656-677, 1932.


\bibitem[Fl\"ugge-Lotz and Fl\"ugge(1973)]{Flugge:1973}
Irmgard Fl\"uigge-Lotz and Wilhelm Fl\"ugge
\newblock {Ludwig Prandtl in the nineteen-thirties: reminiscences}.
\newblock \emph{Ann. Rev. Fluid Mech,},5 :\penalty0 1--8, 1973.



\bibitem[Föppl(1912)]{Foeppl:1912}
Otto Föppl.
\newblock {Ergebnisse der aerodynamischen Versuchsanstalt von Eiffel,
  verglichen mit den Göttinger Resultaten}.
\newblock \emph{Zeitschrift für Flugtechnik und Motorluftschiffahrt},
  3:\penalty0 118--121, 1912.

\bibitem[Fritsch(1928)]{Fritsch:1928}
Walter Fritsch.
\newblock {Der Einfluss der Wandrauhigkeit auf die turbulente Geschwindigkeitsverteilung in Rinnen}.
\newblock \emph{Abhandlungen aus dem Aerodynamischen Institut der Technischen Hochschule Aachen},  8:1928.

\bibitem[Gilbert(2006)]{Gilbert:2006}
Martin Gilbert.
\newblock {Kristallnacht: Prelude to Destruction}.
\newblock  Harper Collins Publishers, 2006.


 \bibitem[G\"ortler(1942)]{Goertler:1942}
H.  G\"ortler.
\newblock {Berechnungen von Aufgaben der freien Turbulenz auf Grund eines neuen N\"aherungsansatzes}.
\newblock \emph{Zeitschrift für Angewandte Mathematik und Mechanik (ZAMM)},22:244-254, 1942


\bibitem[Goldstein(1969)]{Goldstein:1969}
Sydney Goldstein.
\newblock {Fluid Mechanics in the First Half of this Century}.
\newblock \emph{Annual Review of Fluid Mechanics}, 1:\penalty0 1--28, 1969.

\bibitem[Grossmann et al.(2004)]{Grossmann:2004}
Siegfried Grossmann and Bruno Eckhardt and Detlef Lohse.
\newblock {Hundert Jahre Grenzschichtphysik}.
\newblock \emph{Physikjournal}, 3:\penalty0 10, 31--37, 2004.



\bibitem[Gruschwitz(1931)]{Gruschwitz:1931}
E.~Gruschwitz.
\newblock {Die trubulente Reibungsschicht bei Druckabfall und Druckanstieg}.
\newblock \emph{Ingenieur-Archiv}, 2:\penalty0 321--346, 1931.

\bibitem[Gruss and Ruerup(2011)]{Gruss:2011}
Peter Gruss and Reinhard R\"urup 
\newblock {Denkorte. Max-Planck-Gesellschaft und Kaiser-Wilhelm-Gesellschaft, Br\"uche und Kontinuit\"aten 1911Ð2011}.
\newblock \emph{Peter Gruss and Reinhard R\"urup (eds.) Dresden: Sandstein Verlagr}, 2011.


\bibitem[Hager(2003)]{Hager:2003}
W.~H. Hager.
\newblock {Blasius: A life in research and education}.
\newblock \emph{Experiments in Fluids}, 34:\penalty0 566--571, 2003.

\bibitem[Hahn et~al. (1904)]{HahnHerglotzEtAl:1904}
Hans Hahn, Gustav Herglotz, and Karl Schwarzschild.
\newblock {Über das Strömen des Wassers in Röhren und Kanälen}.
\newblock \emph{Zeitschrift für Mathematik und Physik}, 51:\penalty0 411--426,
  1904.
 
\bibitem[Hamel(1943)]{Hamel:1943}
H. Hamel
\newblock {Streifenmethode und \"ahnlichkeitsbetrachtungen zur turbulenten Bewegung}.
\newblock \emph{Abh. preuss. Akad. Wiss. physik.-math.}, 8,  1943.
  


\bibitem[Heim et al. (2009)]{Heim:2009}
Susanne Heim, Carola Sachse and Mark Walker (eds.)
\newblock {The Kaiser Wilhelm Society under National Socialism}.
\newblock {Cambridge University Press,Cambridge, New York, 2009.}

\bibitem[Heisenberg (1948)]{Heisenberg:1948}
W. Heisenberg
\newblock {Zur statistischen Theorie der Turbulenz.
\newblock \emph{Zeitschrift für Physik}, 124:\penalty0 628--657,
  1948.
  
  \bibitem[Heisenberg (1958E)]{Heisenberg:1958E}
W. Heisenberg
\newblock {On the Statistical Theory of Turbulence.
\newblock \emph{NACA-TM-1431},1958.

\bibitem[Hensel(1989)]{Hensel:1989}
Susann Hensel.
\newblock \emph{Mathematik und Technik im 19. Jahrhundert in Deutschland.
  Soziale Auseinandersetzungen und philosophische Problematik}, chapter Die
  Auseinandersetzungen um die mathematische Ausbildung der Ingenieure an den
  Technischen Hochschulen in Deutschland Ende des 19. Jahrhunderts, pages
  1--111.
\newblock Vandenhoeck und Ruprecht, Göttingen, 1989.

\bibitem[Hopf(1910)]{Hopf:1910}
Ludwig Hopf.
\newblock \emph{Hydrodynamische Untersuchungen: Turbulenz bei einem Flusse.
  Über Schiffswellen. Inaugural-Dissertation}.
\newblock Barth, Leipzig, 1910.

\bibitem[Jakob and Erk(1924)]{JakobErk:1924}
M.~Jakob and S.~Erk.
\newblock {Der Druckabfall in glatten Rohren und die Durchflussziffer von
  Normaldüsen}.
\newblock \emph{Forschungsarbeiten auf dem Gebiete des Ingenieurwesens}, 267,
  1924.

\bibitem[Kármán(1957)]{Karman:1957}
Theodore von Kármán
\newblock {Aerodynamcis:Selected Topics in the Light of Their Historical Development}.
\newblock \emph{Cornell University Press}, Ithaca, New York, 1957.


\bibitem[Kármán and Howard(1938)]{KarmanHoward:1938}
Theodore von Kármán and Leslie Howard
\newblock {On the Statistical Theory of Isotropic Turbulence}.
\newblock \emph{Proc. R. Soc. Lond. A},164\penalty0 192-215, 1938.



\bibitem[Kempf(1929)]{Kempf:1929}
Günther Kempf.
\newblock {Neue Ergebnisse der Widerstandsforschung}.
\newblock \emph{Werft, Reederei, Hafen}, 10:\penalty0 234--239, 1929.

\bibitem[Kempf(1932)]{Kempf:1932}
Günther Kempf.
\newblock {Weitere Reibungsergebnisse an ebenen glatten und rauhen Flächen}.
\newblock \emph{G. Kempf, E. Foerster (Hrsg.): Hydromechanische Probleme des
  Schiffsantriebs. Hamburg}, pages 74--82, 1932.


\bibitem[Kuethe (1988)]{Kuethe:1988}
A.~M.~Kuethe
\newblock {THE FIRST TURBULENCE MEASUREMENTS: A Tribute to Hugh L. Dryden}.
\newblock \emph{Ann. Rev. Fluid Mech.}, 20:\penalty0 1--3, 1988.

\bibitem[Klein(1910)]{Klein:1910}
Felix Klein.
\newblock {Über die Bildung von Wirbeln in reibungslosen Flüssigkeiten}.
\newblock \emph{Zeitschrift für Mathematik und Physik}, 59:\penalty0 259--262,
  1910.


\bibitem[Lorentz(1897)]{Lorentz:1897}
Hendrik~Antoon Lorentz.
\newblock {Over den weerstand dien een vloeistofstroom in eene cilindrische
  buis ondervindt}.
\newblock \emph{Versl. K. Akad. Wet. Amsterdam}, 6:\penalty0 28--49, 1897.

\bibitem[Lorentz(1907)]{Lorentz:1907}
Hendrik~Antoon Lorentz.
\newblock {Über die Entstehung turbulenter Flüssigkeitsbewegungen und über den
  Einfluss dieser Bewegungen bei der Strömung durch Röhren}.
\newblock \emph{Hendrik Antoon Lorentz: Abhandlungen über theoretische Physik.
  Leipzig: Teubner}, 1:\penalty0 43--71, 1907.


\bibitem[Lumley and Panofsky (1964)]{LumleyPanofsky:1964}
John~Leask~Lumley, Hans~A.~Panofsky
\newblock {The structure of atmospheric turbulence}.
\newblock \emph{Interscience Monographs and Texts in Physics and Astronomy, New York: Wiley},1964.

\bibitem[Maier(2002)]{Maier:2002}
Helmut Maier (ed.).
\newblock \emph{Rüstungsforschung im Nationalsozialismus. Organisation, Mobilisierung und Entgrenzung der Technikwissenschaften}.
\newblock Wallstein, Göttingen, 2002.

\bibitem[Maier(2007)]{Maier:2007}
Helmut Maier.
\newblock \emph{Forschung als Waffe. R\"stungsforschung in der Kaiser-Wilhelm-Gesellschaft und das Kaiser-Wilhelm-Institut f\"ur Metallforschung 1900--1945/48}.
\newblock Wallstein, Göttingen, 2007.


\bibitem[Manegold(1970)]{Manegold:1970}
Karl-Heinz Manegold.
\newblock \emph{Universität, Technische Hochschule und Industrie: Ein Beitrag
  zur Emanzipation der Technik im 19. Jahrhundert unter besonderer
  Berücksichtigung der Bestrebungen Felix Kleins}.
\newblock Duncker und Humblot, Berlin, 1970.

\bibitem[Meier(2006)]{Meier:2006}
Gerd E.~A. Meier.
\newblock {Prandtl's Boundary Layer Concept and the Work in Göttingen}.
\newblock \emph{G. E. A. Meier, K. R. Sreenivasan (eds.): IUTAM Symposium on
  One Hundred Years of Boundary Layer Research. Proceedings of the IUTAM
  Symposium held at DLR-Göttingen, Germany, August 12-14, 2004. Dordrecht:
  Springer}, pages 1--18, 2006.

\bibitem[Munk(1917)]{Munk:1917}
Max Munk.
\newblock {Bericht über Luftwiderstandsmessungen von Streben. Mitteilung 1 der
  Göttinger Modell-Versuchsanstalt für Aerodynamik}.
\newblock \emph{Technische Berichte. Herausgegeben von der Flugzeugmeisterei
  der Inspektion der Fliegertruppen. Heft Nr. 4 (1. Juni 1917)}, pages 85--96,
  Tafel XXXX--LXIII, 1917.
  
\bibitem[Nagib et al. (2007)]{Nagib:2007}
Hassan M Nagib, Kapil A Chauhan and Peter A Monkewitz.
\newblock { Approach to an asymptotic state for zero pressure gradient turbulent boundary layers}.
\newblock \emph{Phil. Trans. R. Soc. A}, 755 --  770,
2007.


\bibitem[Nikuradse(1926)]{Nikuradse:1926}
Johann Nikuradse.
\newblock {Untersuchungen über die Geschwindigkeitsverteilung in turbulenten
  Strömungen}.
\newblock \emph{Forschungsarbeiten auf dem Gebiete des Ingenieurwesens}, 281,
  1926.

\bibitem[Nikuradse(1930)]{Nikuradse:1930}
Johann Nikuradse.
\newblock {Über turbulente Wasserströmungen in geraden Rohren bei sehr grossen Reynoldsschen Zahlen}.
\newblock \emph{Vorträge aus dem Gebiete der Aerodynamik und verwandter Gebiete (Aachen 1929). Herausgegeben von A. Gilles, L. Hopf und Th. v. Kármán. Berlin: Springer, 1930}, 63-69.

\bibitem[Nikuradse(1932)]{Nikuradse:1932}
Johann Nikuradse.
\newblock {Gesetzmässigkeiten der turbulenten Strömung in glatten Rohren}.
\newblock \emph{Forschungsarbeiten auf dem Gebiete des Ingenieurwesens}, 356,
  1932.

\bibitem[Nikuradse(1933)]{Nikuradse:1933}
Johann Nikuradse.
\newblock {Strömungsgesetze in rauhen Rohren}.
\newblock \emph{Forschungsarbeiten auf dem Gebiete des Ingenieurwesens}, 3611,
  1933.

\bibitem[Noether(1921)]{Noether:1921}
Fritz Noether.
\newblock {Das Turbulenzproblem}.
\newblock \emph{Zeitschrift für Angewandte Mathematik und Mechanik (ZAMM)},
  pages 125--138, 218--219, 1921.
  
\bibitem[O'Malley Jr.(2010)]{Malley:2010}
Robert E. O'Malley Jr..
\newblock {Singular Perturbation Theory: A Viscous Flow out of Göttingen}.
\newblock \emph{Annu. Rev. Fluid Mech.},
  pages 1--17, 42, 2010.
 
  
\bibitem[Oswatitsch and Wieghardt(1987)]{Oswatitsch:1987}
K.~Oswatitsch and K.~Wieghardt.
\newblock {Ludwig Prandtl and his Kaiser-Wilhelm-Institut}.
\newblock \emph{Annu. Rev. Fluid Mech.},
  pages 1--26, 19, 1987.

\bibitem[Prandtl(1905)]{Prandtl:1905}
Ludwig Prandtl.
\newblock {Über Flüssigkeitsbewegung bei sehr kleiner Reibung}.
\newblock \emph{Verhandlungen des III. Internationalen Mathematiker-Kongresses,
  Heidelberg}, pages 484--491, 1905.
\newblock LPGA 2, 575-584.

\bibitem[Prandtl(1910)]{Prandtl:1910}
Ludwig Prandtl.
\newblock {Eine Beziehung zwischen Wärmeaustausch und Strömungswiderstand der
  Flüssigkeiten}.
\newblock \emph{Physikalische Zeitschrift}, 11:\penalty0 1072--1078, 1910.

\bibitem[Prandtl(1914)]{Prandtl:1914}
Ludwig Prandtl.
\newblock {Der Luftwiderstand von Kugeln}.
\newblock \emph{Nachrichten der Gesellschaft der Wissenschaften zu Göttingen,
  Mathematisch-physikalische Klasse}, pages 177--190, 1914.
\newblock (LPGA 2, 597-608).

\bibitem[Prandtl(1921)]{Prandtl:1921}
Ludwig Prandtl.
\newblock {Bemerkungen über die Entstehung der Turbulenz}.
\newblock \emph{Zeitschrift für Angewandte Mathematik und Mechanik (ZAMM)},
  1:\penalty0 431--436, 1921.
  
  
  \bibitem[Prandtl(1921{\natexlab{a}})]{Prandtl:1921a}
Ludwig Prandtl.
\newblock \emph{Ergebnisse der Aerodynamischen Versuchsanstalt zu Göttingen.
  München, Berlin: Oldenbourg}, 1:1921{\natexlab{a}}.
  
  

\bibitem[Prandtl(1922)]{Prandtl:1922}
Ludwig Prandtl.
\newblock {Bemerkungen über die Entstehung der Turbulenz}.
\newblock \emph{Physikalische Zeitschrift}, 23:\penalty0 19--25, 1922.

\bibitem[Prandtl(1925E)]{Prandtl:1925E}
Ludwig Prandtl.
\newblock {Aufgaben der Str\"omungsforschung:Tasks of air flow research}.
\newblock \emph{Die Naturwissenschaften. Vol. 14, No. 16, pp. 355-358, 1925.; NACA Technical Memorandum 365}, 1926.


\bibitem[Prandtl(1925)]{Prandtl:1925}
Ludwig Prandtl.
\newblock {Bericht über Untersuchungen zur ausgebildeten Turbulenz}.
\newblock \emph{ZAMM}, 5:\penalty0 136--139, 1925.
\newblock LPGA 2, 714-718.

\bibitem[Prandtl(1949E)]{Prandtl:1949E}
Ludwig Prandtl.
\newblock {Bericht über Untersuchungen zur ausgebildeten Turbulenz: Report on Investigation of Developed Turbulence}.
\newblock \emph{NACA-TM-1231},1949.


\bibitem[Prandtl(1926)]{Prandtl:1926}
Ludwig Prandtl.
\newblock {Bericht über neuere Turbulenzforschung}.
\newblock \emph{Hydraulische Probleme. Berlin: VDI-Verlag}, pages 1--13, 1926.
\newblock LPGA 2, 719-730.

\bibitem[Prandtl(1926b)]{Prandtl:1926b}
Ludwig Prandtl.
\newblock {Klein und die Angewandten Wissenschaften}.
\newblock \emph{Sitzungsberichte der Berliner Mathematischen Gesellschaft}, pages 81--87, 1926.
\newblock LPGA 2, 719-730.

\bibitem[Prandtl(1927{\natexlab{a}})]{Prandtl:1927}
Ludwig Prandtl.
\newblock {Über den Reibungswiderstand strömender Luft}.
\newblock \emph{Ergebnisse der Aerodynamischen Versuchsanstalt zu Göttingen.
  München, Berlin: Oldenbourg}, 3:\penalty0 1--5, 1927{\natexlab{a}}.
\newblock LPGA 2, 620-626.

\bibitem[Prandtl(1927{\natexlab{b}})]{Prandtl:1927a}
Ludwig Prandtl.
\newblock {Über die ausgebildete Turbulenz}.
\newblock \emph{Verhandlungen des II. Internationalen Kongresses für Technische
  Mechanik. Zürich: Füßli}, pages 62--75, 1927{\natexlab{b}}.
\newblock LPGA 2, 736-751.
\newblock NACA-TM-435/ version in English.

\bibitem[Prandtl(1930)]{Prandtl:1930}
Ludwig Prandtl.
\newblock {Vortrag in Tokyo}.
\newblock \emph{Journal of the Aeronautical Research Institute, Tokyo, Imperial
  University}, 5:65:\penalty0 12--24, 1930.
\newblock LPGA 2, 788-797.

\bibitem[Prandtl(1931)]{Prandtl:1931}
Ludwig Prandtl.
\newblock \emph{Abriss der Strömungslehre}.
\newblock Braunschweig, Vieweg, 1931.


\bibitem[Prandtl(1932)]{Prandtl:1932}
Ludwig Prandtl.
\newblock {Zur turbulenten Strömung in Rohren und längs Platten}.
\newblock \emph{Ergebnisse der Aerodynamischen Versuchsanstalt zu Göttingen},
  4:\penalty0 18--29, 1932.
\newblock LPGA 2, 632-648.

\bibitem[Prandtl(1933)]{Prandtl:1933}
Ludwig Prandtl.
\newblock {Neuere Ergebnisse der Turbulenzforschung}.
\newblock \emph{Zeitschrift des Vereines Deutscher Ingenieure}, 77:\penalty0
  105--114, 1933.
\newblock LPGA 2, 819-845.

\bibitem[Prandtl(1933E1)]{Prandtl:1933E1}
Ludwig Prandtl.
\newblock {Neuere Ergebnisse der Turbulenzforschung: Recent results of turbulence research}.
\newblock \emph{NACA Technical Memorandum 720.; Zeitschrift des Vereines deutscher Ingenieure. Vol. 7, No. 5, pp. 105-114, February 4, 1933}, 1933.


\bibitem[Prandtl(1933E2)]{Prandtl:1933E2}
Ludwig Prandtl.
\newblock {Herstellung einwandfreier Luftstrome (Windkanale):Attaining a steady air stream in wind tunnels}.
\newblock \emph{Handbuch der Experimentalphysik. Vol. 4, Part 2, pp. 65-106, 1932; NACA Technical Memorandum 726}, 1933.


\bibitem[Prandtl(1934)]{Prandtl:1934}
Ludwig Prandtl.
\newblock {Anwendung der turbulenten Reibungsgesetze auf atmosphärische
  Strömungen}.
\newblock \emph{Proceedings of the Fourth International Congress of Applied
  Mechanics, Cambridge}, pages 238--239, 1934.
\newblock LPGA 3, 1098-1099.

\bibitem[Prandtl(1938)]{Prandtl:1938}
L. Prandtl.
\newblock {Beitrag zum Turbulenzsymposium}.
\newblock \emph{Proceedings of the Fifth International Congress on Applied
  Mechanics, Cambridge Mass.}, edited by J.P. Den Hartog and H. Peters, John Wiley, New York :\penalty0 340--346, 1938.
  \newblock LPGA 2, 856-868.
 
\bibitem[Prandtl(1942a)]{Prandtl:1942a}
Ludwig Prandtl.
\newblock \emph{Führer durch die Strömungslehre}.
\newblock Braunschweig, Vieweg, 1942.

 
 \bibitem[Prandtl(1942b)]{Prandtl:1942b}
Ludwig Prandtl.
\newblock {Zur turbulenten Strömung in Rohren und längs Platten}.
\newblock \emph{Ergebnisse der Aerodynamischen Versuchsanstalt zu Göttingen},
  4:\penalty0 18--29, 1932.
\newblock LPGA 2, 632-648.

 \bibitem[Prandtl(1942c)]{Prandtl:1942c}
Ludwig Prandtl.
\newblock {Bemerkungen zur Theorie der Freien Turbulenz}.
\newblock \emph{Zeitschrift für Angewandte Mathematik und Mechanik (ZAMM)},22:241-243, 1942
\newblock LPGA 2, 869-873.


\bibitem[Prandtl(1945)]{Prandtl:1945}
Ludwig Prandtl.
\newblock {Über die Rolle der Z\"ahigkeit  im Mechanismus  der ausgebildete Turbulenz: The Role of Viscosity in the Mechanism of Developed  Turbulence}.
\newblock \emph{GOAR 3712, DLR Archive}.





\bibitem[Prandtl(1948a)]{Prandtl:1948a}
Ludwig Prandtl.
\newblock { Turbulence}.
\newblock \emph{Albert Betz (ed.), FIAT Review of German Science 1939-1946: Hydro-- and Aerodynamics, Office of Military Government for Germany Field Information Agency Technical}, 55--78, 1948.


\bibitem[Prandtl(1948b)]{Prandtl:1948b}
Ludwig Prandtl.
\newblock {Mein Weg zu den Hydrodynamischen Theorien: My Path to the Hydrodynamic Theories}.
\newblock \emph{Physikalische Bl\"atter}, 3,89--92, 1948.
\newblock LPGA 3, 1604-1608.

\bibitem[Prandtl(1948c)]{Prandtl:1948c}
Ludwig Prandtl.
\newblock \emph{Führer durch die Strömungslehre}.
\newblock Braunschweig, Vieweg, 1948.



\bibitem[Prandtl and Eisner(1932)]{PrandtlEisner:1932}
Ludwig Prandtl and Franz Eisner.
\newblock {Nachtrag zum 'Reibungswiderstand'}.
\newblock \emph{G. Kempf, E. Foerster (Hrsg.): Hydromechanische Probleme des
  Schiffsantriebs. Hamburg}, page 407, 1932.

\bibitem[Prandtl et~al. (1932)]{PrandtlETAL:1932}
Ludwig Prandtl et~al.
\newblock {Erörterungsbeiträge}.
\newblock \emph{G. Kempf, E. Foerster (Hrsg.): Hydromechanische Probleme des
  Schiffsantriebs. Hamburg}, pages 87--98, 1932.

\bibitem[Prandtl and Schlichting(1934)]{PrandtlSchlichting:1934}
Ludwig Prandtl and Hermann Schlichting.
\newblock {Das Widerstandsgesetz rauher Platten}.
\newblock \emph{Werft, Reederei, Hafen}, 21:\penalty0 1--4, 1934.
\newblock LPGA 2, 649-662.

\bibitem[Prandtl and Reichardt(1934)]{Prandtl-Reichardt:1934}
Ludwig Prandtl and Hans Reichardt.
\newblock {Einfluss von Wärmeschichtung auf Eigenschaften einer turbulenten Strömung}.
\newblock \emph{Deutsche Forschung}, 15:\penalty0 110--121, 1934.
\newblock LPGA 2, 846--855.


  
\bibitem[Prandtl and Wieghardt(1945)]{Prandtl-Wieghardt:1945}
Ludwig Prandtl and Karl Wieghardt.
\newblock {Über ein neues Formelsystem für die ausgebildete Turbulenz}.
\newblock \emph{Nachrichten der Akademie der Wissenschaften zu Göttingen, Mathematisch-physikalische Klasse}, 6-19, 1945.
\newblock LPGA 2, 874--887.




\bibitem[Rayleigh(1887)]{Rayleigh:1887}
Lord Rayleigh.
\newblock {On the stability or instability of certain fluid motions II}.
\newblock \emph{Proceedings of the London Mathematical Society}, 19:\penalty0
  67--74, 1887.
\newblock Reprinted in Scientific Papers by John William Strutt, Baron
  Rayleigh, vol. III (1887-1892). Cambridge: Cambridge University Press, 1902,
  pp. 17-23.

\bibitem[Reichardt(1933)]{Reichardt:1933}
H.~ Reichardt.
\newblock {Die quadratischen Mittelwerte der L{\"a}ngsschwankungen in der turbulenten Kanalstr{\"o}mung}.
\emph{Zeitschrift für Angewandte Mathematik und Mechanik (ZAMM)},3:177--180, 1933

\bibitem[Reichardt(1934)]{Reichardt:1934}
H.~ Reichardt.
\newblock {Berichte aus den einzelnen Instituten. Physikalisch-Chemisch-Technische Institute}.
\emph{Naturwissenschaften},22:351, 1934.


\bibitem[Reichardt(1935)]{Reichardt:1935}
H.~ Reichardt.
\newblock {Die Torsionswaage als Mikromanometer}.
\emph{Zschr. f. Instrumentenkunde},55:23-33, 1935.

\bibitem[Reichardt(1948E)]{Reichardt:1948E}
H.~ Reichardt.
\newblock {The Torsion Balance as a Micromanometer}.
\emph{NRC-TT-84, 1948-11-13}.

\bibitem[Reichardt(1938a)]{Reichardt:1938a}
H.~ Reichardt.
\newblock {Vortr\"age aus dem Gebiet der Aero- und Hydrodynamik. \"Uber das Messen turbulenter L\"angs- und Querschwankungen}.
\emph{Zeitschrift für Angewandte Mathematik und Mechanik (ZAMM)},18:358-361, 1938

\bibitem[Motzfeld(1938)]{Motzfeld:1938}
H.~Motzfeld .
\newblock {Vortr\"age aus dem Gebiet der Aero- und Hydrodynamik. Frequenzanalyse turbulenter Schwankungen}.
\emph{Zeitschrift für Angewandte Mathematik und Mechanik (ZAMM)},18: 362-365, 1938


\bibitem[Reichardt(1941)]{Reichardt:1941}
H.~ Reichardt.
\newblock  \"Uber die Theorie der  freien Turbulenz}.
\emph{Zeitschrift für Angewandte Mathematik und Mechanik (ZAMM)},21:257 -- 264, 1941

\bibitem[Reichardt(1942)]{Reichardt:1942}
H.~ Reichardt.
\newblock  \"Gesaetm\"assigkeiten der freien Turbulenz}.
\emph{VDI Forschungsheft },414, 22 pages 1942

\bibitem[Reichardt(1944)]{Reichardt:1944}
H.~ Reichardt.
\newblock {Impuls- und Wärmeaustausch in freier Turbulenz}.
\emph{Zeitschrift für Angewandte Mathematik und Mechanik (ZAMM)},24:268 -- 272, 1944



\bibitem[Reichardt(1951E)]{Reichardt:1951E}
H.~ Reichardt.
\newblock {On the Recording of Turbulent Longitudinal and Transverse Fluctuations}.
\emph{NACA-TM-1313},1951.

\bibitem[Reichardt(1938b)]{Reichardt:1938b}
H.~ Reichardt.
\newblock {Messungen turbulenter Spannungen}.
\emph{Naturwissenschaften},26:404--408, 1938.


\bibitem[Rotta(1990)]{Rotta:1990}
Julius~C. Rotta.
\newblock \emph{Die Aerodynamische Versuchsanstalt in Göttingen, ein Werk
  Ludwig Prandtls. Ihre Geschichte von den Anfängen bis 1925}.
\newblock Vandenhoeck und Ruprecht, Göttingen, 1990.

\bibitem[Rotta(2000)]{Rotta:2000}
Julius~C. Rotta.
\newblock {Ludwig Prandtl und die Turbulenz}.
\newblock \emph{Gerd E. A. Meier (ed.), Ludwig Prandtl, ein Führer in der
  Strömungslehre: Biographische Artikel zum Werk Ludwig Prandtls}, pages
  53--123, 2000.

\bibitem[Sachse and Walker(2005)]{SachseWalker:2005}
Carola Sachse and Mark Walker (eds.)
\newblock {Politics and Science in Wartime:Comparative International Perspectives on Kaiser Wilhelm Institutes}.
\newblock \emph{Osiris} 20, University of Chicago Press, 2005.
  


\bibitem[Saffman(1992)]{Saffman:1992}
P.~G. Saffman.
\newblock \emph{Vortex Dynamics}.
\newblock Cambridge University Press, Cambridge, 1992.

\bibitem[Schiller(1921)]{Schiller:1921}
Ludwig Schiller.
\newblock {Experimentelle Untersuchungen zum Turbulenzproblem}.
\newblock \emph{Zeitschrift für Angewandte Mathematik und Mechanik (ZAMM)},
  1:\penalty0 436--444, 1921.

\bibitem[Schlichting(1933)]{Schlichting:1933}
Hermann Schlichting.
\newblock Zur Entstehung der Turbulenz bei der Plattenströmung.
\newblock \emph{Nachrichten der Gesellschaft der Wissenschaften zu Göttingen},
  pages 181--208, 1933.

\bibitem[Schlichting(1949)]{Schlichting:1949}
Hermann Schlichting.
\newblock Lecture series "boundary layer theory", part ii: Turbulent flows.
\newblock \emph{NACA, TM No. 1218}, 1949.
\newblock Translation of "Vortragsreihe" W.S. 1941/42,
  Luftfahrtforschungsanstalt Hermann Göring, Braunschweig.
  
\bibitem[Schmaltz(2005)]{Schmaltz:2005}
Florian Schmaltz.
\newblock \emph{Kampfstoff-Forschung im Nationalsozialismus. Zur Kooperation von Kaiser-Wilhelm-Instituten, Militär und Industrie}.
\newblock Wallstein, Göttingen, 2005.
  
  
\bibitem[Schultz-Grunow(1940)]{Schultz-Grunow:1940}
Fritz Schultz-Grunow.
\newblock {Neues Reibungswiderstandsgesetz für glatte Platten}.
\newblock \emph{Luftfahrtforschung}, 17:\penalty0 239--246, 1940.

\bibitem[Schultz-Grunow(1941E)]{Schultz-Grunow:1941E}
Fritz Schultz-Grunow.
\newblock {New Frictional Resistance Law for Smooth Plates}.
\newblock \emph{NACA-TM-986}, 1941.

\bibitem[Simmons, Salter, and Taylor(1938)]{Simmons-Salter-Taylor:1938}
L. F. G. Simmons, C. Salter, G. I. Taylor 
\newblock {An Experimental Determination of the Spectrum of Turbulence}.
\newblock \emph{ R. Soc. Lond. A},165:\penalty0 73--89, 1938.


\bibitem[Sommerfeld(1935)]{Sommerfeld:1935}
Arnold Sommerfeld.
\newblock {Zu L. Prandtls 60. Geburtstag am 4. Februar 1935}.
\newblock \emph{ZAMM}, 15:\penalty0 1--2, 1935.


\bibitem[Spalding(1991)]{Spalding:1991}
Arnold Sommerfeld.
\newblock {Kolmogorov's Two-Equation Model of Turbulence}.
\newblock \emph{Proc. Roy. Soc. Math. Phys. Sci.}, 434:\penalty0 211--216, 1991.


\bibitem[Schuh(1945)]{Schuh:1945}
H. Schuh
\newblock {Die Messungen sehr kleiner Windschwankungen (Windkanalturbulenz)}.
\newblock \emph{Untersuchungen und Mitteilungen der deutschen Luftfahrtforschung}, 6623, 1945.



\bibitem[Schuh(1946)]{Schuh:1946}
H. Schuh
\newblock {Windschwankungsmessungen mit Hitzdr\"ahten}.
\newblock \emph{AVA Monographien}, D1 4.3.

\bibitem[Tani(1977)]{Tani:1977}
Itiro Tani.
\newblock {History of Boundary-Layer Theory}.
\newblock \emph{Annual Review of Fluid Mechanics}, 9:\penalty0 87--111, 1977.

\bibitem[Tietjens(1925)]{Tietjens:1925}
Oskar Tietjens.
\newblock {Beiträge zur Entstehung der Turbulenz}.
\newblock \emph{ZAMM}, 5:\penalty0 200--217, 1925.

\bibitem[Taylor(1923)]{Taylor:1923}
G.I. Taylor.
\newblock {Stability of a Viscous Liquid Contained between Two Rotating Cylinders}.
\newblock \emph{Phys. Trans. Royal Soc.},223:\penalty0 289--343, 1929.


\bibitem[Taylor(1935)]{Taylor:1935}
G.I. Taylor.
\newblock {Turbulence in a Contracting Stream}.
\newblock \emph{ZAMM}, 15:\penalty0 91--96, 1935.

\bibitem[Taylor(1935a)]{Taylor:1935a}
G.I. Taylor.
\newblock {Statistical Theory of Turbulence}.
\newblock \emph{Proc. R. Soc. Lond. A},151:\penalty0 421--444, 1935.

\bibitem[Taylor(1935b)]{Taylor:1935b}
G.I. Taylor.
\newblock {Statistical Theory of Turbulence II}.
\newblock \emph{Proc. R. Soc. Lond. A},151:\penalty0 444--454, 1935.

\bibitem[Taylor(1935c)]{Taylor:1935c}
G.I. Taylor.
\newblock {Statistical Theory of Turbulence. III. Distribution of Dissipation of Energy in a Pipe over Its Cross-Section}.
\newblock \emph{Proc. R. Soc. Lond. A},151:\penalty0 455--464, 1935.

\bibitem[Taylor(1935d)]{Taylor:1935d}
G.I. Taylor.
\newblock {Statistical Theory of Turbulence. IV. Diffusion in a Turbulent Air Stream}.
\newblock \emph{Proc. R. Soc. Lond. A},151:\penalty0 465--478, 1935.

\bibitem[Taylor(1936)]{Taylor:1936}
G.I. Taylor.
\newblock {Correlation Measurements in a Turbulent Flow through a Pipe}.
\newblock \emph{Proc. R. Soc. Lond. A},157:\penalty0 537--546, 1936.


\bibitem[Taylor(1938a)]{Taylor:1938a}
G.I. Taylor.
\newblock {The Spectrum of Turbulence}.
\newblock \emph{Proc. R. Soc. Lond. A},164:\penalty0 476--490, 1938.

\bibitem[Taylor(1938b)]{Taylor:1938b}
G.I. Taylor.
\newblock {Some Recent Developments in the Study of Turbulence}.
\newblock \emph{Proceedings of the Fifth International Congress on Applied
  Mechanics, Cambridge Mass.}, edited by J.P. Den Hartog and H. Peters, John Wiley, New York :\penalty0 294--310, 1938.
  
\bibitem[Tollmien(1926)]{Tollmien:1926}
Walter Tollmien.
\newblock {Berechnung turbulenter Ausbreitungsvorgänge}.
\newblock \emph{ZAMM}, 6:\penalty0 468--478, 1926.

\bibitem[Tollmien(1929)]{Tollmien:1929}
Walter Tollmien.
\newblock {Über die Entstehung der Turbulenz}.
\newblock \emph{Nachrichten der Gesellschaft der Wissenschaften zu Göttingen},
  pages 21--44, 1929.

\bibitem[Trischler(1994)]{Trischler:1994}
Helmuth Trischler.
\newblock {Self-mobalization or resistance? Aeronautical Research and National Socialism}.
\newblock \emph{Science, Technology and National Socialism,  Monika Renneberg and Mark Walter, editors, Cambridge:Cambridge, 1994}, 72--87.


\bibitem[Vogel-Prandtl(1993)]{Vogel-Prandtl:1993}
Johanna Vogel-Prandtl.
\newblock \emph{Ludwig Prandtl. Ein Lebensbild. Erinnerungen. Dokumente}.
\newblock Max-Planck-Institut für Strömungsforschung, Göttingen, 1993.
\newblock (= Mitteilungen aus dem MPI für Strömungsforschung, Nr. 107).

\bibitem[von Karman(1932)]{Karman:1932}
Theodore von Karman.
\newblock {Theorie des Reibungswiderstandes}.
\newblock \emph{G. Kempf, E. Foerster (Hrsg.): Hydromechanische Probleme des
  Schiffsantriebs. Hamburg}, pages 50--73, 1932.
\newblock CWTK 2, 394-414.

\bibitem[von Kármán(1921)]{Karman:1921}
Theodore von Kármán.
\newblock {Über laminare und turbulente Reibung}.
\newblock \emph{ZAMM}, 1:\penalty0 233--252, 1921.
\newblock CWTK 2, 70-97.

\bibitem[von Kármán(1924)]{Karman:1924}
Theodore von Kármán.
\newblock {Über die Oberflächenreibung von Flüssigkeiten}.
\newblock \emph{Theodore von Kármán und T. Leci-Civita (Hrsg.): Vorträge aus
  dem Gebiete der Hydro- und Aerodynamik (Innsbruck 1922). Berlin: Springer},
  pages 146--167, 1924.
\newblock CWTK 2, 133-152.

\bibitem[von Kármán(1930{\natexlab{a}})]{Karman:1930}
Theodore von Kármán.
\newblock {Mechanische Ähnlichkeit und Turbulenz}.
\newblock \emph{Nachrichten von der Gesellschaft der Wissenschaften zu
  Göttingen, Mathematisch-Physikalische Klasse}, pages 58--76,
  1930{\natexlab{a}}.
\newblock CWTK 2, 322-336.

\bibitem[von Kármán(1930{\natexlab{b}})]{Karman:1930a}
Theodore von Kármán.
\newblock {Mechanische Ähnlichkeit und Turbulenz}.
\newblock \emph{Proceedings of the Third International Congress of Applied
  Mechanics, Stockholm}, 1930{\natexlab{b}}.
\newblock CWTK 337-346.

\bibitem[von Kármán(1934)]{Karman:1934}
Theodore von Kármán.
\newblock {Turbulence and Skin Friction}.
\newblock \emph{Journal of the Aeronautical Sciences}, 1:\penalty0 1--20, 1934.
\newblock CWTK 3, 20-48.

\bibitem[von Kármán(1967)]{Karman:1967}
Theodore von Kármán.
\newblock \emph{The Wind and Beyond}.
\newblock Little, Brown and Company, Boston/Toronto, 1967.
\newblock (with Lee Edson).

\bibitem[von Mises(1921)]{Mises:1921}
Richard von Mises.
\newblock {Über die Aufgaben und Ziele der angewandten Mathematik}.
\newblock \emph{Zeitschrift für Angewandte Mathematik und Mechanik (ZAMM)},
  1:\penalty0 1--15, 1921.
  
  
  
\bibitem[Weizs\"acker(1948)]{Weizs\"acker:1948}
C. F. von Weizs\"acker.
\newblock {Das Spektrum der Turbulenz bei grossen Reynoldsschen Zahlen}.
\newblock \emph{Zeitschrift für Physik}, 124, 614--627, 1948

  
  

\bibitem[Wieghardt(1941)]{Wieghardt:1941}
Karl Wieghardt.
\newblock {Zusammenfassender Bericht \"uber Arbeiten zur statistischen Turbulenztheorie}.
\newblock \emph{Luftfahrt-Forschung}, FB 1563, 1941.

\bibitem[Wieghardt(1942E)]{Wieghardt:1942E}
Karl Wieghardt.
\newblock {Correlation of data on the statistical theory of turbulence}.
\newblock \emph{NACA-TM-1008},1942.


\bibitem[Wieghardt(1942)]{Wieghardt:1942}
Karl Wieghardt.
\newblock {Erhöhung des turbulenten Reibungswiderstandes durch
  Oberflächenstörungen}.
\newblock \emph{ZWB}, FB 1563, 1942.

\bibitem[Wieghardt(1943)]{Wieghardt:1943}
Karl Wieghardt.
\newblock {Über die Wandschubspannung in turbulenten Reibungsschichten bei
  veränderlichem Aussendruck}.
\newblock \emph{ZWB}, UM 6603, 1943.

\bibitem[Wieghardt(1944)]{Wieghardt:1944}
Karl Wieghardt.
\newblock {Zum Reibungswiderstand rauher Platten}.
\newblock \emph{ZWB}, UM 6612, 1944.

\bibitem[Wieghardt and Tillmann(1944)]{WieghardtTillmann:1944}
Karl Wieghardt and W.~Tillmann.
\newblock {Zur turbulenten Reibungsschicht bei Druckanstieg}.
\newblock \emph{ZWB}, UM 6617, 1944.

\bibitem[Wieghardt and Tillmann(1951E)]{WieghardtTillmann:1951E}
Karl Wieghardt and W.~Tillmann.
\newblock {On the turbulent friction layer for rising pressure}.
\newblock \emph{NACA-TM-1314}, 1951.

\bibitem[Wieghardt(1947)]{Wieghardt:1947}
Karl Weghardt.
\newblock {Der Rauhigkeitskanal des Kaiser Wilhelm-Instituts f\"r  Str\"omungsforschung in G\"ottingen }.
\newblock \emph{AVA-Monographien} D1 3.3, 1947.



\bibitem[Wieselsberger(1914)]{Wieselsberger:1914}
Carl Wieselsberger.
\newblock {Der Luftwiderstand von Kugeln}.
\newblock \emph{Zeitschrift für Flugtechnik und Motorluftschiffahrt},
  5:\penalty0 140--145, 1914.
  
  
  



\end{thebibliography}
\end{document}